\documentclass[12pt]{article}
\usepackage{amsmath}
\usepackage{graphicx,psfrag,epsf}
\usepackage{enumerate}
\usepackage{natbib}
\usepackage{url} 

\newcommand{\blind}{0}

\usepackage{amsmath,amssymb}
\usepackage{amsthm}
\usepackage{multirow}
\usepackage{natbib}
\usepackage{bm}
\usepackage{hyperref}
\usepackage[all]{xy}
\usepackage{graphicx,caption, subcaption}
\usepackage{float}
\usepackage{color}

\usepackage{xr}

\newtheorem{thm}{Theorem}[section] 

\newtheorem{definition}{Definition}[section]
\newcommand{\bed}{\begin{definition}}
\newcommand{\eed}{\end{definition}}

\newcommand{\rom}[1]{\uppercase\expandafter{\romannumeral #1\relax}}

\usepackage{bbm}

\newcommand{\bitem}{\begin{itemize}}
\newcommand{\eitem}{\end{itemize}}

\newcommand{\beqn}{\begin{equation}}
\newcommand{\eeqn}{\end{equation}}
\newcommand{\balign}{\begin{align}}
\newcommand{\ealign}{\end{align}}

\usepackage{amssymb}
\newcommand{\beq}{\begin{equation}}
\newcommand{\eeq}{\end{equation}}

\newcommand{\diag}{\mathrm{diag}}

\allowdisplaybreaks

\usepackage{enumitem}
\usepackage{makecell}

\usepackage{listings}
\begin{document}
	
\def\spacingset#1{\renewcommand{\baselinestretch}%
	{#1}\small\normalsize} \spacingset{1}


\if0\blind
{
	\title{\bf Co-citation and Co-authorship Networks of Statisticians}
	\author{Pengsheng Ji$^*$, Jiashun Jin$^{**}$, Zheng Tracy Ke$^\dagger$, Wanshan Li$^{**}$ \\
		\vspace{0.4 em} 
	University of Georgia$^*$, Carnegie Mellon University$^{**}$ and Harvard University$^\dagger$} 
 \date{}

	\maketitle
} \fi

\if1\blind
{
	\bigskip
	\bigskip
	\bigskip
	\begin{center}
		{\LARGE\bf Co-citation and Co-authorship Networks of Statisticians}
	\end{center}
	\medskip
} \fi

\bigskip
\begin{abstract} 
We collected and cleaned a large data set on publications in statistics.   The data set 
consists of the coauthor relationships and citation relationships  
of  $83,331$ papers published in $36$ representative 
journals in statistics, probability, and machine learning, spanning $41$ 
years. The data set allows us to construct many different networks, and 
motivates a number of research problems about the research patterns and trends, research impacts, and network topology  
of the statistics community. 
In this paper we focus on (i) using the citation relationships to estimate the research interests of authors, and (ii) using the coauthor relationships to 
study the network topology.  

Using  co-citation networks we constructed, we discover a  ``statistics triangle", 
reminiscent of the statistical philosophy triangle   \citep{Efron-Fisher}.  
We propose new approaches to constructing the 
``research map" of statisticians, as well as the ``research trajectory" for 
a given author to visualize his/her research interest evolvement. 
Using co-authorship networks we constructed, 
we discover a multi-layer community tree and produce a Sankey diagram to visualize the author migrations in different sub-areas. We also propose several new 
metrics for research diversity of individual authors.

We find that ``Bayes", ``Biostatistics",  and ``Nonparametric" 
are three primary areas in statistics. We also identify 
$15$ sub-areas, each of which can be viewed as a weighted average 
of the primary areas, and identify several underlying reasons for the formation of 
co-authorship communities.  We also find that the research interests of statisticians have evolved significantly in the 41-year time window we studied: some areas (e.g., biostatistics, high-dimensional data analysis, etc.) have become increasingly more popular. The research diversity of statisticians may be lower than we might have expected. For example, for the personalized 
networks of most authors, the $p$-values of the proposed  significance tests are relatively large.

\end{abstract} 


\noindent%

\noindent%

\vfill

\tableofcontents

\bigskip\bigskip

\noindent
{\bf Data link:}  \url{http://zke.fas.harvard.edu/MADStat.html}\\
 or \url{https://github.com/ZhengTracyKe/MADStat}

\vfill

\newpage
\spacingset{1.43} 

\section{Introduction} \label{sec:intro}  
In the past decades,  the size of the scientific community has grown substantially.  
 The rapid growth of the scientific community motivates many interesting Big Data projects, and one of them is  how to use the vast volume of publications of a  scientific field to delineate a complete picture of the research habits, trends, and impacts of this field. These studies are useful for examining 
national and global scientific publication-related activities, ranking  
universities, and making decisions of funding, 
promotions, and awards.

There are two main approaches to studying scientific publications,  the subjective approach and the quantitative approach. The subjective approach is more traditional, but it is time-consuming and susceptible to bias.  The quantitative approach (which uses statistical  tools for analyzing such data) is  comparably inexpensive, fast, objective, and transparent, and will  play an increasingly more important role \citep{Silverman2016}.  

From a statistical standpoint, most existing quantitative approaches are overly simple, using preliminary metrics (e.g.,  counts of papers or citations) for analysis. The h-index  and journal impact factor  are examples of more sophisticated approaches, but they are still not principled statistical methods. Statistical modeling of publication data is a significantly underdeveloped area, where we have only a small number of interesting papers, sparsely scattered over the spectrum, and typically, each focusing on only a specific problem.  
   
On the other hand, this can also be viewed as a golden opportunity for statisticians.  The publication data provide a valuable 
data resource,    important problems in science and social science,  and interesting Big Data projects that demand sophisticated statistical tools.   
Having seen such an opportunity,  Hall encouraged statisticians to take on a more active role in such research  \citep{Hall2011}.  
Hall's viewpoint is shared by \cite{Donoho2017},  among others.  In his illuminating paper ``50 Years of Data Science" \citep{Donoho2017},  Donoho predicted that ``science about data science" will become one of the major divisions of data science,  and one task of this division is to evaluate scientific research outputs. 

This paper is a response to the call by Hall and others. We contribute a large-scale high-quality data set on the publications of  statisticians and use it to showcase how modern statistical tools can be employed for analysis of such kind of data.

{\bf A new data set about the publications of statisticians}. 
We present a new data set about the publications of statisticians,   collected and cleaned by ourselves with enormous efforts. The data set consists of coauthor relationships and  citation relationships of  $83K$ research papers published in $36$ representative journals in statistics, probability, machine learning, and related fields, spanning $41$ years. See the table below.  More  information of these journals is presented in Table~B.1 of the supplement. 

\spacingset{1} 
\begin{center}
\centering
\scalebox{0.99}{
\begin{tabular}{cccc}
\hline
  \#journals &    time span & \#authors &  \#papers  \\
\hline
  $36$    &  $1975$-$2015$   &  $47, 311$  &  $83, 331$  \\
\hline
\end{tabular}
}
\end{center}
\spacingset{1.43}

One might think that the data set is easy to obtain, as BibTeX and citation data seem to be easy to download.  Unfortunately, when we need a large-volume, high-quality data set, this is not the case. For example, the citation counts from Google Scholar are not always accurate, and many online resources do not allow for large volume downloads.  Our data were downloaded from a handful of online resources by techniques including but not limited to web scraping. The data set was also carefully cleaned by a combination of manual efforts and computer algorithms we developed. Both data collection and cleaning are sophisticated and time-consuming processes, during which we had to overcome a number of challenges. For a detailed discussion on data collection and cleaning, see  Section~B.2 of the supplement.

{\bf Results, findings, and   challenges}.  
First, we overview the results. Our data set provides rich material for research and motivates many interesting  problems for 
  research trends, patterns, and impacts of the statistics community. In this paper, we focus on two topics: (1) How to use the citation data to estimate the research interests of statisticians, and (2) How to use the coauthorship data to study the network topology of statisticians.

Section~\ref{sec:citee} studies the first topic. How to model the research interests of an author is 
an open problem in bibliometrics.  Our idea is to first use the co-citation relationships to construct a 
{\it citee network} and then model the research interests of the author as the mixed-memberships he/she has over different network communities.  This gives rise to the degree-corrected mixed-membership (DCMM) model \citep{JKL2017}. 
Such a framework allows us to use principled statistical tools to attack problems about research interests.  Specifically, we develop new models, methods, and theory for (i) estimating the research interests of authors, (ii) clustering authors by research interests, (iii) studying how the research interests of an author evolve over time, and (iv) measuring the research interest diversity of individual authors.  
We discover a ``Research Map" (a cloud of points in $\mathbb{R}^2$, each 
representing the research interests of an author), 
which consists of a ``statistics triangle" and $15$ sub-regions.   The vertices of the triangle represent the three primary research areas in statistics: ``Bayes", ``Biostatistics",  and ``Nonparametric", and each sub-region represents an interpretable sub-area in statistics. The relative position of each author to the three vertices represents the weights of his/her research interests in  the three primary areas.  We also develop a new algorithm that allows us to plot the ``research trajectory" on the ``Research Map" for an  author to visualize   
the evolvement of his/her research interests over time, and propose two new metrics to measure the citation diversity of individual authors.   

Section~\ref{sec:coauthorship} studies the second topic, where the focus is  community detection.
We develop new models and methods for  (i)  hierarchical clustering, (ii) dynamic clustering, and (iii) measuring the coauthorship diversity. For (i), we develop a new approach and build a 4-layer community tree with $26$ leaves. Each leaf represents an interpretable co-authorship 
community where the authors may have some ties (e.g., colleagues, advisor-advisee) or share something (e.g., research interests or geological location) in common.  For (ii), we use a Sankey plot to visualize the birth
and growth of some communities and the migration of authors among different communities. 
For (iii), we propose a new idea to measure the research diversity of an author, by constructing 
the so-called ``personalized networks".

Second, we discuss our findings.   
 First, it is debatable what are {\it primary areas}  and 
{\it representative sub-areas} in statistics.   In Sections~\ref{sec:citee}, 
we suggest that  ``Bayes", ``Biostatistics", and ``Nonparametric"
are the three primary areas in statistics, and identify $15$ representative sub-areas. The ``statistics triangle" is reminiscent of Efron's triangle of statistical philosophy \citep{Efron-Fisher}, 
where the three vertices are  ``Bayes", ``Fisherian", and ``Frequentist". 
Note that our triangle is based on data while Efron's triangle  is more philosophical.  Second, in the 41-year time span of our data set, the research community of statistics has undergone significant changes: Some research areas (e.g., biostatistics) have become much more popular. Some research areas (e.g., nonparametric and semiparametric regressions) have significantly shifted the focus (e.g., with a significant surge of interest in high-dimensional data analysis after 2000). 
Last, the research of statisticians may be less diverse than expected: most researchers continue to collaborate with 
the same cluster of people over many years, with a large $p$-value for the significance test over his/her personalized network.

Last, we discuss some  challenges we face.  Getting meaningful results from a large data set is never easy 
(let alone the time and efforts required for obtaining the data set). We  
need  new methods for computing trajectories in Section~\ref{subsec:trajectory} and for constructing hierarchical community tree in 
Section \ref{subsec:tree}.  
We also need  new ideas to relate research interests to 
network mixed-memberships in Section \ref{subsec:triangle} and 
to connect research diversity of an author to a network global testing problem 
on his/her personalized networks in Section~\ref{subsec:personalized}.   

Even with a handful of new approaches we develop, we still face great challenges: how to properly construct the network
and choose the model, how to make inference, and how to interpret the results. To deal with
such challenges, we need many new ideas.
For example, in Section~\ref{sec:citee}, we discover that ignoring some ``old" citations makes the constructed citee network more useful. We also find that, to get meaningful results, it is critical to use a network model that allows for severe degree heterogeneity.  Also, in our study for ``research trajectory", we find that naively applying existing spectral approaches may face challenges, and to overcome the challenge, we  propose {\it dynamic network embedding} as  a new approach to dynamic network analysis. There are many such examples in Sections~\ref{sec:citee}-\ref{sec:coauthorship}.

In summary,  our findings are the combined results of (a) a large-scale high-quality data set we collected, (b) many new approaches we developed, and (c) many new ideas and substantial efforts in data analysis. 
We will make our data set and code available so  researchers can conveniently use our study as a template to study other research communities.

{\bf Contributions, broader impacts, and disclaimers}. 
We have several major contributions.  First, we contribute a high-quality, large-scale data set,  which provides material for research in bibliometrics, statistics,   and data science. 
Second, we set an example for  how quantitative analysis of large 
publication data can be executed. 
We create a template where we showcase how to use modern statistical tools to study a vast volume of publication data. We build large co-authorship  and co-citation networks, propose new network models,  and demonstrate how to use the output to label research areas, identify latent communities, and measure research diversities.  While we  use the statistics community as our object of study in this template, 
our approaches (data collection, research template, methods, and theory) are easily extendable to study other scientific communities (e.g., economics). Third,  while our  focus is on the new data set, we also contribute in methods and theory. We introduce a handful of methods for network data analysis; some are new, and some are carefully adapted from the recent literature. Our approaches to computing research trajectory, building community tree, and 
measuring research diversity are especially novel.  
Last but not the least, as statisticians, we know partial ground truth of our community. 
For this reason, our data set may provide a benchmark for comparing different methods in statistics, machine learning, and especially network analysis, and so largely help the development of methods and theory in these areas.

Our study has (potential) impacts  in science, social science, and even real life. 
It provides an array of ready-to-use and easy-to-extend statistical tools which the administrators, award committee, 
and individuals can use to study the research profile of an individual, an area, or the whole statistics community.   
For example, suppose a committee wishes to learn the research profile of an individual researcher. Our study provides a long list of tools to help characterize 
and visualize the research profile of the researcher: 
his/her research interests and his/her position on the Research Map, his/her research interest trajectory, to which network community he/she belongs,  his/her research diversity in terms of citation and in terms of 
co-authorship, his/her personalized networks,   the importance of his/her research area, his/her research impact and  ranking relative to his/her peers. 
Such information is not available from his/her curriculum vitae or profile on Google Scholar, 
and can be very useful for the award committee or administrators for decision making. 

Our study also provides a useful guide for researchers (especially junior researchers)
in selecting research topics,   looking for references, and building social networks. 
It also helps understand several important problems in social science and science: 
characterizing research evolvement, predicting emerging communities and significant advancement in 
each research area, checking whether the development of different areas is balanced, and identifying unknown biases in publications. We discuss these with more details in Section~\ref{sec:conclusion}. 

For disclaimers,  note that we have to use real names as  our data are about real-world publications, but 
we have not used any information that is not publicly available.  
It is not our intention to rank a researcher (or a paper, or an area) over others.  
While we tried very hard to create a high-quality data set, the time and effort 
one can invest is limited, so is the scope of our study;  as a result, some of our results may have biases. 
Our paper can be viewed as a starting  
point for an ambitious task, where we create a research template with which the researchers in other fields (e.g., economics)  can use statisticians' expertise in data analysis to study their own fields.  
For this reason, the main contributions of our paper are still valid.  See Section~A of the supplement for  
a longer version of the disclaimers.

{\bf Contents}.  
Section~\ref{sec:citee} studies co-citation networks, where the focus is to study how to estimate the research interests of an author and how the research interests evolve over time. Section~\ref{sec:coauthorship} focuses on coauthorship networks. It studies hierarchical and dynamic community detection, and proposes two new diversity measures. Section~\ref{sec:conclusion} is the conclusion.

\section{Learning research interests by co-citation networks} \label{sec:citee}
A good understanding of the research interests of statisticians 
helps understand the research trends, research impacts, and network topology 
of the statistics community, and also helps understand the research profile of individual statisticians. 
For example, suppose we are given an author with a total of 1000 citation counts. 
To decide whether he/she is highly cited, it is crucial to understand his/her 
major areas of interest, because the average citation count for a researcher in 
one area may be a few times higher than that of another.

The citation counts in our data set provide a valuable resource 
to study the research interests. In this section, we consider four problems: (a) how to model the research interests of  individual authors; (b) how to estimate his/her research interests and how to use the estimated research interests for author clustering; (c) how to study the dynamic evolvement of research interests of an author; (d) how to measure the diversity of research interests of an author. We propose new approaches to studying (a)-(d). Below is a sketch of our ideas.

Consider Problem (a) first.   How to model research interests of individual authors is an open problem. 
We observe that two authors being frequently cited together in the same papers (i.e., co-cited)  
indicates that their works are scientifically related and that they share 
some common research interests. Motivated by this, we propose the following 
approach to tackling Problem (a).    First, we use the co-citation relationship to construct 
an undirected network which we call the {\it citee network} (see Section~\ref{subsec:triangle}).  
 We assume that the citee network has $K$ communities,   each representing a primary research 
area in statistics (primary areas can be further divided into sub-areas). 
For author $i$, we model his/her research interest as a weight vector 
$\pi_i \in \mathbb{R}^K$, with $\pi_i(k)$ being the fraction of his/her interest in community $k$, $1 \leq k  \leq K$.  
We further model the citee network with the recent {\it Degree Corrected Mixed-Membership (DCMM)} model, 
where $\pi_i$ are the vectors of mixed-memberships. 
 
In a network, communities  are tight-knit groups of nodes that have more edges within than between \citep{goldenberg2010survey}.  
For example, suppose $K=3$ and we have three communities, each being a primary area in statistics:  ``Bayes", ``Biostatistics", and ``Nonparametric". Suppose for author $i$, 
$\pi_i = (0.5, 0.3, 0.2)'$. In this case, we think author $i$ has $50\%$, $30\%$, and $20\%$ of 
his research interest or impact in these  primary areas, respectively.

The DCMM model is a recent network model \citep{JKL2017, JiZhuMM}. 
It models both severe degree heterogeneity and mixed-memberships and is reasonable for the current setting.  
Let $A \in \mathbb{R}^{n, n}$ be the adjacency matrix of the citee network, where $A(i,j) = 1$ if $i\neq j$ and there is an edge between nodes $i$ and $j$ and $A(i,j) = 0$ otherwise. 
As above, let $\pi_i$ be the $K$-dimensional vector that models the research interests of author $i$, $1 \leq i \leq n$.  For a nonnegative, unit-diagonal matrix $P \in \mathbb{R}^{K, K}$ that models the community structure and parameters $\theta_1, \theta_2, \ldots, \theta_n>0$ that model the degree heterogeneity,  we assume that the upper triangle of $A$   
contains independent Bernoulli  variables, where for any $1\leq i<j\leq n$, 
\vspace*{-.3cm}
\beq \label{DCMM}
\mathbb{P}\bigl(A(i,j) = 1 \bigr) =\theta_i\theta_j\sum_{k,\ell=1}^K \pi_i(k)\pi_j(\ell)P(k,\ell) =  \theta_i\theta_j\cdot \pi_i'P\pi_j.
\vspace*{-.1cm}
\eeq
This provides a reasonable model for the research interests of individual authors,  and addresses an interesting problem in social science and bibliometrics.

Consider Problems (b)-(c).  We first use the mixed-SCORE \citep{JKL2017} to estimate the research interests of individual authors. 
We discover a {\it statistical triangle} and build the {\it Research Map} for statisticians.    
We then develop a new idea to compute the research trajectory of an author. 
To this end,  we need a new clustering algorithm for building the research map, 
and a new algorithm to draw the trajectory.  We now discuss them separately. 

The clustering problem is well-studied  
 (e.g.,  \cite{zhao2011community}, \cite{amini2013pseudo}, among others).   
Unfortunately, these algorithms have 
focused on the DCBM model \citep{BaN2011}.  Compared to the DCMM model in (\ref{DCMM}), 
DCBM requires each $\pi_i$ to be degenerate (one entry is $1$, all other entries are $0$), 
and is not appropriate for the citee network considered here.  Our idea is to combine mixed-SCORE \citep{JKL2017} with classical 
clustering algorithms.  Suppose we have estimated the research interest vectors $\pi_1, \pi_2, \ldots, \pi_n$ by mixed-SCORE, and let $\hat{\pi}_1, \hat{\pi}_2, \ldots, \hat{\pi}_n$ be the 
estimates. We view this step as a dimension reduction 
step, and propose an author clustering algorithm where we directly apply $k$-means to $\hat{\pi}_1, \ldots, \hat{\pi}_n$.  
Compared to existing clustering algorithms, our method works for the DCMM model where we allow mixed-memberships, 
and so is different.  

The problem of estimating the trajectory is related to the problem of dynamic  mixed-membership analysis.  
 Consider a sequence of citee networks, each 
for  a different time window.  We extend the DCMM model for static networks in (\ref{DCMM}) to dynamic networks, where $\pi_i$ may vary with time.  In such a setting, how to 
estimate  $\pi_i$    is largely an open problem. Related works include  \cite{kim2018review} and \cite{liu2018global}, but these papers focus on settings where each static network 
satisfies the MMSB model (a special  DCMM where we do not allow degree heterogeneity). For this reason,   it is unclear how to extend their approaches to our setting. 
The approach of naively applying mixed-SCORE to each individual network in our setting does not work well either;  
see Section \ref{subsec:trajectory}.

We  propose the {\it dynamic network embedding} as a new approach to analyzing dynamic DCMM.  
For each author in our data set, the approach produces a {\it research trajectory} which visualizes how his/her research interests evolve over time.  
Compared with the approach where we naively apply mixed-SCORE to each network in our setting separately, two approaches are the same for the first time window, but are significantly different for all other time windows; the new approach is more satisfactory both numerically and theoretically.

Consider Problem (d). How to measure the diversity of the research interests of individual authors is a problem of great interest. Using the {\it research trajectory} developed for Problem (c),  
we propose two diversity metrics: One measures the {\it significance} of research interest expansion of an author and the other measures his/her {\it persistence} of research interest expansion.  Compared with 
other diversity metrics, our metrics are new, for they are based on our proposed new approach to estimating the research trajectories. 

Below, Sections~\ref{subsec:triangle}, \ref{subsec:trajectory}, 
and \ref{subsec:citee-diversity} discuss Problem (b), (c), and (d) respectively. 
Note that Problem (a) is already fully addressed.

\subsection{Estimation of research interests, author clustering} \label{subsec:triangle}

We construct a citee network using the co-citations during 1991-2000. 
We limit the time to 1991-2000, for later we will use this network as a reference network to study the research trajectories of selected authors.  For each year $t$, $1991 \leq t \leq 2000$,  define a  year-$t$ weighted network where each node is an author, and for any two nodes $i$ and $j$, the weight of the edge between them is the number of times that the papers by author $i$ published between year $t-9$ to $t$ and the papers by author $j$ published between year $t-9$ and $t$ have been cited {\it together} in a paper by another author published in year $t$.  This results in a weighted adjacency matrix for year $t$. 
Summing the adjacency matrices for  $t = 1991, 1992, \ldots, 2000$ gives rise to a   weighted network. Let the degree of node $i$ be the sum of weights of edges between node $i$ and the other nodes.  We remove all nodes with a degree smaller than $60$, and define a symmetric unweighted network using the remaining nodes, where two nodes have an edge if and only if the weight between them in the previous network is no less than 2. We call the giant component of this network the citee network for 1999-2000, which has 2,831 nodes (these nodes form a subset of  most active and most cited authors).

 There are different ways to construct the citee network (we have studied many options  and recommend the one above).   
 We restricted to ``fresh" citations only (a citation from one paper to the other is considered ``fresh" if the two publication times are no more than 10 years apart). We have removed  low-degree nodes and low-weight edges in the intermediate weighted graph  to reduce noise.  In Section~C.3 of the supplement, we have also studied the case where the threshold $60$ is replaced by $50$ and 
$70$, and observed similar results (e.g., similar triangle and research map for statisticians).    
Thresholding the edge weights is a common practice. It may cause some information loss. But since the goal is to identify active communities, it is unclear how 
such a loss may affect the results. 
Also, just as in different fields of science, the average citations (per paper or author) can vary dramatically in different areas.    
For this reason, we may  threshold the edge weights adaptively with different thresholds for different areas. However, 
it is not immediately clear how to implement such an approach.  We leave these studies to the future.

We wish to use this citee network to study the research interests  of individual authors.  
We model this network with the aforementioned DCMM model \eqref{DCMM}. Under this model,  each of the $K$ communities can be interpreted as a research area, and the research interest of author $i$ is modeled by the mixed-membership vector $\pi_i \in \mathbb{R}^K$. How to estimate $\pi_i$ is known as the problem of  mixed-membership estimation, where we use the method mixed-SCORE \citep{JKL2017}.  The approach uses {\it SCORE embedding} which embeds all authors to a  low dimensional space and provides a way to visualize the research interest of each author.  Specifically, let $\hat{\xi}_1, \ldots \hat{\xi}_K\in\mathbb{R}^n$ be the first $K$ eigenvectors of the adjacency matrix. Each node $i$ is embedded into a $(K-1)$-dimensional space by the vector
\vspace*{-.3cm}
\begin{equation} \label{Definehatr} 
\hat{r}_i = \bigl[\hat{\xi}_2(i) /\hat{\xi}_1(i), \;  \hat{\xi}_3(i) /\hat{\xi}_1(i),  \; \ldots,  \hat{\xi}_K(i) / \hat{\xi}_1(i)\bigr], \qquad 1\leq i\leq n.  
\vspace*{-.3cm} 
\end{equation} 
Now, first, the embedded points are approximately contained in a {\it simplex with $K$ vertices} in $\mathbb{R}^{K-1}$, where each vertex represents a community. 
Second, each embedded point $\hat{r}_i$ is approximately a convex combination of the vertices: $\hat{r}_i \approx \sum_{k=1}^K w_i(k) v_k$, where $v_1, v_2, \ldots, v_K$ are the vertices of the simplex. 
The weight vector $w_i$ is an order-preserving transformation of $\pi_i$, in the sense that $w_i\propto \pi_i\circ b$, where $\circ$ is the Hadamard product and $b\in\mathbb{R}^K$ is a positive vector (not depending on $i$). Therefore, if an embedded point $\hat{r}_i$ is close to one vertex, then $w_i$ is nearly degenerate (with only one nonzero entry that is $1$), and node $i$ is a pure node (i.e., node $i$ is called a pure node of community $k$ if $\pi_i(k) = 1$ and $\pi_i(\ell) = 0$ for all $\ell \neq k$). If $\hat{r}_i$ is deeply in the interior of the simplex, then all entries of $w_i$ are bounded away from $0$ and node $i$ is  highly mixed;  
see \cite{JKL2017} for more discussions.  
 
{\bf Why $K = 3$ is the most reasonable choice}.  To use mixed-SCORE, we need to decide $K$, which is unknown.  
First, we use the scree plot of the adjacency matrix to determine the range of $K$ as $[2,6]$. 
Second, we implemented mixed-SCORE for each $K \in\{ 2, 3, \ldots, 6\}$ and investigated the goodness of fit, by checking whether the rows of $\hat{R}$ fit the aforementioned $(K-1)$-dimensional simplex structure (it is hard to visualize the simplex when $K \geq 4$, so we 
plot two coordinates of $\hat{r}_i$'s at a time to visualize a projection of the simplex to $\mathbb{R}^2$).
Last, 
for each $K$, we manually check the large-degree pure nodes in each community and see whether the results fit with our knowledge of the statistics community. The above analysis suggests $K = 3$ as the best choice. 
See Section~C.2 of the supplement for details.

\spacingset{0.9}
\begin{figure}[tb!]
\centering
\includegraphics[width=.95\textwidth, trim=5 18 0 0, clip=true]{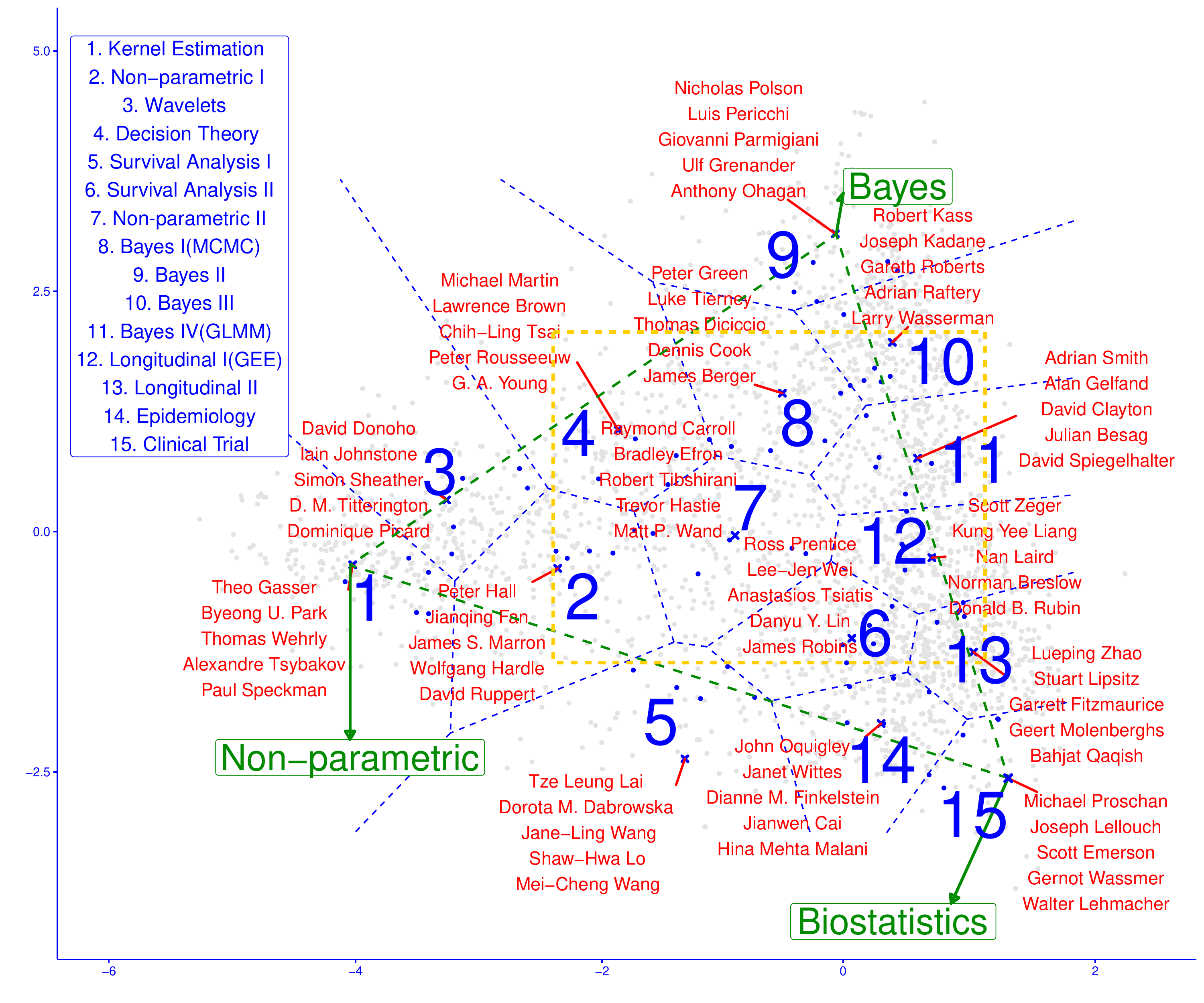}
\caption{\small The research map.   Each grey dot represents a 2-dimensional SCORE embedding vector $\hat{r}_i$, $1 \leq i \leq n$,   and the $15$ clusters and Voronoi diagram are obtained by applying the $K$-means algorithm  
to $\hat{r}_1, \hat{r}_2, \ldots, \hat{r}_n$.  The dashed green line represents the triangle, where 
 the vertices represent the $3$ primary areas.    In each cluster, the cluster center is also presented (blue crosses), together with 5 authors with highest degrees (blue dots). The results are based on citations: it is possible that an author does not work in an area, but have many citations in that area. }
\label{fig:simplex2}
\end{figure}
\spacingset{1.43}

{\bf The statistics triangle}.  Since $K = 3$, the simplex in SCORE embedding is a triangle, each vertex representing (perceivably) a primary statistical research area. See Figure~\ref{fig:simplex2}. To interpret these areas,  we apply mixed-SCORE to the citee network with $K = 3$, and obtain an estimate for the  membership vectors $\pi_1, \pi_2, \ldots, \pi_n$ by $\hat{\pi}_1, \hat{\pi}_2, \ldots, \hat{\pi}_n$.  
We divide all the nodes into three groups: If the largest entry of $\hat{\pi}_i$ is the $k$th entry, then node $i$ is assigned to group $k$, $1 \leq k \leq 3$.  In Section~C of the supplement, we investigate the research interests of authors in each group, using the topic weights estimated from abstracts of their papers. It suggests that the three vertices represent three primary research areas: ``Bayes," ``biostatistics," and 
``nonparametric statistics." 
This triangle is reminiscent of the {\it statistics philosophy triangle} by \cite{Efron-Fisher},   where the three vertices are ``Bayes", ``Fisherian", and ``frequentist". Efron argued that they are the three major philosophies in statistics, and  most statistics methodologies (e.g., bootstrap) can be viewed as weighted averages of these three philosophies.
Different from Efron's triangle, our statistics triangle is  data-driven.

{\bf The research map}.  Perceivably,  we can further split each primary area into sub-areas, and a convenient approach is to use SCORE embedding. For each author $i$  in the citee network, $1 \leq i \leq n$, since $K = 3$,   $\hat{r}_i$ can be viewed as a point in $\mathbb{R}^2$. 
The distance between authors in this space is a measure of closeness of their research areas. 
Therefore, it makes sense to further cluster the authors into sub-areas by applying the $K$-means algorithm to $\{\hat{r}_i\}_{i=1}^n$. We have tried the $K$-means algorithm with $L = 10, 11, \ldots, 20$ clusters,  and picked $L = 15$ due to that the result is most reasonable. We then apply the $K$-means with $L = 15$ and obtain $15$ clusters, each of which can be interpreted  as a sub-area in statistics after a careful  investigation of  the research works by representative authors in the cluster (while we try very hard to find a reasonable label for each cluster,  we should not expect that a simple label is able to explain the research interests of all authors in the cluster).

Figure \ref{fig:simplex2} shows the 15 clusters and their labels,  which we call the {\it research map} of the citee network.  In this map,  each point represents $\hat{r}_i$ for some node $i$, $1 \leq i \leq n$, and the two axes are the two entries of $\hat{r}_i$, respectively. 
The statistics triangle is illustrated by the dashed green lines, 
where the three vertices are estimated by mixed-SCORE and represent the three primary areas   ``Bayes," ``Biostatistics," and ``Nonparametric."  We also present the   
Voronoi diagram for the clusters  (boundaries are illustrated by dashed blue lines), 
and the names for the 5 authors with the largest degrees in each cluster.

For each author, his/her position on the research map illustrates the weight his/her citation has in each of the three primary areas.   
For example, Raymond Carroll and Bradley Efron are  located deeply in the interior of the triangle, suggesting that their citations between 1991 and 2000 have substantial weights in each of the three primary areas. Authors who are located around each corner of the triangle include Nicholas Polson (``Bayes"), Michael Proschan (``Biostatistics"), and Theo Gasser (``Nonparametric"),  suggesting that their citations between 1991 and 2000 are mostly from one community. 
Note that, since the results are based on the citee network, the areas from which an author attracts citations may not be exactly the same as the areas he/she works on. For example, though Donald B. Rubin rarely works in {\it Longitudinal I (GEE)}, he is clustered to GEE for  he is cited together with quite a few authors in GEE  (e.g. Scott Zeger, Nan Laird, and Daniel F. Heitjan).

\subsection{Evolvement of author research interests} \label{subsec:trajectory}
The research map in Figure~\ref{fig:simplex2} was established using the co-citations during 1991-2000. We now study how individual authors' research interests evolve between 2001 and 2015, and propose  {\it dynamic network embedding} as a new approach.  
For each author, the approach produces a trajectory on the research map  to visualize  his/her research interest 
evolvement.  

\spacingset{1}
\begin{table}[htb]
\centering 
\scalebox{.7}{
\begin{tabular}{cccccccccccccccccccccc}
\hline
Window &1 & 2 & 3 & 4 & 5 & 6 & 7 & 8 & 9 & 10 & 11 & 12 & 13 & 14 & 15 &16 & 17 & 18 & 19 & 20 & 21\\
\hline
Start  & 91 &92&93&94&95&96&97&98&99&00&
01&02&03&04&05&06&07&08&09&10&11\\
End & 00&01&01&02&03&04&04&05&06&07&
07&08&09&10&10&11&12&13&13&14&15\\
\hline
Length & 10 & 10 & 9 & 9 & 9 & 9 & 8 & 8 & 8 & 8 & 7 & 7 & 7 & 7 & 6 & 6 & 6& 6 & 5 & 5 & 5\\
\hline
\end{tabular}}
\spacingset{0.9}
\caption{\small The 21 time windows we use to  study the research trajectories. For example, the first window is from 1991 to 2000, covering a $10$-year time period.}
\label{tab:windows} 
\end{table}
\spacingset{1.43}

We consider $21$ time windows (see Table \ref{tab:windows}) and construct a citee network for each of them.  
As the numbers of papers published per year are steadily increasing, 
we use gradually smaller windows so the average node degrees of all $21$ citee 
networks are roughly the same.  
We use the citee network for the first window (1991-2000) 
as the reference network for our study below. This network is the same as the citee network that we use to study the statistics triangle  and  the research map in Figure \ref{fig:simplex2}. Recall that this network has 2,831 nodes. We restrict each of the other 20 networks to the same set of nodes.    We propose a {\it dynamic DCMM model} by extending the (static) DCMM model \eqref{DCMM}. 
Consider $T$ citee networks for the same set of $n$ nodes, and   
let $A_1, A_2,   \ldots, A_T$ be the adjacency matrices. Let $P \in \mathbb{R}^{K,K}$ be the time-invariant community structure matrix, and let $\theta_i^{(t)}>0$ and 
$\pi_i^{(t)}\in \mathbb{R}^K$ be the degree parameter and mixed membership vector of node $i$ at time $t$, $1\leq i\leq n$, $1\leq t\leq T$. Write $\theta_t=\mathrm{diag}(\theta_{1t},\ldots,\theta_{nt})$ and $\Pi_t=[\pi_{1t}, \ldots,\pi_{nt}]'$. 
Given $\{(\theta_{t}, \Pi_{t}\}_{t=1}^T$,
we assume $A_1, A_2, \ldots, A_T$ are independently generated. Also,  the upper triangle of $A_t$  
contains independent Bernoulli  variables satisfying 
\vspace*{-.3cm}
\beq \label{dynamic-DCMM}
\mathbb{P}\bigl(A_t(i,j) = 1 \bigr) = \theta^{(t)}_{i}\theta_{j}^{(t)}\cdot (\pi_{i}^{(t)})'P(\pi_{j}^{(t)}),  \qquad 1\leq i< j\leq n.
\vspace*{-.3cm}
\eeq
Here, we assume $A_1, A_2, \ldots,A_T$ are independent given $\{(\theta_t, \Pi_t\}_{t=1}^T$, but this can be relaxed to 
allow for weak dependence. Also, to allow flexible temporal dependence in $\{(\theta_t, \Pi_t\}_{t=1}^T$, we do not impose any extra conditions on them.

How to estimate $\pi_i^{(t)}$ is known as the problem of dynamic mixed membership estimation.   
Existing works include \cite{kim2018review, liu2018global}. However, 
these works focus on the dynamic MMSB model (a special  dynamic DCMM)  where it is required 
$\theta_i^{(t)}\equiv \alpha_t$ for all $1\leq i \leq n$ at each time $t$. It is therefore unclear 
how to extend their ideas to our setting. 

Alternatively, one may use naive mixed-SCORE (i.e., we apply mixed-SCORE to each network in the sequence separately). Unfortunately, the approach is also unsatisfactory. 
One challenge is that the estimates $\{\hat{\pi}_i^{(t)}\}_{1\leq i\leq n}$ for each time window $t$ are up to 
an unknown permutation among the $K$ communities. 
Since we have $T$ different time windows, 
we have a large number of possible combinations of such permutations, 
and it is unclear how to pick the right one.  
The other challenge is that, each $A_t$ is constructed for a relatively short time period, 
and can be very sparse. In such cases, spectral decomposition of $A_t$ may be rather noisy, 
and the naive mixed-SCORE may perform unsatisfactorily.

We propose {\it dynamic network embedding} as a new approach  to dynamic mixed membership estimation. Note that the network $A_1$ from the first window was used in Section~\ref{subsec:triangle} to build a ``research map" for all the authors. This motivates us to treat $A_1$ as a reference network and project all the other networks onto this ``research map." 
Let $\hat{\lambda}_1,\hat{\lambda}_2,\ldots,\hat{\lambda}_K$ be the $K$ largest eigenvalues (in magnitude) of $A_1$, and let $\hat{\xi}_1, \hat{\xi}_2, \ldots, \hat{\xi}_K$ be the corresponding eigenvectors. For each $1\leq t\leq T$ and each node $1\leq i\leq n$, define a $(K-1)$-dimensional vector $\hat{r}_i^{(t)}$ by ($e_i$: the $i$th standard basis vector of $\mathbb{R}^n$)
\vspace*{-.3cm}
\beq\label{dynamic-SCORE}
\hat{r}_i^{(t)}(k) = [\hat{\lambda}_1(e_i' A_t \hat{\xi}_{k+1})] /  [\hat{\lambda}_{k+1}(e_i' A_t \hat{\xi}_1)],\qquad 1\leq k\leq K-1. 
\vspace*{-.3cm} 
\eeq
Now, for each time $t$, we obtain the low-dimensional embedding $\{\hat{r}_i^{(t)}\}_{1\leq i\leq n}$ of all $n$ nodes, and for each node $i$, we obtain the embedded ``trajectory" as $(\hat{r}_i^{(1)}, \hat{r}_i^{(2)}, \ldots, \hat{r}_i^{(T)})$. For $t=1$, $\hat{r}_i^{(1)}$ coincides with the SCORE embedding \eqref{Definehatr}. It implies that the starting point of each embedded trajectory is always the position of this author in the ``research map."
For $t>1$, the proposed embedding is different from the SCORE embedding \eqref{Definehatr} for $A_t$. Note that in \eqref{Definehatr}, we use the eigenvectors of $A_t$ to construct the embedding at $t$, while in \eqref{dynamic-SCORE}, we use the eigenvectors and eigenvalues of $A_1$ to construct the embeddings for all $t$.  

We now explain how the approach overcomes the two challenges aforementioned.  
First, the new approach utilizes the same $(\hat{\xi}_1, \hat{\xi}_2,\ldots,\hat{\xi}_K)$ to obtain the embeddings for all $t$, so that these networks are projected to the same low-dimensional space. Consequently, the projected points $\hat{r}_i^{(t)}$ are automatically aligned across time. Second, in spectral projection and its variants (e.g., SCORE), the data to project (rows of $A_t$) and the projection directions (eigenvectors of $A_t$) are {\it dependent} of each other. On the contrary, in \eqref{dynamic-SCORE}, the data to project, $A_te_i$, and the projection direction, $\hat{\xi}_k$, are {\it independent} of each other, for any $t\geq 2$. Thus, the projected points are much less noisy. In the preliminary theoretical analysis, we find that $\hat{r}_i^{(t)}$ has a sharp large-deviation bound even when $A_t$ is very sparse and when $\hat{\xi}_k$ is only a moderately good estimate of the population eigenvector of $A_1$.

We explain why the approach is reasonable. Define a population counterpart of \eqref{dynamic-SCORE}. In     
model \eqref{dynamic-DCMM}, 
let $\Theta^{(t)}=\mathrm{diag}(\theta_1^{(t)}, \ldots,\theta_n^{(t)})$, $\Pi^{(t)}=[\pi_1^{(t)}, \ldots,\pi_n^{(t)}]'$, and $\Omega_t=\Theta^{(t)}\Pi^{(t)}P(\Pi^{(t)})'\Theta^{(t)}$, $1\leq t\leq T$.  
Let $\Lambda=\diag(\lambda_1,\lambda_2, \ldots,\lambda_K)$ and $\Xi=[\xi_1,\xi_2,\ldots,\xi_K]$, 
where $\lambda_k$ is the $k$-th largest (in magnitude) eigenvalue of $\Omega_1$ and $\xi_k$ is the corresponding eigenvector.  
For  $1\leq t\leq T$ and $1\leq i\leq n$, define $r_i^{(t)}\in\mathbb{R}^{K-1}$ by 
\vspace*{-.3cm}
\beq \label{dynamic-SCORE-0}
r_i^{(t)}(k) =  [\lambda_1(e_i' \Omega_t \xi_{k+1})] / [\lambda_{k+1}(e_i' \Omega_t \xi_1)],\qquad 1\leq k\leq K-1,
\vspace*{-.3cm}
\eeq
\begin{thm} \label{thm:simplex}
Consider the dynamic DCMM model \eqref{dynamic-DCMM}. For each $1 \leq t \leq T$,  letting $M_t=P(\Pi^{(t)})'\Theta^{(t)}\Xi\Lambda^{-1}\in\mathbb{R}^{K,K}$, we suppose $\mathrm{rank}(M_t) = K$ and $\min_{1\leq k\leq K}\{M_t(1,k)\}>0$.  
Let $v^{(t)}_k=\frac{1}{M_t(k,1)}[M_t(k,2),M_t(k,3), \cdots, M_t(k,K)]'$, $1\leq k\leq K$, and let ${\cal S}_t\subset\mathbb{R}^{K-1}$ be the simplex with $K$ vertices $v_1^{(t)}, \ldots,v_K^{(t)}$. For all  $1\leq t\leq T$, first, 
each $r_i^{(t)}$ is contained in the simplex ${\cal S}_t$. If $i$ is a pure node of community $k$ ($\pi_i^{(t)}=e_k$), then $r_i^{(t)}$ is located on the vertex $v_k^{(t)}$. If $i$ is not a pure node of any community, then $r_i^{(t)}$ is in the interior of ${\cal S}_t$ (including the edges and faces, but not any of the vertices). 
Second, each $r_i^{(t)}$ is a convex combination of $v_1^{(t)}, v_2^{(t)},\ldots,v_K^{(t)}$, denoted by $r_i^{(t)}=\sum_{k=1}^K w_i^{(t)}(k)v_k^{(t)}$. The coefficient vector $w_i^{(t)}\in\mathbb{R}^K$ satisfies that $w_i^{(t)}=(\pi_i^{(t)}\circ h_t)/\| (\pi_i^{(t)}\circ h_t) \|_1$, where $\circ$ is the Hadamard product and $h_t\in\mathbb{R}^K$ is a positive vector that does not depend on $i$. 
\end{thm} 

Theorem~\ref{thm:simplex}  is proved in the supplement. By Theorem~\ref{thm:simplex}, in the noiseless case, the embedded data cloud $\{r_i^{(t)}\}_{1\leq i\leq n}$ at every $t$ form a low-dimensional simplex, similar to that in \cite{JKL2017}. We can then borrow the idea there and estimate $\pi_i^{(t)}$ from the embedded data cloud via a simplex vertex hunting algorithm. This explains the rationale of our procedure. To focus on real data analysis, we relegate more detailed analysis of the approach to a forthcoming manuscript 
We now apply the procedure to our data set.

\spacingset{0.9}
\begin{figure}[tb!]
\centering
\includegraphics[width=.68\textwidth, trim=0 18 0 0, clip=true]{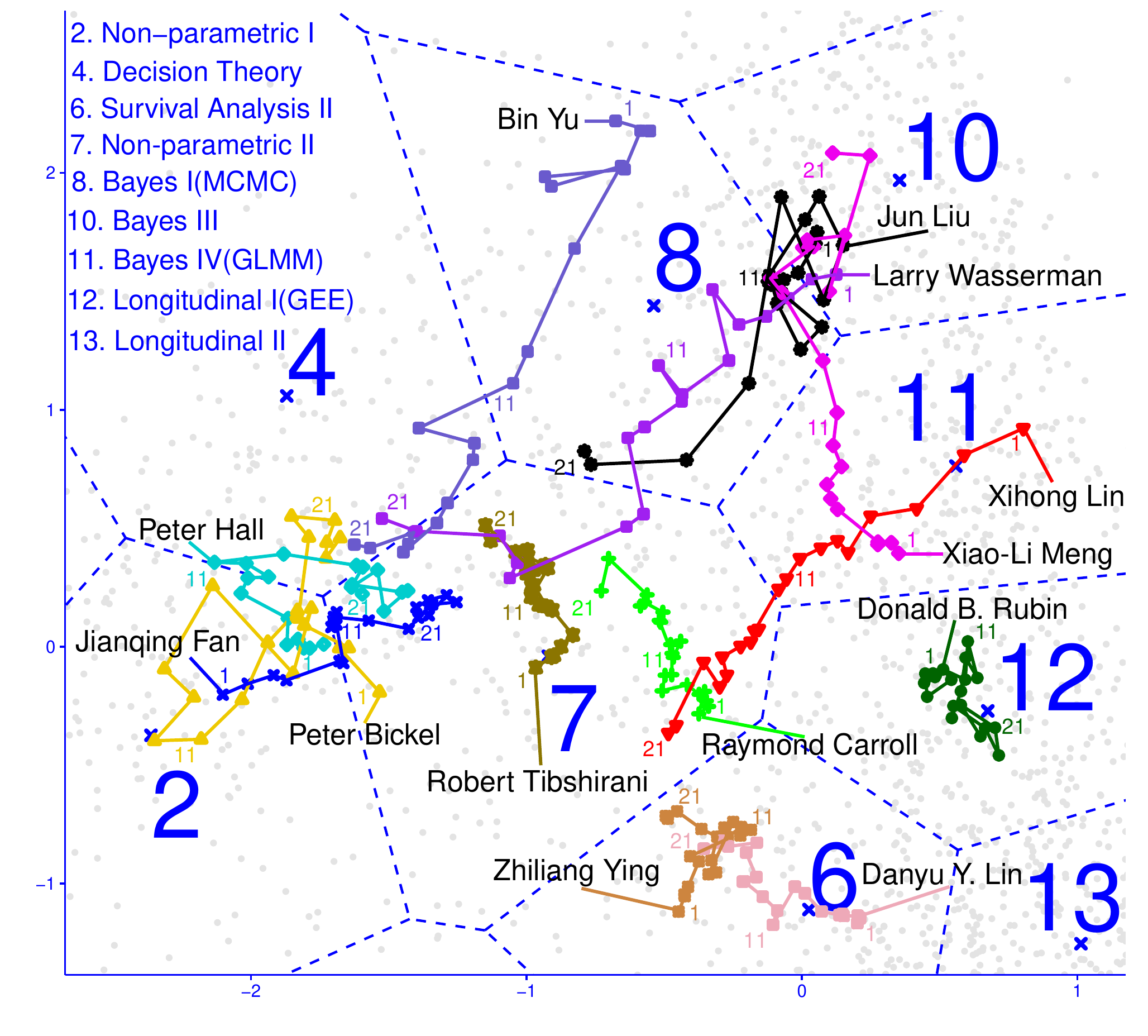}
\caption{\small Research trajectories of representative authors (this is a zoomed-in view of the region in Figure \ref{fig:simplex2} within the dashed yellow square, with the same Voronoi diagram).  Each trajectory has 21 knots,  corresponding to the 21 time windows in Table \ref{tab:windows} (knots 1, 11, and 21 are marked with 1, 11, and 21, respectively).  
The starting point (marked with 1) is the same as the author's position in Figure \ref{fig:simplex2}. 
For interpretation, we selected some authors we are familiar with, but 
we can plot the trajectory for any author with a reasonably long publication history in our data set.  
The results are based on citations: it may happen that an author (e.g., D. Rubin) does not work in an area, but have many citations in that area. }
\label{fig:trajectory_all}
\end{figure}
\spacingset{1.43}

{\bf Research trajectories for individual authors}.   
Recall that we have constructed a 2831-node citee network for each of the 21 time windows in Table~\ref{tab:windows}. 
Applying \eqref{dynamic-SCORE}, we get an embedding $\hat{r}_i^{(t)}$ for each author $i$ at each time $t$. 
Viewing $\hat{r}_i^{(t)}$ as a point on the research map,  we have 21 points for author $i$, each corresponding to a time window.  Connecting these time-ordered points gives rise to the research trajectory of author $i$, which visualizes how the research interests of author $i$ evolve over time.  
The starting point of his/her research trajectory  is the same as his/her position in the research map in Figure \ref{fig:simplex2}.

In Figure~\ref{fig:trajectory_all}, we present the research trajectories of a handful of representative authors in statistics. 
For better visualization,  note that the whole region covered by Figure~\ref{fig:trajectory_all} is the zoom-in of the rectangular region bounded by dashed yellow lines in Figure~\ref{fig:simplex2}.   Since all of these authors happen to be in the reference citee network,  
the starting point of each author's trajectory is the same as his/her position on the 
research map in Figure \ref{fig:simplex2}.  We have the following observations: (a) A few authors (e.g., Xihong Lin, Jun Liu, Xiao-Li Meng, Larry Wassermann, and Bin Yu)  exhibit a significant change of research interest from 2000 to 2015, suggesting that they persistently tried to 
broaden their research horizon and scope of interest. (b) The research trajectories of Peter Bickel, Raymond Carroll, Jianqing Fan, Peter Hall and Robert Tibshirani stayed in the regions of  {\it Decision Theory}   and  {\it Non-parametric I and II}, and the research trajectories of  Danyu Lin, Donald Rubin and Zhiliang Ying stayed in the regions of  {\it Survival Analysis II} and {\it Longitudinal I (GEE)}.  A  possible reason is that the research areas of these authors in 1991-2000 continued to be  ``hot areas" for the time period 2000--2015.  (c)  
The two subregions, {\it Non-parametric I and II},  are among the most ``popular" research areas between 1991 and 2015.  
Research leaders (e.g., Peter Bickel, Jianqing Fan, Peter Hall, and Robert Tibshirani) who  worked in these areas in 1990s continued to work in these research areas in 2000-2015. 
At the same time, research leaders who used to work on some seemingly distant areas or in distant  regions (e.g. Xihong Lin, Jun Liu, Larry Wasserman, and Bin Yu) gradually migrate to the center of these two regions.  
These two sub-areas highly overlap with the research area of {\it high-dimensional data analysis},  which was one of the most rapidly growing areas in statistics between 2000 and 2015. The claim is confirmed by investigating more authors in these two subregions. 


\subsection{Diversity of author research interests}\label{subsec:citee-diversity}
The research trajectories in Section~\ref{subsec:trajectory} suggest  that research interests of some authors may vary more significantly than those of others. This motivates us to propose some metrics 
for research diversity of individual authors. 
Recall that the 21 knots for the trajectory of author $i$ are $\hat{r}_{i}^{(1)},\ldots,\hat{r}_i^{(21)}$. We introduce two diversity metrics: $E_i = \|\hat{r}_i^{(21)}-\hat{r}_i^{(1)}\|$ and $M_i = \max_{2\leq k \leq 21}\|\hat{r}_i^{(t)}  - \hat{r}_i^{(1)}\|$, 
\spacingset{0.9}
\footnote{Here,  $\hat{r}_i^{(t)}$ are defined by (2.5) through the leading eigenvalues and  
eigenvectors $(\hat{\lambda}_k, \hat{\xi}_k)$ of $A_{t_0}$ with $t_0 = 1$. 
Since we use the first one in the $21$ networks as the reference, $t_0 = 1$ is 
the most natural choice.  For robustness check, we have also studied the case of $t_0 \in \{2, 5, 10\}$;  see Section~C.4 of the supplement. The results are largely similar to those in this section.}   
\spacingset{1.43}
where $E_i$ is called $\textit{se}_{-}\textit{distance}$  (distance between the starting point and the ending point) and $M_i$ is called $\textit{max}_{-}\textit{distance}$ (maximum distance between a point and the starting point).  
A large $E_i$ suggests that the research areas for author $i$ in 2011-2015 (the last time window) are significantly different from his/her  research areas in 1991-2000, and a large $M_i$ suggests that the research areas  for author $i$ in at least some of the time windows are significantly different from his/her research areas in 1991-2000.  

\spacingset{0.9}
\begin{figure}[tb!]
\centering
\includegraphics[width=0.55\textwidth, height=.42\textwidth, trim=10 18 0 0, clip=true]{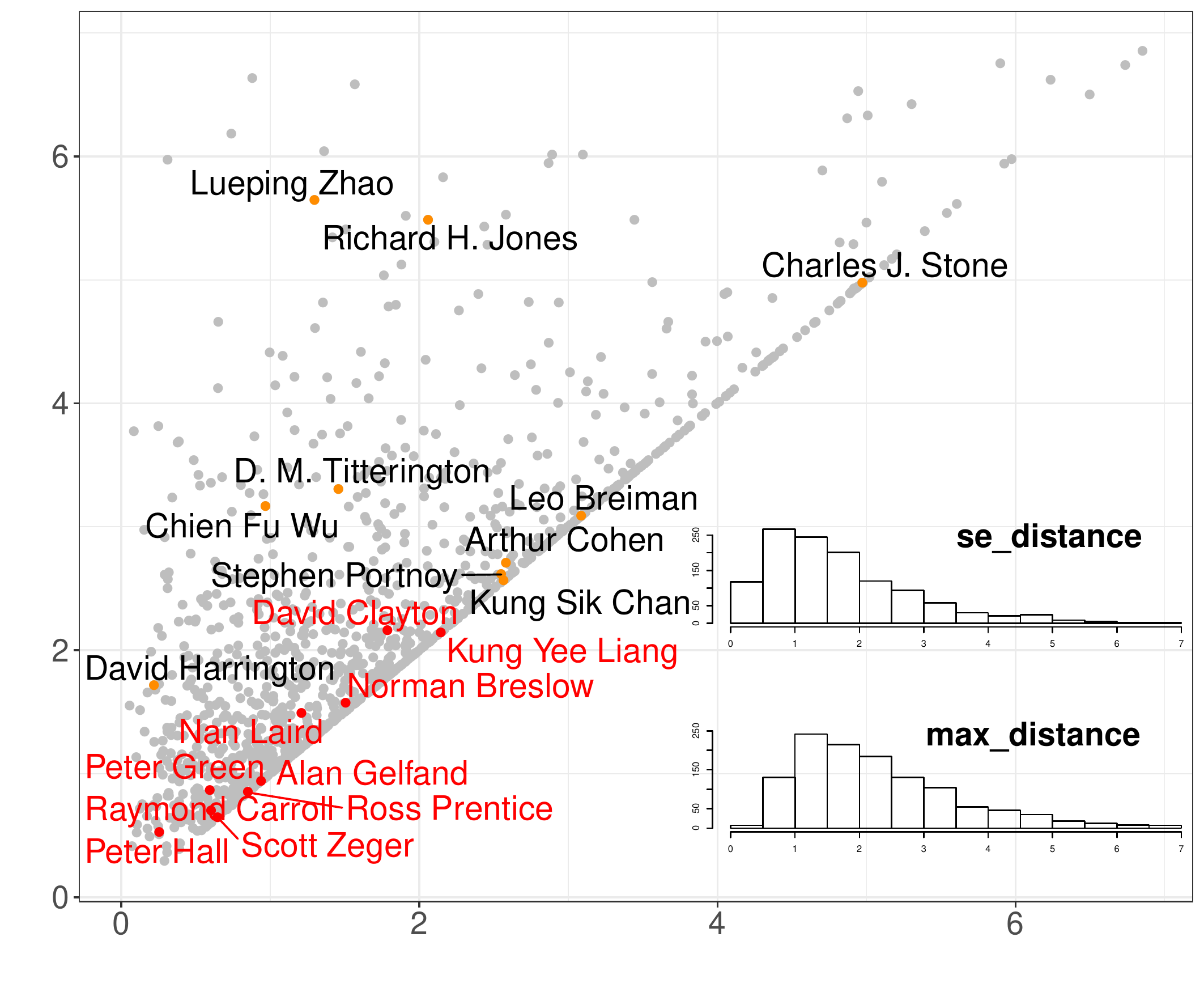}
\caption{\small The two diversity metrics of 1,202 authors ($x$-axis:  $\mathrm{se}_{-}\mathrm{distance}$; $y$-axis: $\mathrm{max}_{-}\mathrm{distance}$). The red dots represent the 10 highest-degree authors. The  orange dots represent (among the top 200 highest-degree nodes) the 5 authors with the largest  $\mathrm{se}_{-}\mathrm{distance}$ and the 5 authors with the largest differences between $\mathrm{max}_{-}\mathrm{distance}$  and $\mathrm{se}_{-}\mathrm{distance}$.}\label{fig:trajectory_distance}
\end{figure}
\spacingset{1.43}

Figure~\ref{fig:trajectory_distance} presents the two metrics for a total of 1,202 authors.   The reference network has 2,831 nodes in total, but in the 21 citee networks (each for a different time window)  
only 1202 authors are always in the giant component, so we present only the $E_i$ and $M_i$ for these 1,202 authors. In this figure,  the 10 highest-degree nodes are marked with red dots, where their names are also presented in red. Also, among the 200 authors who have the largest degrees,  
the 5 authors who have the largest $E_i$ values (Charles J. Stone, Leo Breiman, Arthur Cohen, Kun Sik Chan, Stephen Portnoy) are marked with orange dots,  and the $5$ authors who have the largest $(M_i - E_i)$ values (Luoping Zhao, Richard H. Jones, Chien Fu Wu, D.M. Titterington, David Harrington)
are also marked with orange dots.

For author $i$,  if both $M_i$ and $E_i$ are large, we call the changes of the research areas of author $i$ {\it significant and persistent (SP)}, and for short, author $i$ is an SP type. 
If  $M_i$ is large but $E_i$ is relatively small, we call the changes of the research areas of author $i$ {\it significant but not persistent (SnP)}, and for short, author $i$ is an SnP type. 
For the 20 authors whose names are showed in the figure,  Charles J. Stone 
 has the largest $E_i$ value and is seen to be an SP type, 
and Lueping Zhao has the largest $M_i$ value and is seen to be an SnP type.

\section{Learning communities from coauthorship networks} \label{sec:coauthorship} 
The study of coauthorship patterns and community structures in an academic society is an interesting topic \citep{New2004}.  
The co-author relationship in our data set provides a valuable resource to 
study the community structure, which is the focus of this section. 
Compared to the co-citation relationship (focus of Section \ref{sec:citee}), 
the co-author relationship is quite different in nature:  Citations are  primarily driven by scientific relevance, but collaborations may be driven  by many factors (e.g.,  geographical proximity, academic genealogy, cultural ties).  Therefore,  the study below may shed new insight which we do not see in Section \ref{sec:citee}.  
We focus on the following problems: (a) hierarchical community detection (and especially interpretation of 
different communities),  (b) evolvement of communities,  and (c) diversity measure of individual authors. We discuss these in Sections~\ref{subsec:tree}-\ref{subsec:personalized} separately.

\subsection{Estimation of the hierarchical community structure} \label{subsec:tree}    
Compared to the citee networks, the effect of mixed-memberships in co-authorship networks 
is notably less significant;  see Section~D.5 of the supplement for detailed discussion.  So  instead of focusing on the mixed-memberships as in Section \ref{sec:citee}, we focus on the problem of {\it recursive community detection}: 
We think that the co-authorship network has many communities (each is a research sub-area in statistics),  
and the sub-areas may have a tree structure. The goal is to (possibly recursively) cluster the authors into these sub-areas.  

A popular strategy to recursive community detection is as follows:  First, we partition the network into $K_0$ groups, for a small integer $K_0<K$, where $K$ is the total number of communities. This gives rise to $K_0$ subnetworks restricted to each group. Next, for each subnetwork, we test whether it has only one community (null hypothesis) or  multiple communities (alternative hypothesis). If the null hypothesis is rejected, this subnetwork is further split. The algorithm stops when the null hypothesis is accepted in every subnetwork. The output is a hierarchical tree, with each leaf being an estimated community. 

As the mixed-membership effect here is less significant than that in citee networks, 
it is reasonable to use the DCBM model \citep{BaN2011}. Compared with the DCMM model in (\ref{DCMM}), DCBM is a special case 
where we require all vectors $\pi_i$ to be degenerate (i.e., one entry is $1$, all other entries are $0$), and so   
 the nodes partition to non-overlapping communities ${\cal C}_1, {\cal C}_2, \ldots,{\cal C}_K$. 
Let $A\in \{0,1\}^{n\times n}$ be the symmetrical adjacency matrix of a coauthorship network, where $A(i,j)=1$ if and only if authors $i$ and $j$ have co-authored papers in the range of interest. In DCBM, we assume 
\vspace*{-.2cm}
\beq \label{DCBM}
\mathbb{P}(A(i, j)=1)=\theta_i\theta_j P_{k\ell},\qquad\mbox{if } i\in{\cal C}_k, \; j\in  {\cal C}_\ell, \mbox{ for all }1\leq k,\ell\leq K.  
\vspace*{-.15cm}
\eeq
where $(P, \theta_1,\theta_2,\ldots,\theta_n)$ are the same as those in \eqref{DCMM}. 
In this subsection, we assume both the whole network and subnetworks satisfy the DCBM. 
A more careful modeling for the hierarchical structure is possible (e.g., \cite{li2020hierarchical}). But since our primary focus here is to analyze a valuable new data set,  we leave this to the future.

There are many interesting works on recursive community detection (e.g., \cite{li2020hierarchical}), but they focused on the stochastic block models, a special case of the DCBM model in \eqref{DCBM} that does not allow degree heterogeneity. It is unclear how to extend their methods to our settings. 
%
%
We propose a new algorithm for recursive community detection, consisting of a community detection module and a hypothesis testing module. Both modules are able to properly deal with severe degree heterogeneity. We now discuss them separately.

The community detection module clusters the nodes in a network into $K_0$ communities, for a given $K_0\geq 2$. We use the following algorithm. For a tuning parameter $c_0>0$, let $I_n$ be the identity matrix,  let $\hat{\mu}_k$ be the $k$-th largest eigenvalue (in magnitude) of $A + c_0 I_n$, and let $\hat{\xi}_k$ be the corresponding eigenvector, $1 \leq k \leq K_0$.  
Define a matrix $\hat{R} \in \mathbb{R}^{n, K_0-1}$ by $\hat{R}(i, k) =  \hat{\xi}_{k+1}(i)/ \hat{\xi}_1(i)$.  
For a threshold $t>0$, we apply 
element-wise truncation on $\hat{R}$
and obtain  
a matrix $\hat{R}^* \in \mathbb{R}^{n, K_0-1}$ by 
$\hat{R}^*(i,k) = \mathrm{sgn}(\hat{R}(i,k)) \cdot \min\{|\hat{R}(i,k)|,\, t\}$,  $1 \leq i \leq n, 1 \leq k \leq K_0-1$. 
We then apply the $k$-means algorithm to the rows of $\hat{R}^*$, assuming there are $\leq K_0$ clusters. 
There are two tuning parameters $(c_0, t)$. We set $c_0=1$ and $t=\log(n)$.

The approach extends SCORE \citep{Jin2015}, where $c_0=0$.   
Recall that we call $\hat{\xi}_k$ the $k$-th largest eigenvector of $A$ if it corresponds to the $k$-th largest (in magnitude)  
eigenvalue of $A$. SCORE uses the first $K$ eigenvectors of $A$ for clustering, but unfortunately, 
the estimated network is dis-assortative (a network is assortative if 
for any pair of communities, they have more edges within than between \citep{RefAssort1}).   For co-authorship networks, 
such a result is hard to interpret.  Note that  
for an assortative network, a negative eigenvalue is more likely to be spurious than a positive one.
This motivates the above approach, where we replace $A$ by $A + c_0 I_n$: the term $c_0 I$ penalizes 
the rankings of negative eigenvalues,    so the set of first $K$ eigenvectors of $A + c_0 I_n$ is different from those of $A$.  
How to choose $c_0$ is an interesting problem. We find all estimated networks for  $c_0 \geq 1$ are assortative, 
so we choose $c_0$ as $1$ for convenience.  The asymptotic consistency of the proposed approach is similar to that of the original SCORE.

Given a cluster (subnetwork), the hypothesis testing module determines whether the cluster should be further split. To abuse the notation a little bit, let $A$ be the adjacency matrix of the network formed by restricting nodes and edges to the set of nodes in the current cluster. As  before, we assume $A$ follows a DCBM model with $K_0$ communities and test the null hypothesis $K_0=1$. 
We use the Signed-Quadrilateral (SgnQ) test by \cite{JKL2019}.  Define $\hat{\eta}=\frac{1}{\sqrt{{\bf 1}_n'A{\bf 1}_n}} A{\bf 1}_n\in\mathbb{R}^n$ and $A^*=A-\hat{\eta}\hat{\eta}'\in\mathbb{R}^{n,n}$. The SgnQ test statistic is
\vspace*{-.2cm}
\beq \label{SQ}
\psi_n =\frac{1}{\sqrt{2}}\biggl( \frac{\sum_{i_1,i_2,i_3,i_4 \text{(distinct)}}A^*_{i_1i_2}A^*_{i_2i_3}A^*_{i_3i_4}A^*_{i_4i_1}}{2 (\|\hat{\eta}\|^2-1)^2} -1\biggr). 
\vspace*{-.08cm}
\eeq
It was showed in \cite{JKL2019} that under mild conditions, $\psi_n\to N(0,1)$ in the null hypothesis. This asymptotic normality holds even when the network has severe degree heterogeneity. Then, we can compute the $p$-value conveniently and use it to set the stopping rule of the recursive algorithm (e.g., when $p$-value is $\geq 0.05$, a cluster will not be split). 

{\bf The coauthorshp network (36 journals)}.  
We build a coauthorship network using all the data in 36 journals during 1975-2015 as follows: 
Each node is an author; there is an edge between two nodes if they have coauthored at least $m_0$ papers 
in the data range.  As we wish to focus on (a) the subset of long-term active researchers, and (b) solid collaborations, 
choosing $m_0 = 1$ would be too low (see Ji and Jin 2016)): we may include too many edges between active researchers and non-actives ones (e.g., a Ph.D advisee who joined industry and stopped publishing in academic journals).   
We take $m_0 = 3$ and focus on the giant component, which has 4,383 nodes. 
Taking $m_0 = 2$ may also be a reasonable choice, but the network is comparably denser and larger  (10,741 nodes), 
and so requires more time and efforts 
to interpret the results (as we need to check each identified community one by one manually).  
Below, we present the result for $m_0 = 3$, 
and leave the results for $m_0 = 2$ to Section~D.6 of the supplement, where we see the results of two cases are largely consistent.

\begin{figure}[tb!]
\centering
\includegraphics[width=.95\textwidth]{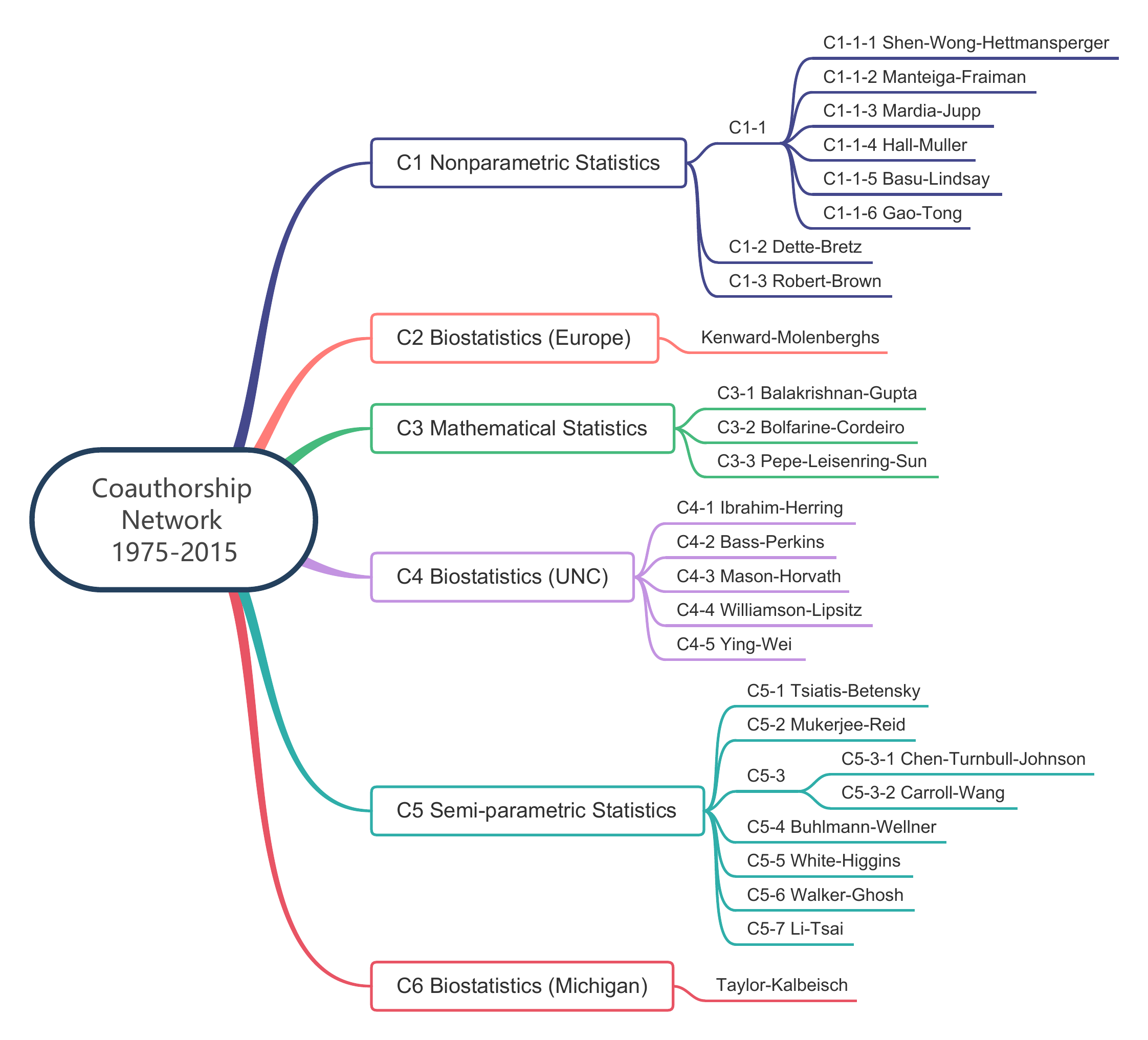}
\caption{\small The community tree for coauthorship network. Each rightmost leaf community is labeled with the last names of 2 or 3 authors, selected by node betweenness and closeness.  For each leaf, the representative nodes are shown in Table~\ref{tab:cd_names_all} (and Tables~D.4-D.6 in the supplement). 
} \label{fig:coau_tree}
\end{figure}

We now apply our proposed algorithm. Note that the community detection module still requires an input of $K_0$.  
Similar to that in Section~\ref{subsec:triangle}, we choose $K_0$  by 
combining the scree plot, goodness-of-fit, and evaluation of output communities (details are in Section~D.4  of the supplement). 
Since we use the eigenvectors of $(A+I_n)$ for community detection, the scree plot contains the absolute eigenvalues of $(A + I_n)$ instead of those of $A$. The stopping rule of the recursive algorithm is set as follows: Either the SgnQ $p$-value is $>0.001$ or the community has $\leq 250$ nodes. The output is a hierarchical community tree in Figure~\ref{fig:coau_tree}.

{\bf The hierarchical community tree.} 
First, we investigate the 6 communities in the first layer. To help for interpretation, we apply topic modeling on paper abstracts (see Section~D of the supplement, especially Figure~D.6). Combining the topic modeling results with a careful read of the large-degree nodes in each community, we propose to label 
these communities as in Table \ref{tb:hierarchical_CommName}, where we also list some comments on each community.
\spacingset{0.95}\footnote{In Section~\ref{subsec:triangle}, ``Bayes" is one of the three vertices of the statistics triangle. Here,  Bayes continues to play an important role, but it splits into multiple communities and so the word ``Bayes" does not appear in the community labels. }  
\spacingset{1.43}

\spacingset{1}
\begin{table}[tb!]
\centering
\scalebox{.8}{
\begin{tabular}{p{0.35\textwidth}p{0.85\textwidth}}
 Community & Description\\
 \hline
C1. Non-parametric Statistics & Decision theory, non-parametric methods, high-dimensional statistics  \\
C2. Biostatistics (Europe) & Biostatisticians from Europe, and their close collaborators\\
C3. Mathematical Statistics & Testing, computational statistics, probability, and other classical topics in probability and statistical theory\\
C4. Biostatistics (UNC) &  Survival analysis, longitudinal data analysis, Biostatisticians from University of North Carolina (UNC) and  collaborators\\
C5. Semi-parametric Statistics & Semiparametric methods, machine learning, variable selection, biostatistics\\
C6. Biostatistics (UM) & Biostatisticians from University of Michigan (UM) and close collaborators\\
\hline
\end{tabular}}
\spacingset{0.9}
\caption{\small The communities C1, C2, $\ldots$, C6 and a brief description for each community.} 
\label{tb:hierarchical_CommName}
\end{table}
\spacingset{1.43}



Next, we look at the other layers of the tree. The stopping rule of recursive partition is that either the SgnQ $p$-value is $>0.001$ or the community size is $\leq 250$, but there are a few exceptions in Figure~\ref{fig:coau_tree}: (a) C6 has 264 nodes, but its giant component has no more than 250 nodes. We thus keep C6 unchanged. (b) The second largest component of C4 contains 60 nodes which form a tight-knit group. While these nodes are not in the giant component, we keep them as a separate community C4-5. (c) C3-1 has 311 nodes and its $p$-value $\approx 0$. However, after we further split it into 2 sub-communities by SCORE, one sub-community contains only 8 nodes, and the other has a $p$-value $0.1$. We thus keep C3-1 unchanged.

\spacingset{0.9}
\spacingset{1.43}

For each leaf community (i.e., the community corresponding to a leaf in the tree),    
we provide a manual label using two commonly used centrality measures, the  {\it betweenness} \citep{freeman1977set} and the {\it closeness} \citep{bavelas1950communication}. For a node in a community, its {\it betweenness} is defined as the number of 
pairs of nodes in the same community that are connected through this node via the shortest path  
(therefore, a node with a large betweenness plays an important role in bridging other nodes), and the {\it closeness} of the node is defined as the reciprocal of the sum of distances from all other nodes in the same community to this node. Given a leaf community, we use the last names of the two nodes with largest betweenness and the one node with largest closeness to label the community (of course, if the latter happens to be one of the former,  we will not use the same name twice).  As a result,  each leaf community  is labeled with the last names of either two or three authors (not necessarily in alphabetical order). 
Table~\ref{tab:cd_names_all}
presents a few representative nodes for each leaf community.  
More information of each leaf community is in Tables~D4-D.6 of the supplement.

\spacingset{.9}
\renewcommand\theadalign{bc}
\renewcommand\theadgape{\Gape[3pt]}
\renewcommand\cellgape{\Gape[3pt]}
\setlength{\tabcolsep}{3pt}

\begin{table}[tb!]
\centering
      \scalebox{0.62}{
        \begin{tabular}{|c|c|c|c|c|l|}
        \hline
        ID & Name & \#Authors & p-value& Representative Authors\\
\hline
C1-1-1 & \makecell{Shen-Wong-\\-Hettmansperger} & 144 &  0 &  \makecell[l]{
Hannu Oja, Harvard Rue, Friedrich Gotze, Wei Pan,  \emph{Thomas P. Hettmansperger}, 
\\ Jun Liu, \emph{Xiaotong Shen}, Douglas A. Wolfe, Ishwar Basawa, Leonhard Held
}\\
 \hline
C1-1-2 & \makecell{Manteiga-Fraiman} &118 & $.04$ & \makecell[l]{\emph{Wenceslao Gonzalez-manteiga}, Graciela Boente, Juan Antonio Cuesta, Daniel Pena, 
	 \\Antonio Cuevas, \emph{Ricardo Fraiman}, Richard Johnson, Michael Akritas
}\\
\hline
C1-1-3 & \makecell{Mardia-Jupp}& 102& $0$ & \makecell[l]{Christian Genest, Ian Dryden, \emph{Kanti V. Mardia}, Rainer Von Sachs, Wensheng Guo
}\\
 \hline
C1-1-4 & \makecell{Hall-M\"uller} & 331 & $.34$  &\makecell[l]{\emph{Peter Hall}, James S. Marron, Jianqing Fan, Liang Peng, Byeong U. Park,
	\\ \emph{Hans-Georg M\"uller,} M. C. Jones,  Laurens De Haan,  Theo Gasser, Wolfgang Hardle
}\\
\hline
C1-1-5 & \makecell{Basu-Lindsay}& 68 & $.012$ &  \makecell[l]{\emph{Bruce Lindsay}, Dankmar Bohning, Domingo Morales, Leandro Pardo,  Dongwan Shin,
	\\ \emph{Ayanendranath Basu}, Maria Luisa Menendez, Konstantinos Zografos, 
}\\
\hline
C1-1-6 & \makecell{Gao-Tong} &189& $0$  & \makecell[l]{Marc Hallin, Wai Keung Li, David Nualart, David Nott, \emph{Howell Tong}, Vo Anh\\
}\\
\hline
C1-2 & \makecell{Dette-Bretz} & 104 &  $.0049$ & \makecell[l]{\emph{Holger Dette}, \emph{Frank Bretz}, Axel Munk, Tony Hayter, Wei Liu, Henry Wynn\\
 }\\
\hline
C1-3 & \makecell{Robert-Brown} & 249 & $0$ & \makecell[l]{William Strawderman, George Casella, Kerrie Mengersen, \emph{Christian Robert},\\
	 \emph{Lawrence Brown}, Tony Cai, Eric Moulines, Murad Taqqu, Anthony Pettitt
 }\\
\hline  
C2 & \makecell{Kenward-Molenberghs}& 202 & $0$ & \makecell[l]{\emph{Geert Molenberghs}, Emmanuel Lesaffre, Marc Aerts, Christophe Croux, Helena Geys,\\
		 \emph{Mike Kenward}, 	Paddy Farrington,  Byron J. T. Morgan, Ariel Alonso
} \\
\hline
C3-1 & \makecell{Balakrishnan-Gupta}& 311 & 0 &  \makecell[l]{\emph{Narayanaswamy Balakrishnan}, \emph{Arjun Gupta}, Manlai Tang, Yasunori Fujikoshi \\
}\\  
\hline
C3-2 & \makecell{Bolfarine-Cordeiro }& 58 & .0003 & \makecell[l]{\emph{Gauss M. Cordeiro}, \emph{Heleno Bolfarine}, Victor H. Lachos, Reinaldo B. Arellano-valle \\
}\\
\hline
C3-3 & \makecell{Pepe-Leisenring-Sun}& 86 & .0002 & \makecell[l]{\emph{Jianguo Sun}, Govind S. Mudholkar, \emph{Margaret Pepe}, Liuquan Sun, \emph{Wendy Leisenring}, \\
	Yudi Pawitan, Xinyuan Song, Xingwei Tong,  Xian Zhou,  Ziding Feng
}\\
\hline
C4-1 & \makecell{Ibrahim-Herring}& 142& .003 &\makecell[l]{\emph{Joseph Ibrahim}, David Dunson, Hongtu Zhu, Andy Lee, Ming-hui Chen, \\
	Keith E. Muller, Kelvin K. W. Yau, Haitao Chu,	Wing Fung
}\\
\hline
C4-2 & \makecell{Bass-Perkins}& 104 &0 & \makecell[l]{
	Yuval Peres, \emph{Richard Bass}, Zhen Qing Chen, Frank Den Hollander,\\
	 Davar Khoshnevisan,	Donald Dawson,  	Klaus Fleischmann, \emph{Edwin Perkins}, Jay Rosen
 }\\
\hline
C4-3 & \makecell{Mason-Horvath}& 109 & 0 & \makecell[l]{
	\emph{Lajos Horvath}, Josef Steinebach, Miklos Csorgo, Luc Devroye, Piotr Kokoszka,\\
	 Evarist Gine,	Armelle Guillou, Marie Huskova, \emph{David Mason}, Ricardas Zitikis
 }\\
\hline
C4-4 & \makecell{Williamson-Lipsitz}& 120 &.0003&\makecell[l]{\emph{Stuart Lipsitz}, Robert H. Lyles, Enrique Schisterman, Brian Reich, \\
	 \emph{John Williamson}, Peter Diggle,	Nan Laird, Huiman X. Barnhart, Amita Manatunga
}\\
\hline
C4-5 & \makecell{Ying-Wei}& 60 &.008 & \makecell[l]{
	\emph{Lee-jen Wei}, \emph{Zhiliang Ying}, Tze Leung Lai, Danyu Y. Lin, David Siegmund, \\
	Daniel Krewski, Lu Tian,	Tianxi Cai, Louis Gordon, Sin-ho Jung
}\\
\hline   
C5-1 & \makecell{Tsiatis-Betensky}&185&.009& \makecell[l]{Paul Yip, Xiaohua Zhou, \emph{Rebecca Betensky}, John Crowley, Adrian Raftery, \\
	\emph{Anastasios Tsiatis}, Ji Zhu, Richard Huggins,	George Michailidis, John Oquigley
}\\
\hline
C5-2 & \makecell{Mukerjee-Reid}& 193 & 0 & \makecell[l]{\emph{Rahul Mukerjee}, Zhidong Bai, Christos Koukouvinos, Kashinath Chatterjee \\
}\\
\hline
C5-3-1 & \makecell{Chen-Turnbull-\\-Johnson}& 201 &.31 & 
\makecell[l]{\emph{Wesley Johnson}, Brian Caffo, Dongchu Sun, Weichung J. Shih, \emph{Bruce Turnbull},\\
	 Richard Lockhart, 	Richard Simon, \emph{Gemai Chen}, Mathias Drton, Galin L. Jones
}\\
\hline
C5-3-2  & \makecell{Carroll-Wang}& 231 & 0 & \makecell[l]{\emph{Raymond Carroll}, Mitchell Gail, Xihong Lin, Laurence Freedman, Hua Liang,\\
		Jianhua Huang, David Ruppert, Suojin Wang, Kevin W. Dodd, Dean Follmann
}\\
\hline
C5-4 & \makecell{Buhlmann-Wellner}& 166 &.0013 & \makecell[l]{Mark Van Der Laan, Aad Van Der Vaart, \emph{Peter Buhlmann}, Subhashis Ghosal,\\
	 Ram Tiwari, Larry Wasserman,	Bin Yu, Joseph Kadane, Thomas Kneib
}\\
\hline
C5-5 & \makecell{Whilte-Higgins}& 71 &.016 & \makecell[l]{Martin Schumacher, Simon Thompson, John Whitehead, Nicky Best, \emph{Ian White},\\
	 \emph{Julian P. T. Higgins},	Jon Wakefield, Dan Jackson, Sylvia Richardson
 }\\
\hline
C5-6 & \makecell{Walker-Ghosh}& 197 & 0 & \makecell[l]{\emph{Stephen Walker}, \emph{Malay Ghosh}, Alan Gelfand, Pranab Kumar Sen, Robert Kohn,\\
}\\
\hline
C5-7 & \makecell{Li-Tsai}& 159&.034 & \makecell[l]{Lixing Zhu, Robert Tibshirani, Dennis Cook, \emph{Chih-ling Tsai}, \emph{Runze Li},\\
	 Jun Shao, Trevor Hastie, 	Shein-chung Chow, Riquan Zhang, Andreas Buja
}\\
\hline
C6 & \makecell{Taylor-Kalbfleisch}& 264 &0 & \makecell[l]{\emph{Jeremy Taylor}, Xin Tu, Daniel Commenges, Donald R. Hoover, Thomas Ten Have
}\\
\hline
\end{tabular} 
}
\caption{\small The leaf communities and the representative authors (ordered by degree within leaf community). To label a community, two or three authors are selected by node betweenness and closeness; if any of them is also a representative author, we present his/her full name in italics. More details are in Tables~D.4-D.6 of the supplement.}
      \label{tab:cd_names_all} 
\end{table} 
\spacingset{1.43}

The results confirm that there are multiple
factors for the formation of a tightly knit cluster of coauthorship: 
similar research interest,  academic genealogy, friendship, colleague relationship, geological 
proximity, or close cultural ties. Below are some examples. 

{\it Example 1. Similar research interest}.  A number of leaf communities can be interpreted as 
groups of researchers sharing similar research interest.  For example: 
{\it C1-3: Robert-Brown (Decision theory)},  
{\it C1-1-4: Hall-M\"uller (Nonparametric statistics)}, 
{\it C4-2: Bass-Perkins (Probability)},  
{\it C4-5: Ying-Wei (Sequential data analysis)},  
{\it C5-4: B\"uhlmann-Wellner (Theoretical machine learning)}, 
{\it C5-3-2: Carroll-Wang (Semi-parametric statistics)},  
{\it C5-7: Li-Tsai (Variable selection and dimension reduction)}.

{\it Example 2.  Geological and cultural factors}. It is more likely for people who are geologically or culturally close to each other (e.g., colleagues, researchers in neighboring institutes or in the same region or country)  to form 
tightly knit clusters.  For example:  {\it C2: Kenward-Molenberghs (Biostatisticians in Belgium)}, {\it C4-1: Ibrahim-Herring (Statisticians in the North Carolina research triangle)}, and {\it C5-5: White-Higgins (Biostatisticians in the U.K.)}. 
Additionally,  C4-1 also contains a group of statisticians in Hong Kong, China. This group is brought together with the North Carolina group largely due to the collaboration between Joseph Ibrahim (faculty at University of North Carolina (UNC))  and Qi-Man Shao (faculty at the Chinese  University of Hong Kong).  
Our analysis also suggests that the geological and cultural effect plays a more important role in forming clusters among biostatisticians than (say) among theoretical statisticians, and a possible reason is that collaborated research in biostatistics depends more on manpower and data sharing. For example, to comply with the data-sharing policies, it is simply easier for one to collaborate with someone in the same institute/country than with others. 

 {\it Example 3. Academic genealogy}. The academic  advisor-advisee relationship is also a common source of collaboration. For example, the leaf community {\it C1-1-1 Shen-Wong-Hettmansperger}
has a component of 29 nodes, which is largely formed by students of three authors, Wing H Wong, Jun Liu, and Xiaotong Shen; Liu and Shen are former students of Wong. We also note that this leaf community has sub-communities. For example, the network has a component of 24 nodes containing 
Thomas P. Hettmansperger. We did not further split C1-1-1 simply because its size falls below 250.

Recall that we name the first-layer communities, C1, C2, \ldots, C6, using the results of topic learning (see Figure~D.6  and Table~\ref{tb:hierarchical_CommName}). In most cases, the interpretations of umbrellaed leaf communities match with the name of the first-layer community. One exception is ``C3-3 Pepe-Leisenring-Sun." It is under ``C3 Mathematical Statistics" but consists of a group of biostatisticians. After some investigation, we find that this group is brought together with other groups in C3 largely by the author Xingqiu Zhao. She collaborated with both Narayanaswamy Balakrishnan, a hub node of C3, and Jianguo Sun, a hub node of C3-3.

The community tree is constructed by SCORE. To compare with other clustering methods, 
we apply Newman-Girvan's modularity approach (Newman's spectral approximation) \citep{newman2006modularity} 
to the same co-authorship network, and obtain $6$ communities.  We then check the numbers of nodes 
in the intersection between each of these communities and each of $26$ leaves in our tree. The results are in Table~D.7 of the supplement.   We find that for most of the $26$ leaf communities identified by SCORE, the majority of nodes in the community are contained in one of the $6$ communities identified by Newman's approach. Therefore,  at least to some extent,  two clustering results are consistent with each other.

\subsection{Evolvement of coauthorship clusters} \label{subsec:sankey}
Our data set spans a relatively long time period (1975-2015), and it is interesting to study and  visualize how the network communities evolve over time. The Sankey plot is a popular visualization tool for dynamic networks. 
However, to have a nice plot with interpretable results, we face many challenges: (a) the coauthorship network constructed using all data has too many communities (so it is hard to interpret all of them, and the resultant Sankey plot will also be too crowded); (b) 
it is unclear how to determine the number of communities; (c) it is also unclear how to interpret each community.

For (a), we decided to focus on the coauthorship network constructed with only  papers from 4 representative journals, {\it AoS}, {\it Bka}, {\it JASA}, and {\it JRSSB} (the full journal names are in Table~B.1).  Compared to the co-authorship network constructed with the papers in all 36 journals,   research interests of the authors in the current network are more homogeneous.  As a result, the network has many fewer communities and is comparably easier to analyze. We have also spent a lot of efforts in dealing with challenges (b)-(c); see details below. 

{\bf The dynamic coauthorship networks (4 journals).} We consider three time windows in our study: (i) 1975-1997, (ii) 1995-2007, and (iii) 2005-2015. As in many works on dynamic network  analysis \citep{kim2018review},   we let the adjacent time windows be slightly overlapping, so the results on community detection will be much more stable.  For each time period,  
we construct a coauthorship network where each author who has ever published in any of the $4$ aforementioned journals during this time period is a node, and two nodes have an edge if and only if they have coauthored one or more papers. For each network, there are relatively few nodes outside the giant component, so we remove them and consider the giant component only. 
  Denote the resultant coauthorship networks for the three time periods by $G_1, G_2$ and $G_3$, respectively.

\begin{figure}[tb!]
\centering
\includegraphics[width=.83\textwidth]{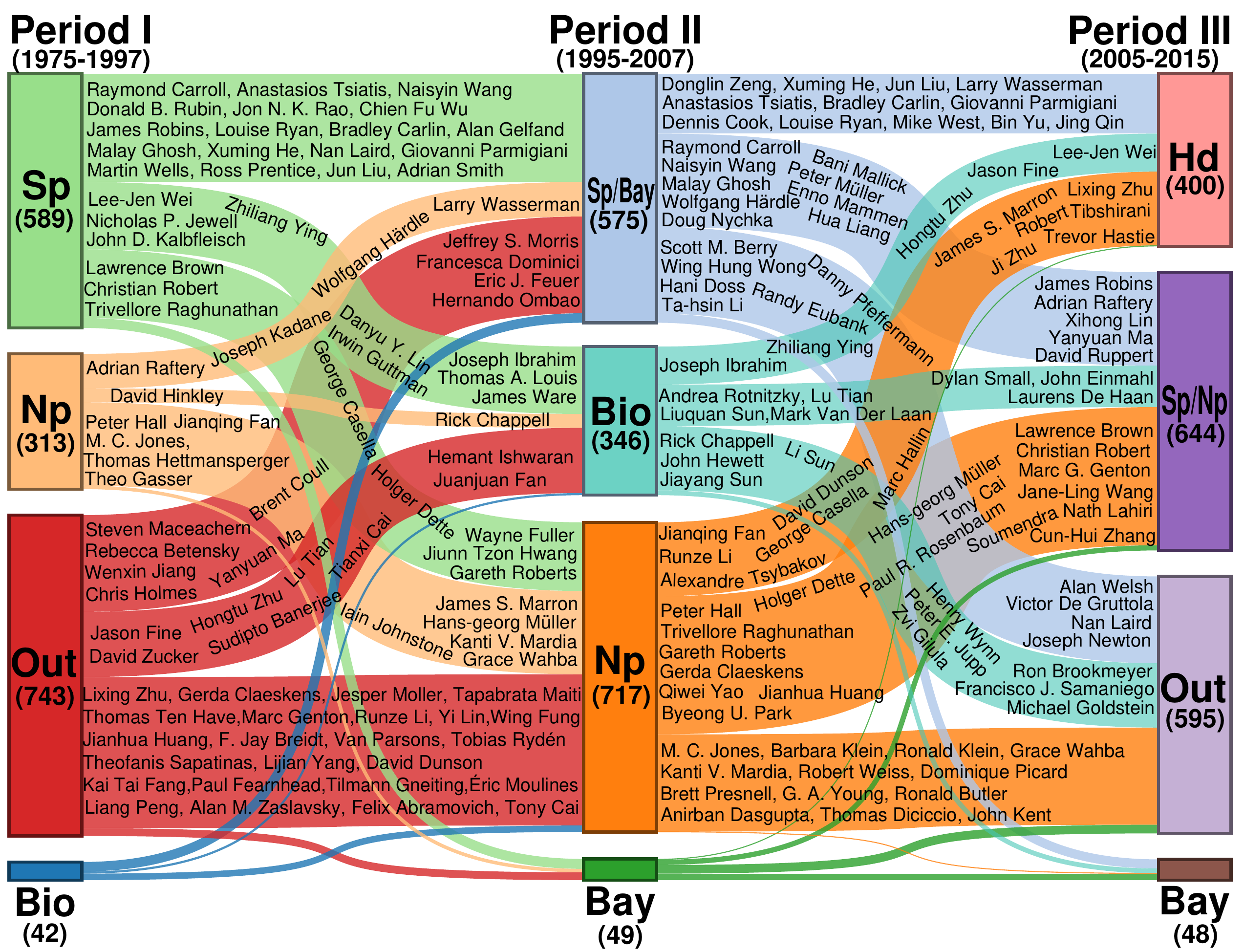}
\caption{\small Evolution of communities in the dynamic coauthorship network (based on papers in 4 journals). The representative authors are selected by average degree in two adjacent networks.}
\label{fig:sankey}
\end{figure}

{\bf The Sankey diagram}. 
By careful investigation, we found that the three networks have 3, 4, and 3 communities respectively.  Once these numbers are determined, we first perform a community detection for each network by applying the modified SCORE described in Section~\ref{subsec:tree},  and then use the  estimated community labels to generate a Sankey diagram; see Figure~\ref{fig:sankey}.    Since the sets of nodes of three networks are different,  we focus on the set $V = (G_1\cap G_2)\cup (G_2\cap G_3)$, which has 1,687 nodes, for the Sankey diagram. 

We explain some notations in Figure~\ref{fig:sankey}.  Consider the network for the time period 1 first.   By similar analysis as before, 
we propose to label the three communities obtained from applying modified SCORE to the network  
by {\it semiparametric statistics (SP)},  {\it nonparametric statistics (NP)}, and {\it Bayes (Bay)}. 
We do not have a separate community for biostatisticians, but a significant number of biostatisticians  (e.g., Jason Fine,  Lu Tian, Hongtu Zhu) are 
outside $V$, and another significant number of them (e.g., Lee-jen Wei, Zhiliang Ying, Joseph Ibrahim, Nicholas P. Jewell) are in SP. 
 Let SP1, NP1, and Bio1 be the intersection of $V$ and each community, respectively.   
We have $V = SP1 \cup NP1 \cup Bio1  \cup O_1$,  where $O_1  = V \setminus G_1$.

The discussion of the third network is similar, except that the estimated communities are interpreted as {\it high-dimensional data analysis (HD)}, {\it nonparametric and semiparametric (NP/SP)}, and {\it Bayes (Bay)}.  Similarly, $V = HD \cup (NP/SP)  \cup Bay3  \cup O_3$,  where $O_3  = V \setminus G_3$. 

Last, consider the second network.   The four communities obtained by applying   SCORE 
can be similarly interpreted as {\it seimparametric statistics and Bayes (SP/Bay)}, {\it nonparametric (NP)}, {\it Bayes (Bay)}, and {\it biostatistics (Bio)}.  We have $V = (SP/Bay) \cup NP2 \cup Bay2 \cup Bio2$, 
where NP2 is the intersection of NP with $G_2$; similar for Bay2 and Bio2. 
Note here  that $V$ is a subset of $G_2$ (but not a subset of $G_1$ or $G_3$), and so $O_2 = V \setminus G_2$ is an  empty set.  See Figure~\ref{fig:sankey} for details.

\spacingset{0.9}
\spacingset{1.43}

The Sankey diagram suggests several noteworthy observations. First, in time period 1, our algorithm suggests that there is no  ``Bio" community, although many biostatisticians (e.g., Jason Fine, Hongtou Zhu, Lu Tian) are outside the set $V$ (recall that $V = (G_1\cap G_2)\cup (G_2\cap G_3)$). 
In time period 2, our algorithm suggests that there is a ``Bio" community, where  
 a significant fraction of the members come from the outside of $V$, and another significant fraction 
(e.g., Lee-jen Wei, Zhiliang Ying, Joseph Ibrahim, Nicholas P. Jewell) come from SP in time period 1.  
Second,  from time period 2 to time period 3, a noticeable point is the rise of the community of {\it high dimensional data analysis (HD)}, which attracts authors from nonparametric statistics (e.g., Jianqing Fan, David Dunson, James  S. Marron, Lixing Zhu), semiparametric statistics and Bayes (e.g., Dongling Zeng, Xuming He, Jun Liu, Larry Wassermann), and biostatistics (e.g., Joseph Ibrahim, Zhiliang Ying, Hongtu Zhu, Jason Fine).  
Last, in all three time periods, there are significant migrations between 
semiparametric statistics and nonparametric statistics. 

Also, as examples, we note that (a) Raymond Carroll, Malay Ghosh, Bruce Lindsay, Ross Prentice, Jon N. K. Rao, James Robins, and Naisyin Wang remain in ``SP" all the time; (b) Peter Hall, Hans-Georg M\"uller remain  in ``NP" all the time; (c) Jianqing Fan, Trevor Hastie,  James S. Marron, Robert Tibshirani stay in ``NP" in time period 1, 2, and migrate to ``HD" in period 3; 
(d) Bradley Carlin, Xuming He,  Jun Liu,  Rahul Mukerjee, Lous Ryan,  Anastasios Tsiatis, and Martin Wells, stay in ``SP" in time period 1, 2 and migrate to ``HD" in  period 3. 
(e) Danyu Y. Lin,  Lee-jen Wei, Zhiliang Ying start from ``SP" in time period 1, migrate to ``Bio" in   period 2, and migrate to ``HD" in  period 3.


\subsection{A new approach to measuring an author's research diversity} \label{subsec:personalized} 
In Section \ref{subsec:citee-diversity}, we have proposed two diversity metrics 
for the research interests of individual authors, using the trajectory. 
In this section, we propose a new approach to measuring research diversity by using the personalized networks and 
a recent tool in network global testing. The approach is quite different from that in Section \ref{subsec:citee-diversity} (and also those in the literature), and  provides new insight on the research diversity of statisticians. 

Fixing a node in a symmetrical network, the {\it personalized network} (also called the ego network) is the subnetwork 
consisting of the node itself and all of its adjacent nodes.  
We construct a coauthorship network similar to that in Section~\ref{subsec:tree} but with $m_0=1$: Every author who ever published a paper  in any of the 36 journals between 1975 and 2015 is a node, and two nodes have an edge if and only if they coauthored one or more papers. Once this large network is constructed,  for every author,  we can obtain a personalized coauthorship network accordingly.

We model each personalized coauthorship network with a DCBM model \eqref{DCMM} with $K$ 
communities. We consider the global testing problem \citep{YFS2018} where we test  
$H_0$: $K = 1$ versus $H_1$:   $K > 1$.  
Viewing each community as a tight-knit group, this is testing whether the 
given personalized coauthorship network has only one or multiple tight-knit groups. 
We approach the testing problem by the SgnQ test \citep{JKL2019} which was already described in Section~\ref{subsec:tree}. 
Let $Q_i$ be the test score $\psi_n$ in \eqref{SQ} for the personalized coauthorship network of author $i$. According to \cite{JKL2019}, when the null hypothesis is true,  $Q_i\to N(0,1)$ as the size of the personalized network grows to $\infty$. We thus calculate the $p$-value by $p_i=\mathbb{P}(N(0,1)\geq Q_i)$ and assign $p_i$ to author $i$.  We propose to use $p_i$ to measure the coauthorship diversity of author $i$: a large $p$-value suggests that  
his/her coauthors form a tightly knit group, and a small $p$-value suggests that 
his/her coauthors are from two or more groups and so he/she is more diverse in coauthorship.

\begin{figure}[tb!]
\centering
\includegraphics[height=0.25\textwidth]{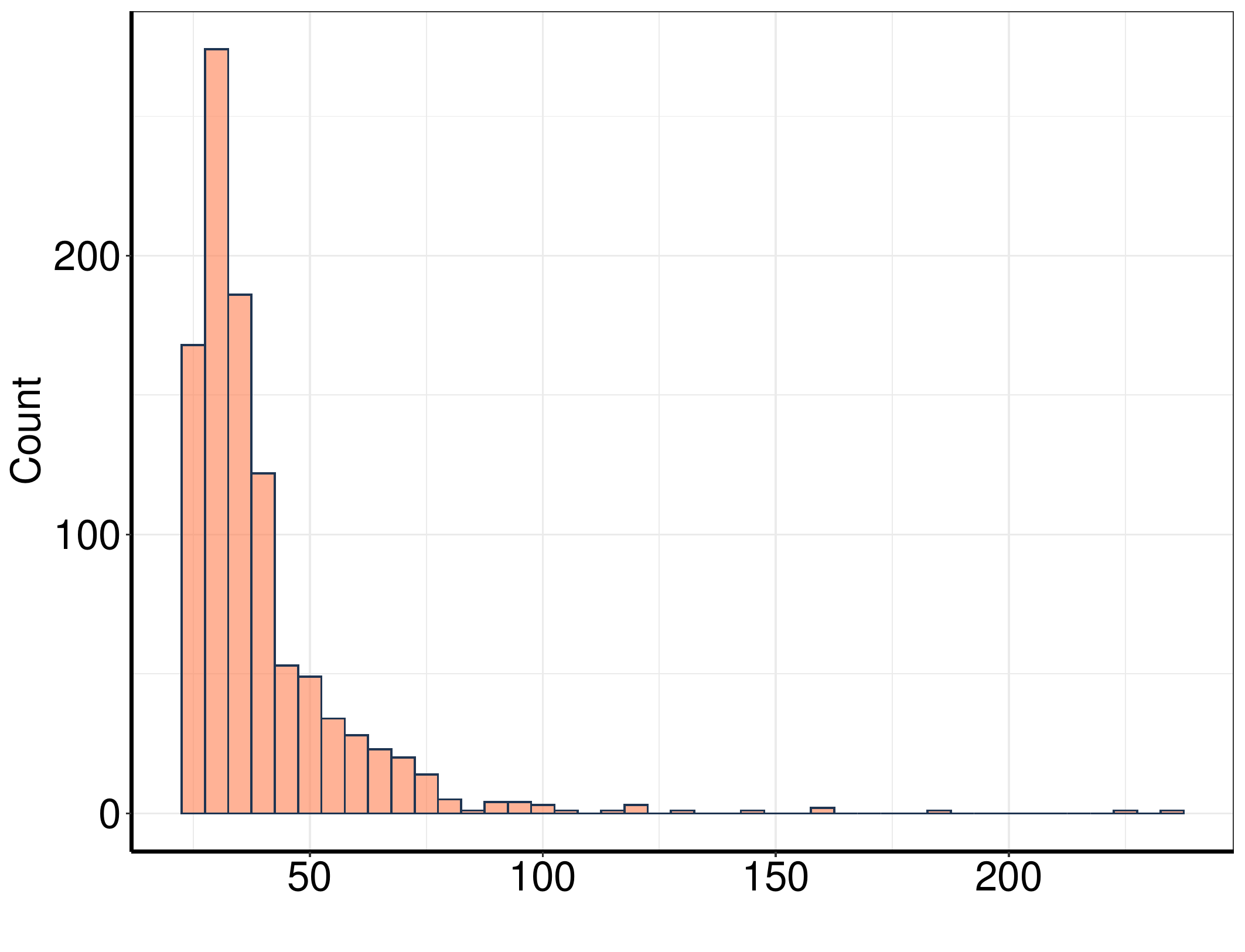} $\quad$
\includegraphics[height=0.25\textwidth]{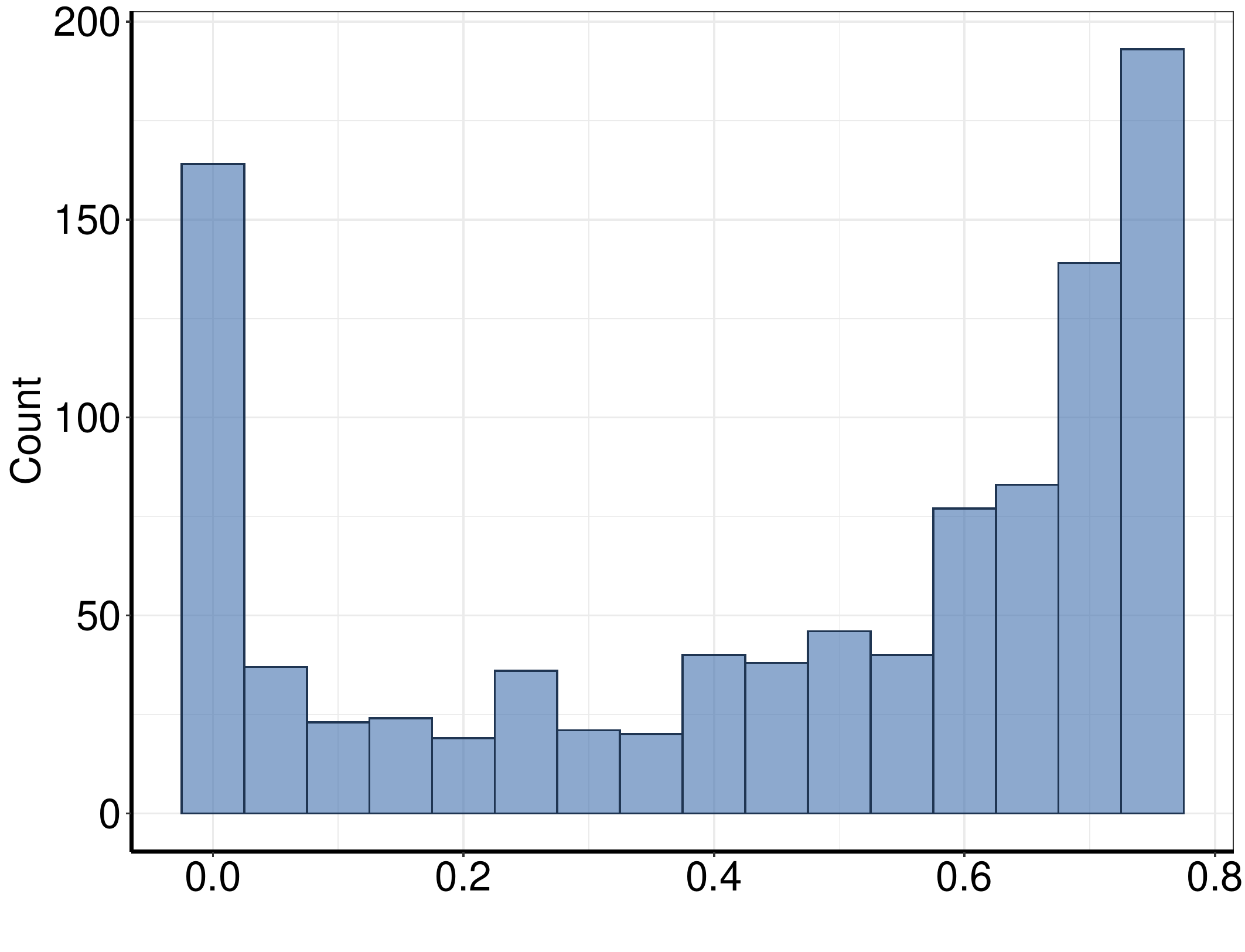}
\caption{\small Left: histogram for the numbers of coauthors of 1,000 authors who have the largest number of coauthors in our data set. Right: histogram for the SgnQ $p$-values for the 1,000 personalized coauthorship networks.  A smaller $p$-value suggests that the personalized 
network is more likely to have multiple tight-knit groups (so the author is more diverse in terms of coauthorship).}
\label{fig:p_Q_coau}
\end{figure}

Figure~\ref{fig:p_Q_coau} presents the results for the personalized coauthorship networks 
of 1,000 authors who have the largest numbers of coauthors in our data set. 
 The left panel presents the histogram for the numbers of coauthors of these 1,000 authors, and the right panel presents the histogram for the $p$-values of their personalized 
 coauthorship networks.  
The $p$-values spread between $0$ and $0.8$, and 190 of them are  
 smaller than $5\%$. Therefore,  for about $80\%$ of these 1,000 authors, their coauthors form a tight-knit group.
 
\spacingset{0.9}
\spacingset{1.43}

\spacingset{1}
\begin{table}[bt!]
\centering
\scalebox{.75}{
\begin{tabular}{lll | lll | lll}
\hline
Name & \#Coau & $p$-value &  Name &  \#Coau  & $p$-value & Name & \#Coau & $p$-value  \\
\hline
Raymond Carroll & 234 & .02 &  Geert Molenberghs & 146 & 0  & Pranab Kumar Sen & 112 & .71 \\
Peter Hall & 222 & .23 &  James S. Marron & 130 & .007  &  Lixing Zhu 	& 103 & .65    \\
Naray. Balakrishnan & 186 & .70 &  Malay Ghosh & 119 & .51 &  David Dunson & 101 & .64  \\
Jeremy Taylor & 159 & 0  &  Emmanuel Lesaffre & 119 & 0   &   Jianqing Fan & 101 &	.38    \\
Joseph Ibrahim & 158 & 0.01 & Xiaohua Zhou & 119 & .31 &  Stuart Lipsitz & 98 & .11\\
\hline
\end{tabular}}
\spacingset{0.9}
\caption{\small Numbers of coauthors and $p$-values of the personalized coauthorship networks for the 15 authors who have the largest numbers of coauthors in our data set (zero $p$-value means $<10^{-6}$).} \label{tb:p_Q_coau}  
\end{table}
\spacingset{1.43}

\begin{figure}[tb!]
\centering
\includegraphics[width=0.4\textwidth, trim=35 0 0 0, clip=true]{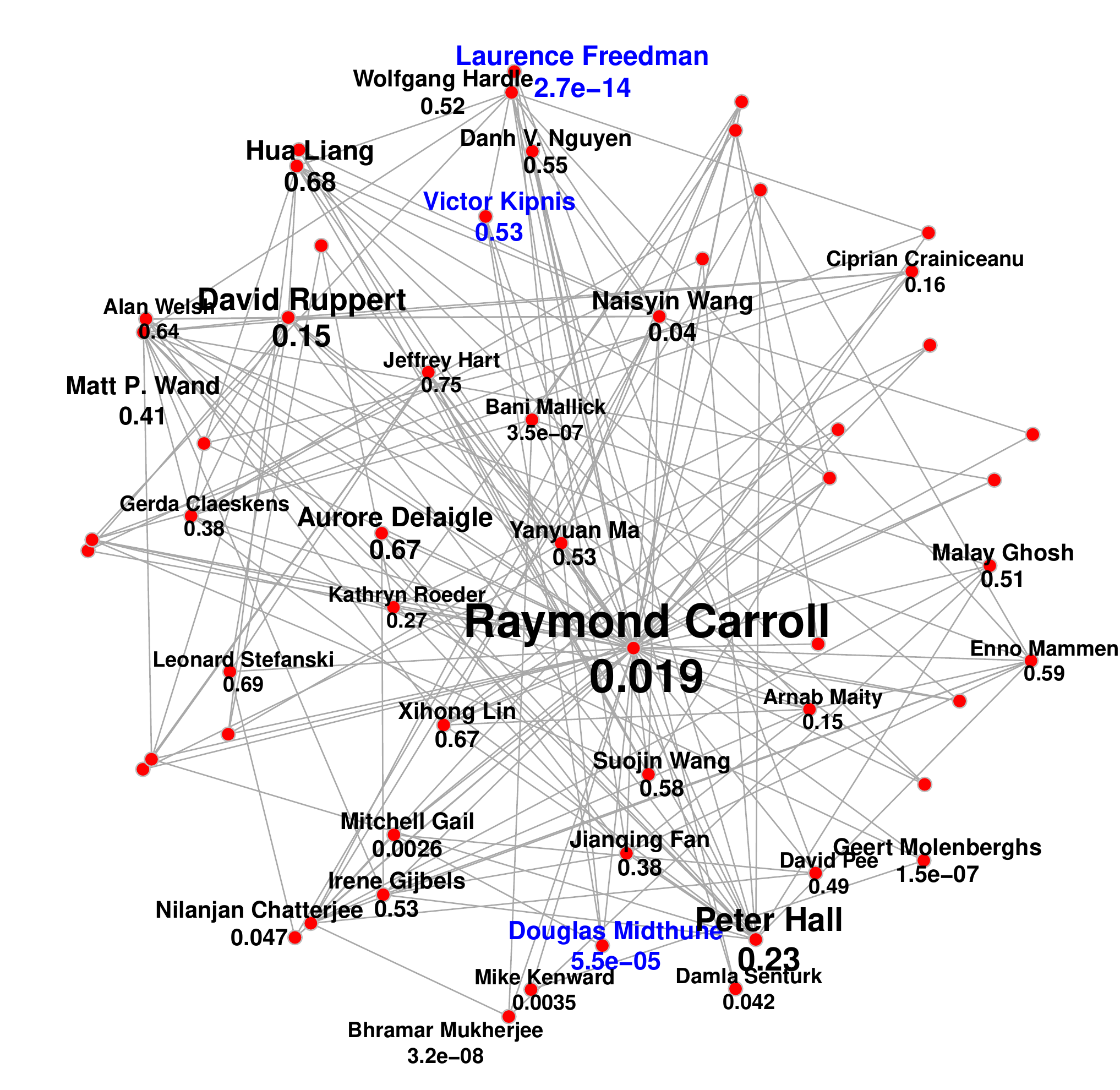}
\includegraphics[width=0.55\textwidth, trim= 10 15 0 0, clip=true]{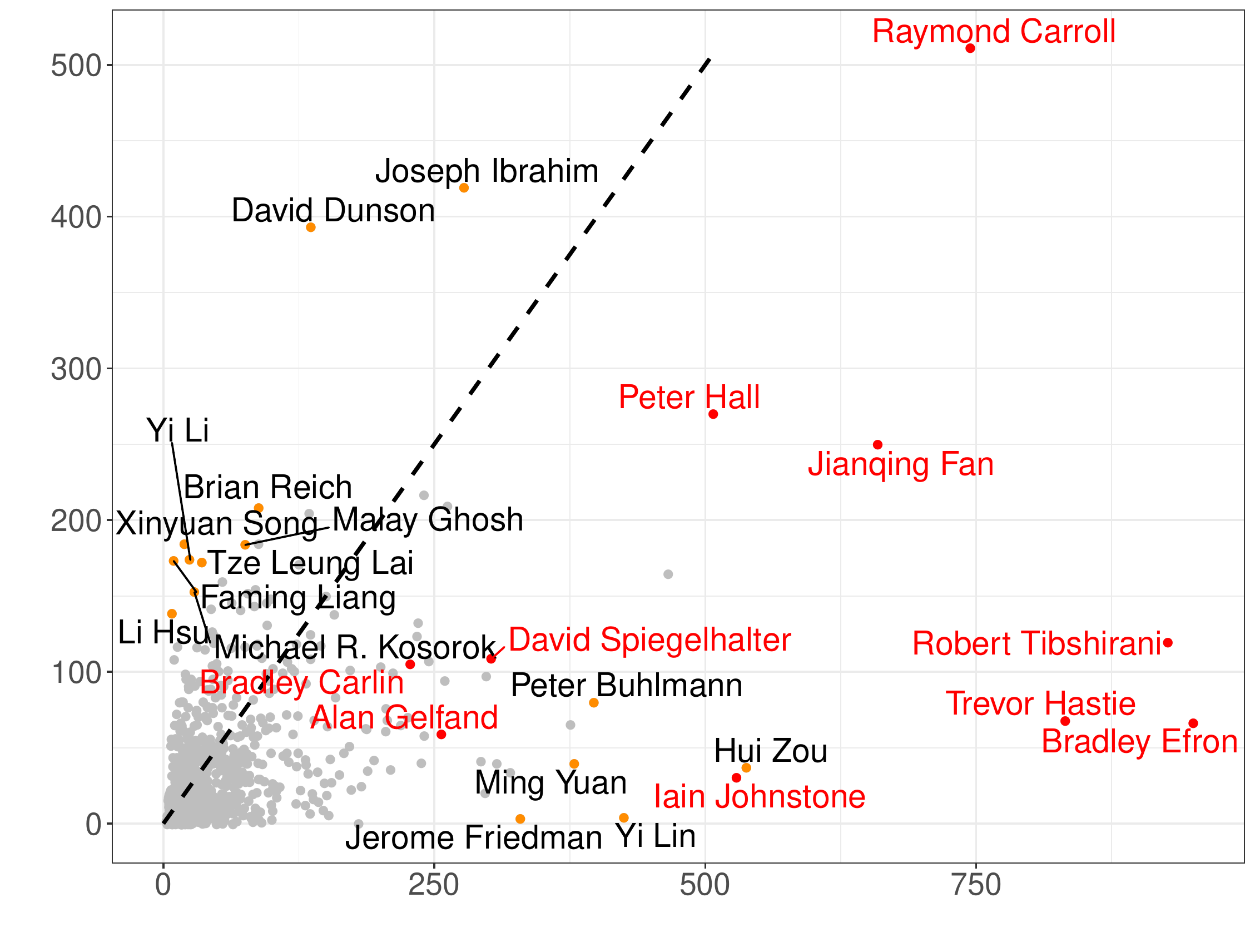}
\caption{\small Left: The personalized coauthorship network of Raymond Carroll (the most collaborative author; see Table \ref{tb:p_Q_coau}). Only nodes with $40$ or more coauthors are shown. Different colors of names indicate two communities identified by SCORE.  Similar plot can be generated for any author whose personalized network is reasonably large ($\geq 50$ nodes, say). Right: The pair SgnQ test statistics $(T_i^{citer}, T_i^{citee})$ on personalized citer and citee networks of 1,000 authors with highest degrees. The red dots correspond to high-degree authors. The yellow dots correspond to authors with either the largest or the smallest values of $(T_i^{citer}-T_i^{citee})$}
\label{fig:test_rc_40}
\end{figure}

Moreover, Table~\ref{tb:p_Q_coau} presents the $p$-values from the SgnQ test for the personalized networks of 15 authors who have the largest numbers of coauthors.  Take the first two authors, for example. They both have a large number of coauthors, but the $p$-value for Raymond Carroll is $0.02$ while the $p$-value for Peter Hall is $0.23$. This suggests that Hall's coauthors are likely to form  
a tight-knit group, while Carroll's coauthors may come from multiple groups. 
To identify such groups, we perform a community detection on Carroll's personalized coauthorship network 
(excluding Carroll \spacingset{0.95}\footnote{We exclude Carroll here for the edges between him and all other nodes contain little information 
of the community structure, but have a significant effect in the spectral domain, which makes the estimated communities by SCORE (a spectral method) less clear.}\spacingset{1.43})   
by SCORE (see Section \ref{subsec:tree} and \cite{Jin2015})  and find that the research areas of a group of coauthors (e.g., Laurence Freedman, Victor Kipnis, Douglas Midthune, etc.---they work or used to work for National Cancer Institute (NCI))  are quite different from those of the other coauthors of Carroll.  This explains why Carroll's network has a relatively small $p$-value. 
See Figure~\ref{fig:test_rc_40} (left panel) for the personalized coauthorship network of Carroll, where the $p$-value of any node presented there is the $p$-value for his/her own personalized coauthorship network.


\spacingset{0.9}
\spacingset{1.43}


{\bf Extension to measuring the diversity of citers and citees}.    
We extend the study to personalized 
citer/citee networks. 
In a citer network, two authors have an edge if they have both cited some other authors. 
In a citee network, two authors have an edge if they have been both cited by some other nodes. 
Similarly as above, we construct a personalized citer network and a personalized citee network for each author $i$. We apply the SgnQ test and denote the two test scores by $T_i^{citer}$ and $T_i^{citee}$, respectively.   
Figure~\ref{fig:test_rc_40} shows the two test scores for 1,000 authors with the largest numbers of coauthors. 
First, 
for most authors (705 out of 1000) the personalized citer network is more diverse than the personalized citee network.   
This is because each author typically focuses on only a few research areas, but his/her work may be cited by researchers from various areas.
Second, there is a group of authors whose $T_i^{citee}$ is much smaller than $T_i^{citer}$, most of whom are theoretical statisticians (e.g., Bradley Efron, Iain Johnstone). This is probably because theoretical papers mainly cite theoretical papers but can be cited by many methodology and applied papers.
Third,  there is a group of authors in biostatistics (e.g., Michael Kosorok, Tze Leung Lai), whose test score for the citee network is much larger than that for the citer network.  This is probably because biostatistics papers cite a variety of methodology papers; another reason is that many citations to papers in biostatistics are from other disciplines not covered by our data set.    
Last,  
for Raymond Carroll, Jianqing Fan, Peter Hall, and Joseph Ibrahim, 
both test scores are relatively large, suggesting that they are diverse both in citer and citee. 


We have proposed 5 metrics for measuring the research interests and diversity: 
two (denoted by A1 and A2) in Section \ref{subsec:citee-diversity} where we measure the diversity using 
the research trajectory computed from the co-citation networks, and three (denoted by
B1-B3) in this section, for the co-authorship, citer, and citee networks, respectively.   
These metrics measure diversity from different angles using different types of networks. 
Also, 
the networks are based on data in different ranges. 
For these reasons, our results on diversity may have some inconsistencies, 
and we must interpret them with caution. For example,  it is not rare that a paper on one research topic may impact several other research topics, so an author who is not diverse in co-authorship 
can be significantly diverse in research impacts. For example, most papers by Donald Rubin's are in  
Bayesian statistics and causal inferences, but he has impacts over many other areas (e.g.,  GEE); see Figures \ref{fig:simplex2}-\ref{fig:trajectory_all}.    
Xihong Lin is regarded as highly diverse in research impact,  but  not regarded as diverse in co-authorship (based on results in our data range); see Figures \ref{fig:trajectory_all} and \ref{fig:test_rc_40}.   
Also, while Approaches A1-A2 and B3 are both for citee networks, A1-A2 are for a dynamic DCMM setting and measures how the membership vector,  $\pi_{it}$, evolve over time,  and B3 considers a (static) DCMM setting and measures 
whether the personalized network has only one or multiple communities.   

For reasons of space, we focus on the network approach in this paper where we model the co-author relationships by networks. As an extension, we may model the co-author relationships
by the more sophisticated hypergraph model (e.g., \cite{JinTensor, FengTensor, KeTensor}). 
In comparison, the literature on the hypergraph approach is much less developed than that of the network approach,  so we leave the study on the hypergraph approach to the future.

\section{Conclusion}  \label{sec:conclusion}

We have several contributions. First, we produce a large-scale high-quality  data set.  
Second, we set an example for how to conduct a data science project that is highly demanding (in data resource, tools, computing, and time and efforts). We showcase this by creating a research template where 
we (a) collect and clean a valuable large-scale data set, (b) identify a list of interesting 
problems   in social science and science, (c) attack these problems by 
developing new tools and by adapting exiting tools, (d) deal with a long array of 
challenges in real data analysis so as to get meaningful results, and (e) use multiple 
resources to interpret the results, from perspectives in science and social science. 
We have also made significant contributions in methods and theory by developing an array 
of ready-to-use tools (for analysis and for visualization).  

Our study has (potential) impact  in social science, science, and real life. 
For example, suppose an administrator (in an university or a funding agency) wants to learn the research profile of a researcher. Our study provides a long list of tools to characterize 
and visualize the research profile of the researcher.  Such information  can be very useful for decision making. 
Our study also provides a useful guide for researchers (especially junior researchers)
in selecting research topics,  looking for references, and building social networks. 

In social science, an important problem  is to study the 
evolvement of a scientific community \citep{rosvall2010mapping}. 
We attack the problem by providing several tools (e.g., research map, research trajectory,  
Sankey plot) for characterizing and visualizing 
the evolvement of the statistical community.  Another important problem is to check whether the development of a research field is balanced (e.g., if some areas are over-studied or under-studied) 
and whether there are unknown biases (e.g., whether scientists have biases when publishing papers 
related to COVID-19) \citep{foster2015tradition}. 
Our study can tell  which areas  have far more  researchers, papers, or citations than others, and so helps check the balance of the field.  Our study is also potentially useful for 
checking unknown biases.

In science, an important problem is how to identify patterns and so to predict new discoveries 
 ahead of time.  For example, in material science, one can use the abstracts 
of published papers to recommend materials for functional applications 
several years ahead of time \citep{tshitoyan2019unsupervised}. We can do similar things with our data 
set to predict emerging new areas and significant advancements. 
 For example,  in \cite{SCC-paper2},  we combine our citation data with the paper abstracts (treated as text data) 
to rank different research topics and identify the most active research topics.   
We find that in the past decade, machine learning has been rising to one of the active research topics in statistics.  

 Though our data set is high quality,  we still need some necessary data preprocessing, and focus on 
networks with sizes much smaller than $47K$. The bottleneck for studying much larger networks  is the time and efforts required to manually label each research area and to interpret the results in each case. For better use of such a valuable data set, 
our hope is that, the data set (which will be publicly available soon) will motivate many lines of researches, so over the years, researchers may continue to use different parts of the data set  for new projects and new discoveries.

For future work, note that our data set provides at least two data resources: 
co-author relationships and citation relationships.  It is noteworthy that most existing works 
in bibliometrics have been focused on one data source 
and one specific problem.   
Our results suggest the following: (a) The two data resources 
provide different information for the same group 
of researchers,  and analysis 
of different data resources may have different results. 
The data resources and the results complement with each other. 
(b) Analysis focusing on only one aspect may have limited insight. 
Combining analysis of different aspects helps paint a more 
complete picture. 
(c) Therefore, it is highly preferable to combine the data resources 
for our study, with a multi-dimensional framework and multi-way
analysis.  In our real data analysis, we have combined the two data resources. For example, in Section~\ref{subsec:personalized},  we use different metrics to measure the diversity of an author, where some metrics are based on
the co-citation data and others are based on the coauthorship data. 
How to combine different data resources more  efficiently is an interesting problem. 
We leave this to the future work.   


\medskip
\noindent

{\bf Acknowledgments}.  The authors thank the Associate Editor and referees
for very helpful comments. They thank Yoav Benjamini, Raymond Carroll,
David Donoho, Yi Li, Jun S. Liu, Xiao-Li Meng, Neil Shephard, Bill Shi and Peter Song for many helpful comments and encouragements. The research of   
J. Jin is supported in part by NSF Grant DMS-2015469, and
the research of Z. T. Ke is supported in part by NSF Grant DMS-1943902.

{\bf Supplemental Material}: 
Supplemental material contains a disclaimer, details of the data set, supplemental data analysis results, and proof of Theorem~\ref{thm:simplex}.   

\spacingset{0} 
\small
\bibliographystyle{chicago}
\bibliography{reference} 

\newpage

\spacingset{1.45}
 
\appendix

\begin{center}
{\LARGE\bf Supplementary Material}
\end{center}

\section{Disclaimer} \label{sec:disclaimer}
It is not our intention to rank a researcher (or a paper, or an area) over others. For example, when we say a paper is “highly cited,” we only mean that the citation counts are high, and we do not intend to judge how important or influential the paper is. Our results on journal ranking are based on journal citation exchanges, but we do not intend to interpret the ranking more than the numerical results we obtain from the algorithms we use.

As our data set is drawn from real-world publications, we have to use real names, but we have not used any information that is not publicly available. For interpretation purposes, we frequently need to suggest a label for a research group or a research area, and we wish to clarify that the labels do not always accurately reflect all the authors/papers in the group. Our primary interest is the statistics community as a whole, and it is not our intention to label a particular author (or paper, or topic) as belonging to a certain community (group, area).

While we try very hard to create a large-scale and high-quality data set, the time and effort one can invest in a project is limited. As a result, the scope of our data set is limited. Our data set focuses on the development of statistical methods and theory in the past 40 years, and covers research papers in 36 journals between 1975 and 2015 (we began downloading data in 2015). These journals were selected from the 175 journals on the {\it 2010 ranked list of statistics journals by the Australian Research Council} (see Section~\ref{subsec:journals}).  Journals on special themes and most journals on econometrics, interdisciplinary research, and applications are not included (see Section 6.1 for detailed description). As a result, papers on econometrics, interdisciplinary research, and applications may be underrepresented.

Due to the limited scope of our data set, some of our results may be biased. For example, for the citations a paper has received, we count only those within our data range, so the resultant citation counts may be lower than the real counts the paper has received. Alternatively, for each paper, we can count the citation by web searching (e.g., Google Scholar, which is known to be not very accurate), or by reference matching (e.g., Web of Science and Scopus). Our approach allows us to perform advanced analysis (e.g., ranking authors/papers by citation counts, reporting the most cited authors and papers, excluding self-citations, and calculating cross-journal citation). For such analysis, it is crucial that we know the title, author, author affiliation, references, and time and place where it is published for each paper under consideration. For each of the two alternative approaches, we can gather such information for a small number of papers, but it is hard to obtain such information for 83,336 papers as in our data set.

A full scope study of a scientific community is impossible to accomplish in one paper. The primary goal of our paper is to serve as a starting point for this ambitious task by creating a template where researchers in other fields (e.g., physics) can use statisticians' expertise in data analysis to study their fields.  
For these reasons, the main contributions of our paper are still valid, 
despite some limitations discussed above.

\section{Data description and data collection} \label{sec:data-supp}  
One of our contributions is creating a high-quality, large-scale data set on the publications in 36 statistics-related journals (see Section~\ref{subsec:dataset} of the main paper).   
We present information of these journals and describe how the 36 journals were selected. We also discuss the challenges we  encountered in data collection and cleaning,  and how we overcame the challenges. 

\subsection{The 36 journals in the data set} \label{subsec:journals}  
Our data set  consists of papers from 36 journals in statistics, probability, 
machine learning, and related fields. 
Table \ref{tab:journal} presents 
some basic information of these journals. 
The impact factors for each journal in  2014 and 2015 are also included.  

The $36$ journals are selected as follows.  We start with the $175$ journals in the 2010 ranked list of statistics journals provided by the Australian Research Council (ARC) \citep{ARC2010}. \footnote{\url{https://www.righttoknow.org.au/request/616/response/2048/attach/3/2010}.}   The list was used for performance evaluation of Australian universities, as part of its program of {\it Excellence in Research for Australia}. 
The 175 journals are divided into four categories:  $A^*$, $A$,  $B$, and $C$.   
For our study, first, we include all 9 Category $A^*$ journals,  where  two of them  (AOP and PTRF) are probability journals.  Second, we include all Category $A$ journals, except the strongly themed journals in applied probability or in engineering (Advances in Applied Probability, Electronic Journal of Probability, Finance and Stochastics, Journal of Applied Probability, Stochastic Processes and their Applications, Theory of Probability and its Applications, Technometrics, Queueing Systems, Random Structures $\&$ Algorithms).  Last,  there are about $50$ journals in  Category B covering a wide range of themes, where we only select the journals on methodology and theory, such as  Australian $\&$ New Zealand Journal of Statistics, Bayesian Analysis, Canadian Journal of Statistics, etc.   We do not include any Category $C$ journals.   

Econometrics journals (e.g., Journal of Business and Economic Statistics by American Statistical  Association)  are not included in our study (they are not on the ARC list).

\spacingset{1.1}
    \begin{table}[htb!]
      \centering
      \scalebox{0.8}
{
        \begin{tabular}{|r|l|l|c|c|c|c|c|}
        \hline
      & & {\small Abbrev.} & {\small Starting} & {\small  $\#$ of}  & {\small  $\#$ of } &   &  \\  
         & Full name of the journal & {\small Name} & {\small Time} & {\small Papers} & {\small Authors} & {\small IF2014} & {\small IF2015} \\
        \hline
        1      & \textit{Ann. Inst. Henri Poincare Probab. Stat.} & AIHPP  & 1984   & 967    & 1152   & 1.27   & 1.099 \\
        2      & \textit{Annals of Applied Statistics} & AoAS   & 2007   & 729    & 1824   & 0.942  & 0.769 \\
        3      & \textit{Annals of Probability} & AoP    & 1975   & 3318   & 2277   & 2.032  & 1.842 \\
        4      & \textit{Annals of Statistics} & AoS    & 1975   & 4168   & 3065   & 1.729  & 1.968 \\
        5      & \textit{Annals of the Institute of Statistical Mathematics} & AISM   & 1975   & 2016   & 2056   & 3.055  & 3.528 \\
        6      & \textit{Australian \& New Zealand Journal of Statistics} & AuNZ   & 1998   & 592    & 968    & 0.509  & 0.62 \\
        7      & \textit{Bayesian Analysis} & Bay    & 2006   & 138    & 314    & 1.519  & 1.031 \\
        8      & \textit{Bernoulli} & Bern   & 1997   & 1065   & 1446   & 1.829  & 1.412 \\
        9      & \textit{Biometrics} & Bcs    & 1975   & 4347   & 5357   & 1.491  & 1.603 \\
        10     & \textit{Biometrika} & Bka    & 1975   & 3359   & 3239   & 2.94   & 2.114 \\
        11     & \textit{Biostatistics} & Biost  & 2002   & 732     & 1575    & 1.642  & 1.336 \\
        12     & \textit{Canadian Journal of Statistics} & CanJS  & 1985   & 1202   & 1542   & 1.676  & 1.41 \\
        13     & \textit{Communications in Statistics-Theory and Methods} & CSTM   & 1976   & 8390   & 8041   & 0.424  & 0.437 \\
        14     & \textit{Computational Statistics \& Data Analysis} & CSDA   & 1983   & 4656   & 6725   & 0.713  & 0.6 \\
        15     & \textit{Electronic Journal of Statistics} & EJS    & 2007   & 703    & 1156   & 1.303  & 0.903 \\
        16     & \textit{Extremes} & Extrem & 2008   & 176    & 262    & 1.5    & 1.68 \\
        17     & \textit{International Statistical Review} & ISRe   & 1975   & 855    & 1128   & 2.081  & 1.711 \\
        18     & \textit{Journal of Computational and Graphical Statistics} & JCGS   & 1997   & 907    & 1488   & 2.319  & 2.038 \\
        19     & \textit{Journal of Machine Learning Research} & JMLR   & 2001   & 1332   & 2362   & 1.544  & 2 \\
        20     & \textit{Journal of the American Statistical Association} & JASA   & 1975   & 5154   & 5686   & 0.939  & 1.676 \\
        21     & \textit{Journal of the Royal Statistical Society} & JRSSB  & 1975   & 1682   & 1882   & 2.742  & 3.125 \\
               & \textit{Series B-Statistical Methodology} &   &    &     &   &      &  \\
        22     & \textit{Journal of Applied Statistics} & JoAS   & 1993   & 2219   & 3798   & 1.18   & 1.058 \\
        23     & \textit{Journal of Classification} & JClas  & 1984   & 435    & 551    & 0.569  & 0.587 \\
        24     & \textit{Journal of Multivariate Analysis} & JMVA   & 1976   & 3574   & 3601   & 2.286  & 2.357 \\
        25     & \textit{Journal of the Royal Statistical Society} & JRSSA  & 1975   & 1117    & 1821   & 4      & 5.197 \\
               & \textit{Series A-Statistics in Society} &   &    &     &   &      &  \\
        26     & \textit{Journal of the Royal Statistical Society} & JRSSC  & 1975   & 1359    & 2282   & 1.753  & 1.615 \\
               & \textit{Series C-Applied Statistics} &   &    &     &   &      &  \\
        27     & \textit{Journal of Statistical Planning and Inference} & JSPI   & 1977   & 6111   & 6372   & 0.818  & 0.869 \\
        28     & \textit{Journal of Time Series Analysis} & JTSA   & 2000   & 692    & 925    & 0.939  & 1.387 \\
        29     & \textit{Journal of Nonparametric Statistics} & JNS    & 1998   & 817    & 1187   & 0.586  & 0.556 \\
        30     & \textit{Probability Theory and Related Fields} & PTRF   & 1986   & 2164   & 1874   & 1.657  & 2.025 \\
        31     & \textit{Statistical Science} & StSci  & 1993   & 564    & 980    & 1.59   & 1.641 \\
        32     & \textit{Scandinavian Journal of Statistics} & ScaJS  & 1977   & 1393   & 1730   & 2.154  & 1.741 \\
        33     & \textit{Statistica Sinica} & Sini   & 1991   & 1685   & 2235   & 0.718  & 0.63 \\
        34     & \textit{Statistics and Computing} & SCmp   & 1993   & 907    & 1518   & 1.032  & 1.155 \\
        35     & \textit{Statistics \& Probability Letters} & SPLet  & 1984   & 7063   & 6670   & 1.382  & 0.952 \\
        36     & \textit{Statistics in Medicine} & SMed   & 1984   & 6743   & 9575   & 2.942  & 2.817 \\
        \hline
        \end{tabular}%
        
}
\caption{For each of the 36 journals, we present the full name, abbreviated name, starting time, total number of authors, total number of papers, and impact factors 
in 2014 and 2015. For each journal, our data set consists of all papers between a certain year (i.e., the starting time) and 2015. The starting time is not necessarily the year the journal was launched.} 
\label{tab:journal}%
\end{table}%

\spacingset{1.45}

\subsection{Data collection and cleaning} \label{subsec:dataset}
One might think that our data sets are easy to obtain, as it seems that BibTeX and citation data are easy to download.  Unfortunately, when we need a large-volume high-quality data set, this is not the case. For example, the citation data by Google Scholar is not very accurate, and many online resources do not allow large volume downloads.  Our data are downloaded using a handful of online resources by techniques including, but not limited to, web scraping. The data set was also carefully cleaned by a combination of manual efforts and computer algorithms we developed. Both data collection and cleaning are sophisticated and time-consuming processes, during which we have encountered a number of challenges.

The first challenge is that, for many papers, we need multiple online resources to acquire the complete information.  For example, to download complete information of a paper, we might need online resources 1, 3, and 5 for paper 1, whereas   online resources 2, 4, and 6  for paper 2. Also, each online resource may have a different system to label their papers. As a result, we also need to carefully match papers in one online resource to the same ones in another online resource. These make the downloading process rather complicated.

The second challenge is name matching and cleaning. For example, some journals list the authors only with the last name and first initial, so it is hard to tell whether “D. Rubin” is Donald Rubin or Daniel Rubin. Also,  the name of the same author may be spelled differently in different papers (e.g.,  “Kung-Yee Liang”  and  “Kung Yee Liang”).    A more difficult case is that different authors may share the same name (e.g., Hao Zhang at Purdue University and Hao Zhang at Arizona State University). To correctly match the names and authors, we have to combine manual efforts with some computer algorithms.

Last, an online resource frequently has internal inconsistencies, syntax errors, encoding issues, etc. We need a substantial amount of time and efforts to fix these issues.

\section{Supplementary results for Section~\ref{sec:citee}}\label{sec:citee-supp}

In this section, we present supplementary results about the citee networks. Section~\ref{subsec:interpretation-citee} describes how to interpret the three vertices in the statistics triangle using the topic modeling on paper abstracts. Section~\ref{subsec:citee_chooseK} explains the choice of $K$ in mixed membership estimation. Section~\ref{subsec:citee_robustness} investigates the robustness of results with respect to the way we construct the citee networks. Section~\ref{subsec:citee_projection} discusses a variant of the dynamic network embedding method and compares the results with those in Section~\ref{subsec:trajectory}.

\subsection{Interpretation of the three vertices}
\label{subsec:interpretation-citee}

In Section~\ref{subsec:triangle}, we constructed a citee network using the co-citation data in 1991-2000 and applied mixed-SCORE \citep{JKL2017} to obtain a {\it statistics triangle}. Our interpretation of three vertices of this triangle were based on estimating a topic model on paper abstracts, which we explain below. 

Topic modeling \citep{BNJ2003} is a popular tool in text analysis. Given $n$ documents written on a vocabulary of $p$ words, a topic modeling algorithm outputs $K$ ``topic vectors" $A_1, A_2,\ldots, A_K\in\mathbb{R}^p$ and $n$ ``weight vectors" $w_1,w_2, \ldots, w_n\in\mathbb{R}^K$. Each $A_k$ is a probability mass function (PMF) on the vocabulary, representing a conceptual ``topic." By investigating those words that correspond to large entries in $A_k$, one can relate a conceptual ``topic" to a real-world topic (e.g., news, sports, etc.). Each $w_i$ is a nonnegative vector whose entries sum up to 1, where $w_i(k)$ is the weight that document $i$ puts on topic $k$. 

In a companion paper \cite{SCC-paper2}, we conducted topic modeling using the abstracts of papers in our data set via the spectral algorithm in \cite{KaW2017}. The results suggested 11 perceivable topics: {\it Bayes (Bayesian statistics)}, {\it Bio/Med. (Biostatistics and medical statistics)}, {\it Clinic. (Clinical trials)}, {\it Exp.Design (Experimental design)}, {\it Hypo.Test (Hypothesis testing)}, {\it Inference (Statistical inference)}, {\it Latent. Var. (Latent variables)}, {\it Mach.Learn (Machine learning)}, {\it Math.Stats. (Mathematical statistics)}, 
{\it Regression (Regression analysis)}, and {\it Time Series (Time series)}. We obtained a vector  
$w_i\in\mathbb{R}^{11}$ for each abstract. Let $\bar{w}$ be the average of $w_i$'s of all the abstracts in our data set. Then, for each author $j$, we obtained a vector $z_j\in\mathbb{R}^{11}$, which is defined as the average of $w_i-\bar{w}$ among all the abstracts $i$ coauthored by this author $j$. We call $z_j$ the centered topic interest of author $j$. 

\spacingset{1.1}
\begin{figure}[htb!]
\centering
\includegraphics[width=0.975\textwidth]{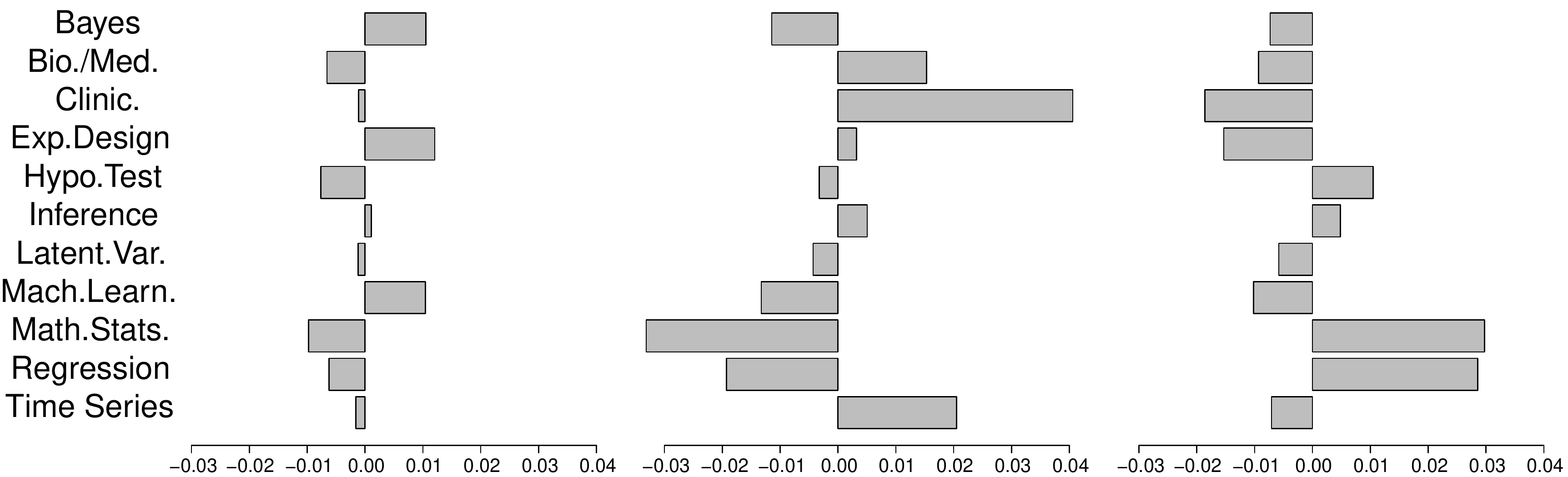}
\caption{The average topic weights of authors in the $3$ communities. From left to right: ``Bayes,"  ``Biostatistics," and ``Nonparametric"  ($x$-axis: the 11 topics; $y$-axis: within-group average of the centered topic weights).  
For each author, the community label is obtained by  mixed-SCORE, and the centered topic weight vector is obtained in \cite{SCC-paper2}.}
\label{fig:bar_mix_9100}
\end{figure}
\spacingset{1.45}

We now use these vectors $z_i$ for authors to interpret the three vertices in the ``statistics triangle."
Let $\hat{\pi}_i$ be the estimated mixed membership vector of node $i$ by mixed-SCORE, $1\leq i\leq n$. 
 We divide all the nodes into 3 groups: If the largest entry of $\hat{\pi}_i$ is the $k$th entry, then node $i$ is assigned to group $k$, $1 \leq k \leq 3$. Recall that the topic modeling results in \citep{SCC-paper2} produce a vector $z_i$ for each author $i$. We now compute $\bar{z}_1, \bar{z}_2, \bar{z}_3$ by taking the the within-group average of $z_i$ for each of the 3 groups, respectively. The three vectors are presented in Figure~\ref{fig:bar_mix_9100}. By this figure, we propose to interpret the three communities from mixed-SCORE as three primary statistical areas: (a) {\it Bayes}.    The topic interests are mainly about ``Bayesian statistics," ``experimental design," and ``machine learning" (including research on EM algorithm and Markov chain Monte Carlo (MCMC)).  (b)  {\it Biostatistics}. The topic interests are mainly in ``biostatistics and medical statistics,"   ``clinical trials,"  and ``times series" (including research on longitudinal data).   (c) {\it Nonparametric statistics}.   The topic interests are in ``mathematical statistics," ``regression," ``hypothesis testing," and ``statistical inference."

Furthermore, 
recall that each author is assigned to one of the three primary areas as above.    
For each primary area, we first select the 30 nodes with the highest degree, and then out of the 30 nodes, we select the 10 with the highest purity. The selected nodes are presented in Table~\ref{tab:mix_9100_repre}, where for each node,  we present the 2 (out of $11$) representative topics where the node has the largest weights (note:  ``time series" includes longitudinal data, ``machine learning" includes EM algorithm and MCMC). The results in Table~\ref{tab:mix_9100_repre} are largely consistent with the above interpretation.

\spacingset{1.1}
 \begin{table}[hbt!]
\hspace*{-1cm}      \scalebox{0.64}{
        \begin{tabular}{ l | l || l | l || l | l }
        \hline
        Name ($\hat{\pi}_i(1)$)  & Representative topics & Name ($\hat{\pi}_i(2)$) & Representative topics & Name ($\hat{\pi}_i(3)$) & Representative topics \\
        \hline
Raftery, Adrian (.73)  & Var.Select., Mach.Learn.  & Zhao, Lueping (.69)        & Bio./Med., Regression  & Gasser, Theo (.77)          & Regression, Inference   \\
Liu, Jun (.68)         & Mach.Learn., Latent.Var.  & Lipsitz, Stuart (.68)      & Clinic., Regression    & Cox, Dennis (.74)           & Regression, Inference   \\
Kadane, Joseph (.67)   & Mach.Learn., Latend.Var. & Fitzmaurice, Garrett (.67) & Clinic., Regression    & Bowman, Adrian (.71)       & Regression, Clinic.      \\
Wong, Wing Hung (.67)  & Mach.Learn., Bayes     & Rotnitzky, Andrea (.67)    & Clinic., Bio./Med.     & Sheather, Simon (.71)      & Regression, Inference       \\
Tierney, Luke (.66)    & Mach.Learn., Inference & De Gruttola, Victor (.65)  & Time Series, Clinic.   & Stute, Winfried (.68)      & Regression, Hypo.Test        \\
Wasserman, Larry (.66)  & Inference, Mach.Learn.  &     Lin, Danyu Y. (.64)        & Time Series, Bio./Med. & Gijbels, Irene (.67)      & Regression, Hypo.Test     \\
Kass, Robert (.65)     & Mach.Learn., Bayes    &  Gail, Mitchell (.63)       & Bio./Med., Inference   & Eubank, Randy (.62)        & Regression, Hypo.Test      \\
 Berger, James (.65)    & Bayes, Latent.Var.     & Pepe, Margaret (.63)       & Bio./Med., Clinic.     & Mammen, Enno (.62)         & Regression, Math.Stats.     \\
Roberts, Gareth (.65)  & Mach.Learn., Time Series  & Prentice, Ross (.63)       & Bio./Med., Time Series & Staniswalis, Joan G. (.59) & Regression, Time Series    \\
Besag, Julian (.62)    & Mach.Learn., Latent.Var.  & Jewell, Nicholas P. (.62)  & Time Series, Bio./Med. & M\"uller, Hans-Georg (.59)   & Regression, Time Series  \\ 
\hline
\end{tabular}%
}
\caption{From left to right: high degree pure nodes in ``Bayes",  ``Biostatistics", and ``Nonparametric".  In each primary area, we first select the 30 nodes with the highest degree, and out of which we then select the 10 with the highest purity.  For each node,  we present the 2 (out of $11$) representative topics where the node has the largest weights (note:  ``time series" includes longitudinal data, ``machine learning" includes EM algorithm and MCMC).}
\label{tab:mix_9100_repre}%
\end{table}%
\spacingset{1.45}

\subsection{The choice of $K$}
\label{subsec:citee_chooseK}
\begin{figure}[htb!]
    \centering
    \includegraphics[height=0.28\textwidth, width=0.4\textwidth, trim=0 10 20 0, clip=true]{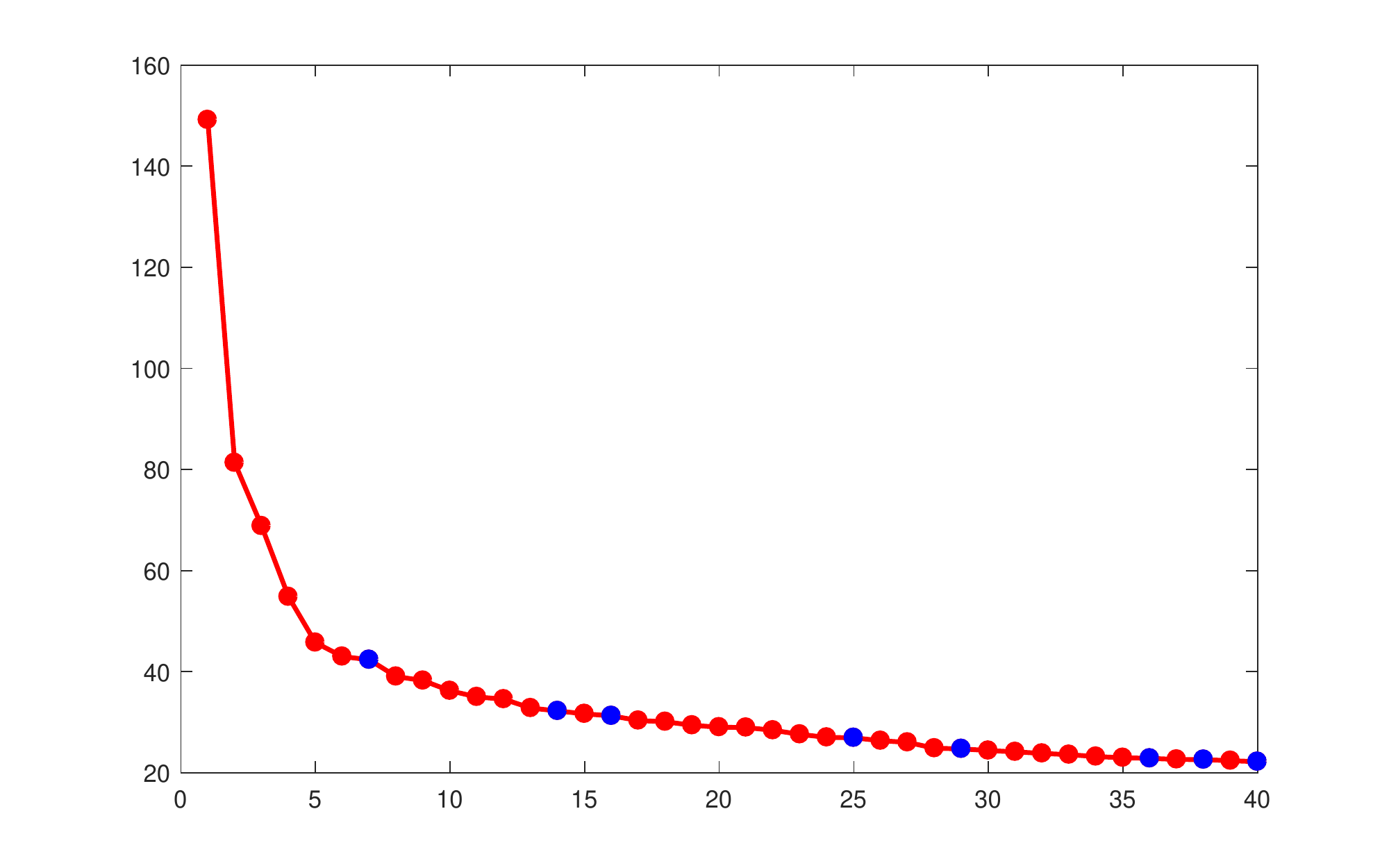}
    \caption{Scree plot of the citee network (red and blue color indicate positive and negative eigenvalues, respectively).}
    \label{fig:citee_scree}
\end{figure}

We explain how we decided $K=3$ for the citee network (1991-2000; 2831 nodes). The scree plot is shown in Figure~\ref{fig:citee_scree}, where $K=4$ is an elbow point. We thus search the value of $K$ in the neighborhood of this elbow point. Additionally, since we believe that the citee network is assortative (a network is assortative if there are more edges within communities than between; in such a network, a negative eigenvalue is more likely to be spurious), we focus on $K\leq 6$ to prevent enrolling negative eigenvalues. We then apply mixed-SCORE to each $K\in \{2,3,\ldots, 6\}$ and check (i) the goodness of fit and (ii) research interests of authors in the estimated communities. We combine (i)-(ii) to make a choice for $K$. 

\spacingset{1.1}
\spacingset{1.45}

\begin{figure}[htb!]
    \centering
    \includegraphics[width=0.35\textwidth]{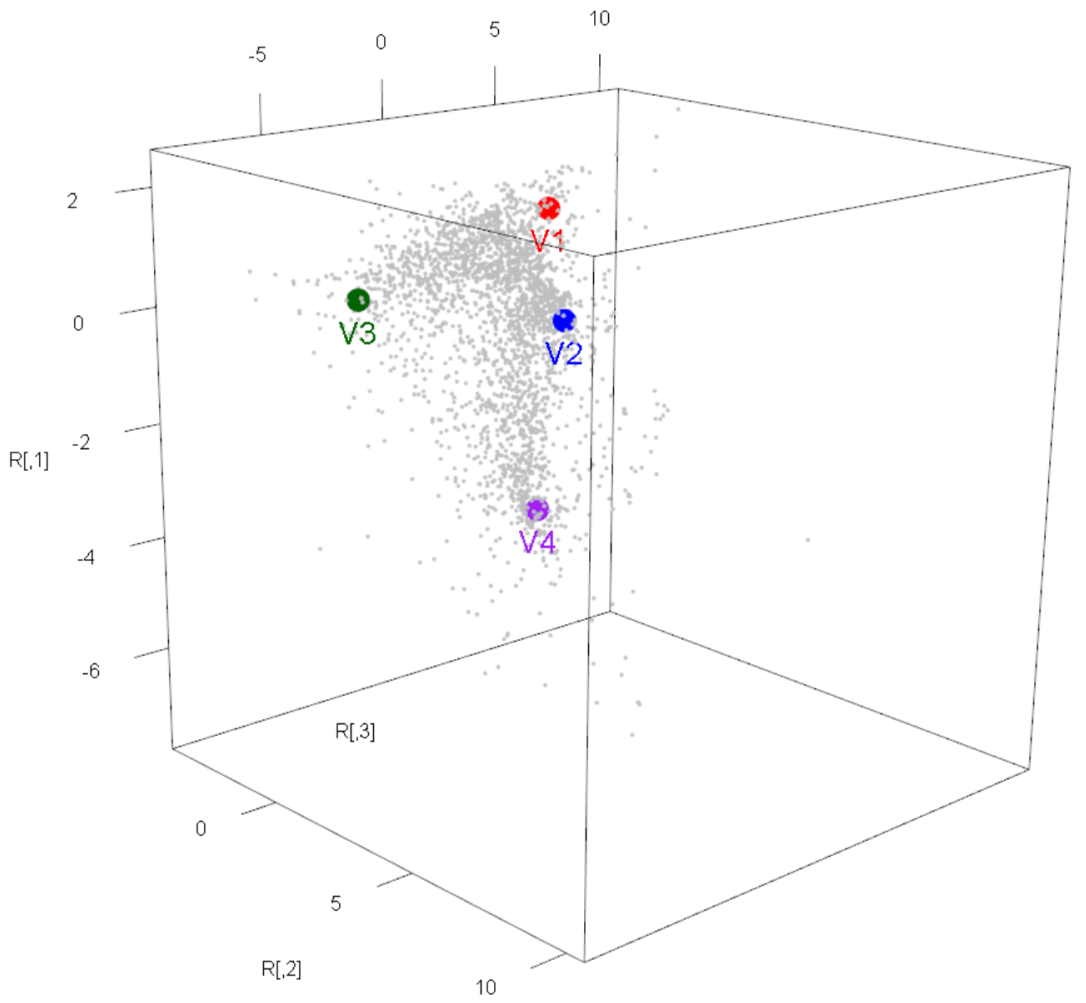} \hspace{1cm}
    \includegraphics[width=0.35\textwidth, height=0.3\textwidth, trim=0 0 0 20, clip=true]{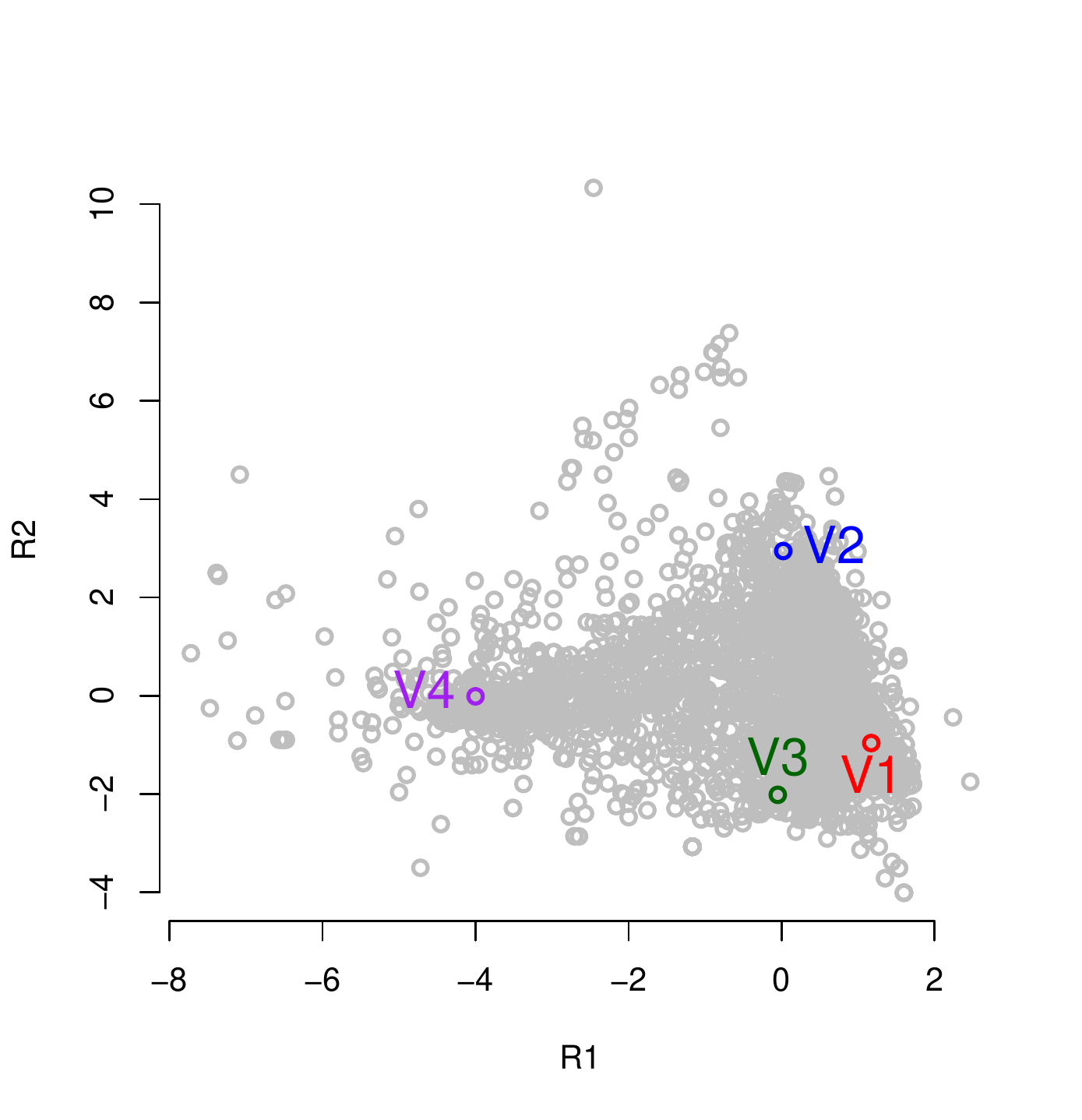}\\
    \hspace{.3cm}
    \includegraphics[width=0.35\textwidth, height=0.3\textwidth, trim=0 0 0 20, clip=true]{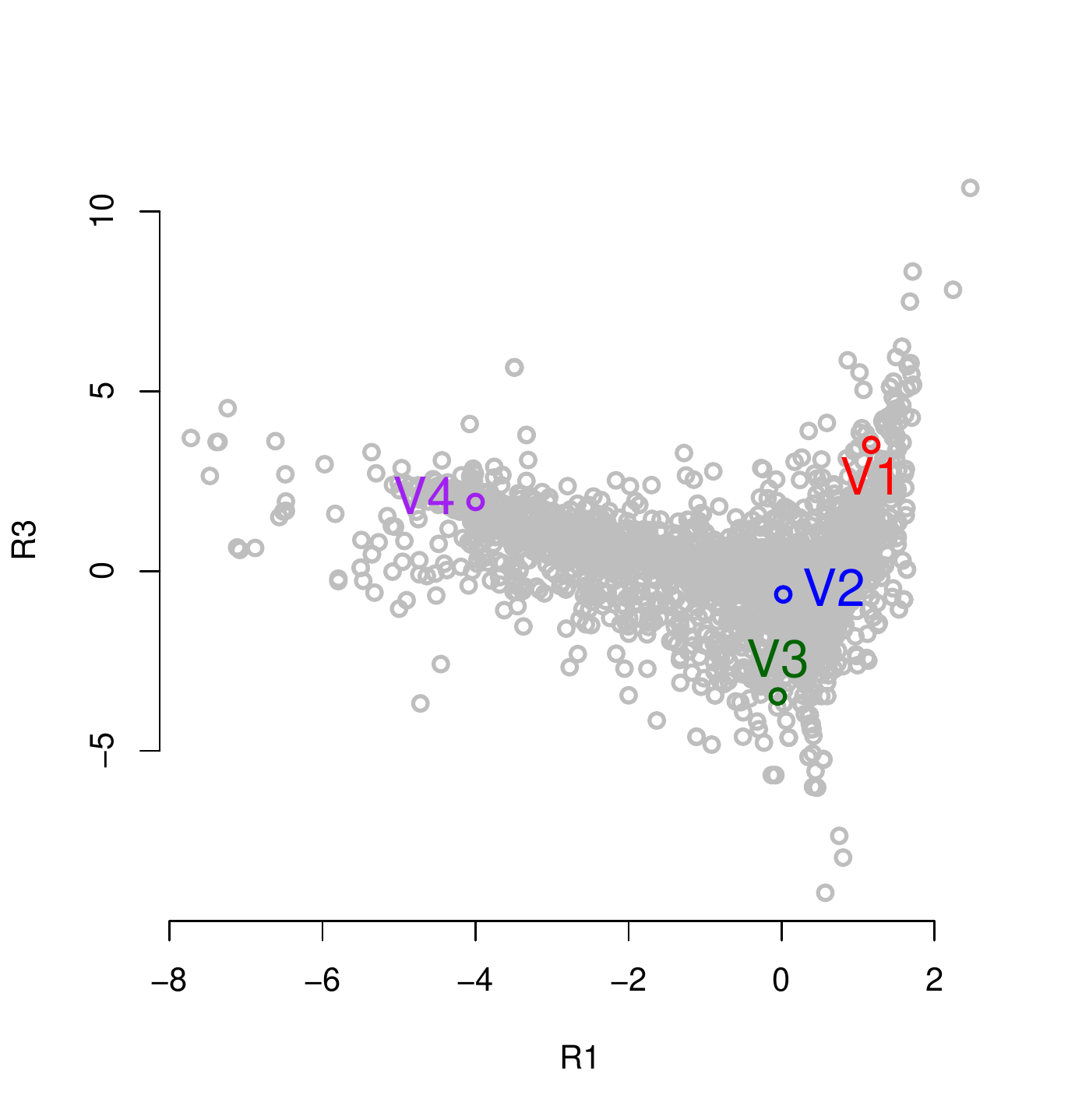}
    \hspace{.6cm}
    \includegraphics[width=0.35\textwidth, height=0.3\textwidth, trim=0 0 0 20, clip=true]{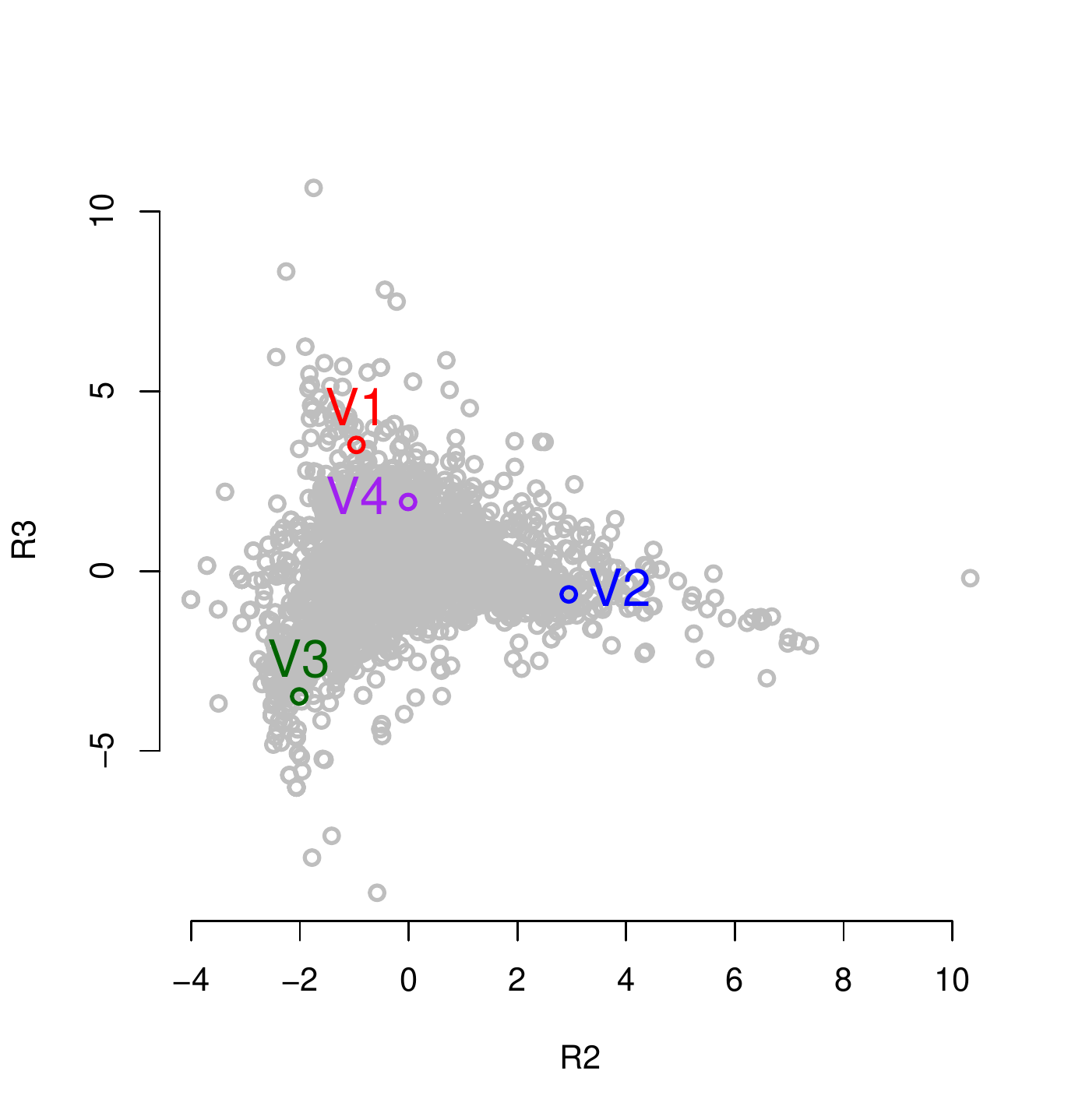}
    \caption{Scatter plots of the matrix $\hat{R}\in\mathbb{R}^{n\times (K-1)}$ for $K=4$. Four vertices are estimated by the SVS algorithm in \cite{JKL2017}. Top left panel: rows of $\hat{R}$ in dimension 3. The other panels: pairwise coordinate plots. }
    \label{fig:citee_Rplot}
\end{figure}

As an illustration, we explain why we preferred $K=3$ to $K=4$. The results for $K=3$ are reported in the main article. We now consider $K=4$ and apply mixed-SCORE to get a matrix $\hat{R}=[\hat{r}_1,\hat{r}_2,\ldots,\hat{r}_n]'\in\mathbb{R}^{n\times 3}$. The theory of mixed-SCORE indicates that the data cloud $\{\hat{r}_i\}_{1\leq i\leq n}$ has approximately the silhouette of a simplex in $\mathbb{R}^3$ with four vertices. We estimate the four vertices using the SVS algorithm in \cite{JKL2017}. In Figure~\ref{fig:citee_Rplot}, the top left panel shows the data cloud and the estimated simplex (tetrahedron). We recall that  for $K=3$, the data cloud fits a triangle very well; however, the data cloud here does not fit a tetrahedron very well. This is also seen from the pairwise coordinate plots in Figure~\ref{fig:citee_Rplot}, which are projections of the data cloud onto 2-dimensional subspaces. All these pairwise coordinate plots have the approximate shape of a triangle, not a quadrilateral. Although it is possible that a tetrahedron becomes a triangle in the projection, we find that the fitting between the data cloud and the estimated simplex is not as good as that for $K=3$.  


\spacingset{1.1}
\spacingset{1.45}

The above analysis alone is still insufficient for us to choose between $K=3$ and $K=4$. We further investigate those pure nodes in each community. For $K=3$, the results can be found in Section~\ref{subsec:triangle}, and the three communities are interpreted as `Bayes', `Biostatistics' and `Non-parametrics', respectively. For $K=4$, the results are in Table~\ref{tb:citee_pureNodes}: Given each of $1\leq k\leq 4$, we first select 300 nodes with highest $\hat{\pi}_i(k)$; out of these 300 nodes, we further select the 10 nodes with largest degrees. Denote by ${\cal C}_1$, ${\cal C}_2$, ${\cal C}_3$, and ${\cal C}_4$ the four communities. According to Table~\ref{tb:citee_pureNodes}, we interpret these communities as ${\cal C}_1$: `{\it Biostatistics I}', ${\cal C}_2$: `{\it Bayes}',  ${\cal C}_3$: `{\it Biostatistics II}', and ${\cal C}_4$: `{\it Nonparametrics}'. 
Comparing them with the three vertices in Figure~\ref{fig:simplex2}, we conclude as follows: when $K$ is increased from 3 to 4,  the `Nonparametrics' and `Bayes' communities are the same, but the `Biostatistics' community is split into two (this is also seen from the top right panel of Figure~\ref{fig:citee_Rplot}, where V1 and V3 are both near the `Biostatistics' vertex of the statistics triangle in Figure~\ref{fig:simplex2}).   
A careful investigation suggests that the difference of author research interests between  `Biostatistics I' and `Biostatistics II' is smaller than their difference from `Nonparametrics' or `Bayes'. Therefore, it is more appropriate to treat the communities from $K=3$ as the `primary areas'.

\spacingset{1.1}
 \begin{table}[hbt!]
 \hspace*{-.5cm}
\scalebox{0.68}{
\begin{tabular}{ lll | lll | lll | lll}
\hline
Name & Deg. & $\hat{\pi}_i(1)$ & Name  & Degree & $\hat{\pi}_i(2)$ &  Name  & Deg. & $\hat{\pi}_i(3)$ & Name & Deg. & $\hat{\pi}_i(4)$\\
\hline
Lueping Zhao	&441	&0.57	&Adrian Raftery	&282	&0.80	&Thomas R. Fleming	&294	 &0.68	&Simon Sheather	&234	&0.84\\
Stuart Lipsitz	&404	&0.66	&Kerrie Mengersen	&197	&0.75	&John Crowley	&262	 & 0.74	&Theo Gasser	&213	&0.92\\
Garrett Fitzmaurice	&324	&0.69	&Jeremy York	&186	&0.78	&Stephen Lagakos	&245	 & 0.68	&D. M. Titterington	&148	&0.85\\
Paul S. Albert	&306	&0.58	&Christian Robert	&178	&0.75	&Terry M. Therneau	&217 & 	0.70	&Byeong U. Park	&121	&1.0\\
Geert Molenberghs	&275	&0.75	&Dave Higdon	&178	&0.74	&Richard Simon	&205	 & 0.70 &Thomas Wehrly	&117	&0.87\\
Bahjat Qaqish	&266	&0.62	&Malay Ghosh	&167	&0.75	&Odd Aalen	&199	 & 0.76	&Roger Koenker	&113	&0.84\\
Emmanuel Lesaffre	&264	&0.74	&Dipak Dey	&165	&0.76	&Patricia Grambsch	&157	 & 0.71	&Alexandre Tsybakov	&112	&0.82\\
John Whitehead	&254	&0.69	&Purushottam Laud	&145	&0.74	&John Oquigley	&142	& 0.86	&Paul Speckman	&106	&0.89\\
John Neuhaus	&227	&0.56 &Augustine Kong	&135	&0.90	&Michael Akritas	&131	 & 0.69	&Alois Kneip	&104	&1.0\\
Walter W. Hauck	&221	&0.57	&Andrew Gelman	&133	&0.73	&Wei Yann Tsai	&129 & 	0.72	&Vincent Lariccia	&104	&0.84\\
\hline
\end{tabular}}
\caption{The relatively pure nodes of each community for $K=4$ in mixed-SCORE. For each $1\leq k\leq 4$, we first selected 300 nodes with highest $\hat{\pi}_i(k)$; out of these 300 nodes, we then output the 10 nodes with highest degrees. The four communities are interpreted as `Biostatistics I', `Bayes', `Biostatistics II', and `Nonparametrics', respectively.} \label{tb:citee_pureNodes}
\end{table}%
\spacingset{1.45}

In the above, we have compared $K=4$ with $K=3$ from (i) the goodness-of-fit of the simplex and (ii) the interpretation of communities. Both (i) and (ii) suggest that $K=3$ is a better choice. We also studied other choices of $K$ in a similar fashion and came up with the final decision of $K=3$.

\subsection{Robustness to the construction of the network}
\label{subsec:citee_robustness}
We had a few preprocessing steps when constructing the citee network using the co-citations in 1991-2000. One of them is first forming a weighted adjacency matrix and then removing all nodes with a degree smaller than 60. We now vary the threshold 60 and investigate the robustness of the statistics triangle. 

\begin{figure}[htb!]
\hspace*{-1.5cm}
    \includegraphics[width=0.6\textwidth]{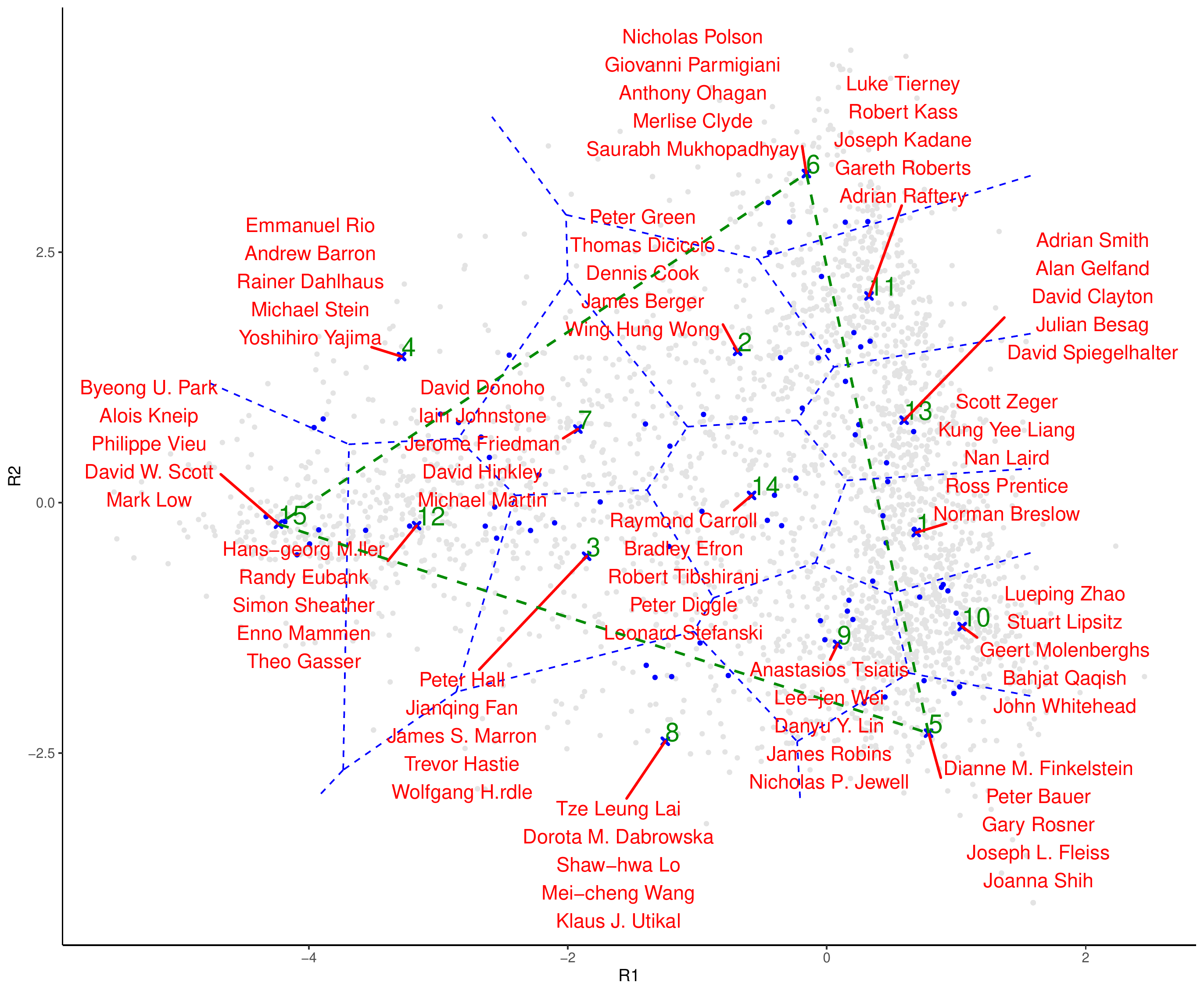}
    \includegraphics[width=0.6\textwidth]{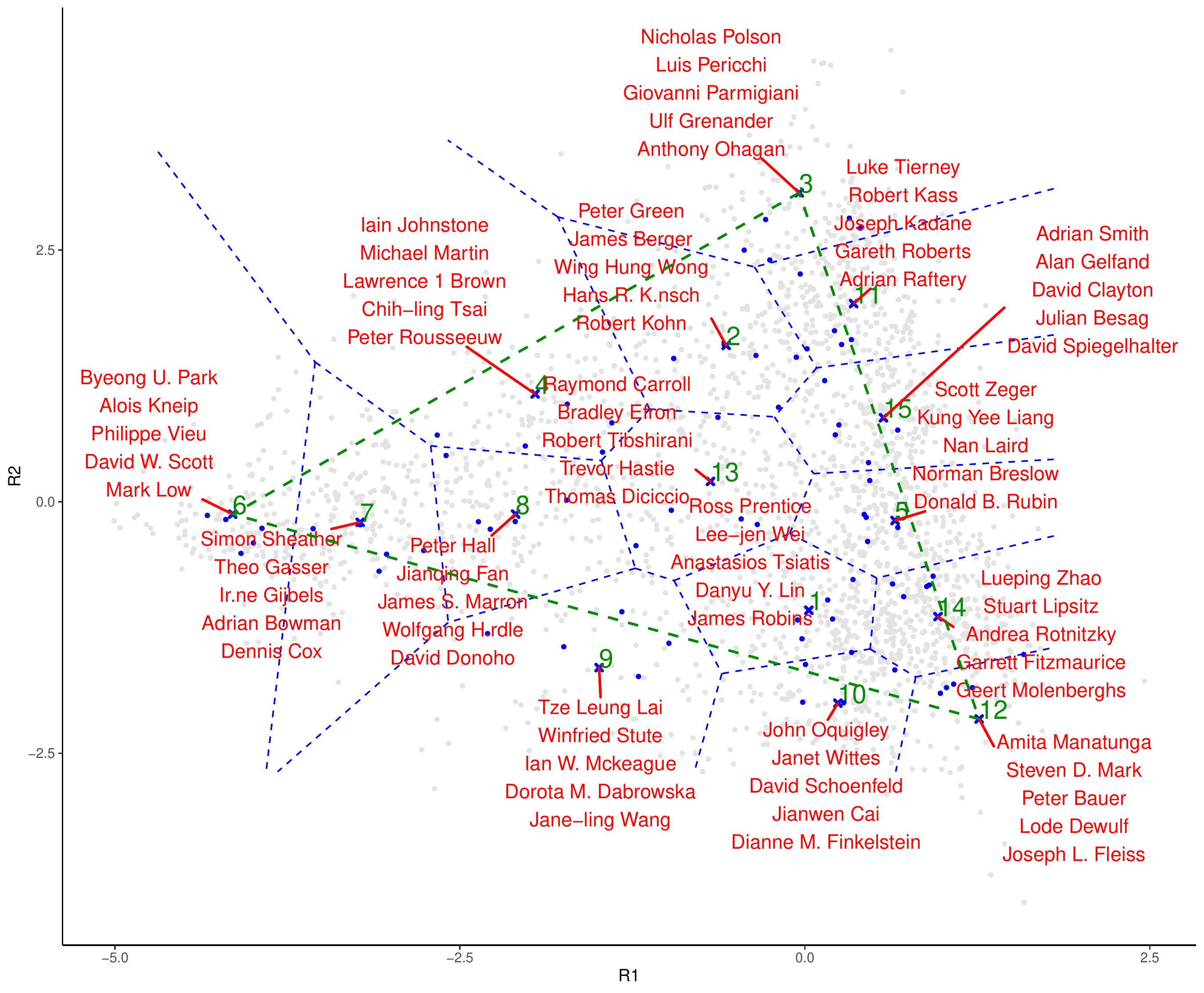} 
    \vspace*{-.8cm}
    \caption{The statistics triangles, when the threshold for constructing the citee network is 50 (left panel) and 70 (right panel), respectively. In each plot, the top vertex, bottom left vertex, and bottom right vertex correspond to `Bayes', `Nonparametrics' and `Biostatistics', respectively. They are similar to the vertices in Figure~\ref{fig:simplex2} (where threshold is 60).}
    \label{fig:citee_newThresh}
\end{figure}

By setting the threshold at 50, 60 and 70, the resultant citee network has 3125, 2831 and 2558 nodes, respectively. We observe that the size of the network only changes moderately as the threshold varies. We then apply mixed-SCORE to all three networks (with exactly the same algorithm parameters) to produce the research maps and statistics triangles. For the threshold 60, the research map and statistics triangle are shown in Figure~\ref{fig:simplex2} of the main article. For the thresholds 50 and 70, the results are shown in Figure~\ref{fig:citee_newThresh}. 

\spacingset{1.1}
\spacingset{1.45}

The shape of the triangle has changed when the threshold changes. This is because the leading eigenvectors have changed, which cause the projected subspace to change. However, it is very clear that the three vertices can always be interpreted as `Bayes', `Nonparametrics' and `Biostatistics', no matter which threshold is used. As the threshold varies, the author clustering results have also changed, but the 15 `sub-areas' identified by author clustering are largely the same. We conclude that the results are robust to the choice of threshold in constructing the citee network.

\subsection{A variant of the dynamic network embedding}
\label{subsec:citee_projection}
In Section~\ref{subsec:trajectory}, we proposed dynamic network embedding as a new approach to drawing research trajectories and calculating diversity metrics for individual nodes. The key idea of this method is using the leading eigenvalues and eigenvectors of $A_1$ to embed the networks at all time points. As a variant of this method, we can also use the leading eigenvalues and eigenvectors of $A_{t_0}$, for some $t_0>1$, to embed the networks at all time points. 
Let $\hat{\lambda}_{k,t_0}$ be the $k$th largest eigenvalue (in magnitude) of $A_{t_0}$, and let $\hat{\xi}_{k,t_0}$ be the associated eigenvector. For each $1\leq i\leq n$ and $1\leq t\leq T$, let 
\beq\label{dynamic-SCORE-variant}
\hat{r}_i^{(t)}(k) = \frac{\hat{\lambda}_{1, t_0} (e_i' A_t \hat{\xi}_{k+1, t_0})}{ \hat{\lambda}_{k+1, t_0}(e_i' A_t \hat{\xi}_{1, t_0}) },\qquad 1\leq k\leq K-1.  
\eeq
This creates a trajectory $\{\hat{r}_i^{(1)}, \hat{r}_i^{(2)}, \ldots,\hat{r}_i^{(T)}\}\subset\mathbb{R}^{K-1}$ for each node $i$. The embedded point $\hat{r}_i^{(t)}$ coincides with the SCORE embedding \citep{Jin2015,JKL2017} at $t=t_0$, but the two embeddings are different at $t\neq t_0$.

\spacingset{1.1}
\begin{figure}[hbt!]
\centering
\includegraphics[width=0.32\textwidth, height=0.28\textwidth, trim=0 10 25 50, clip=true]{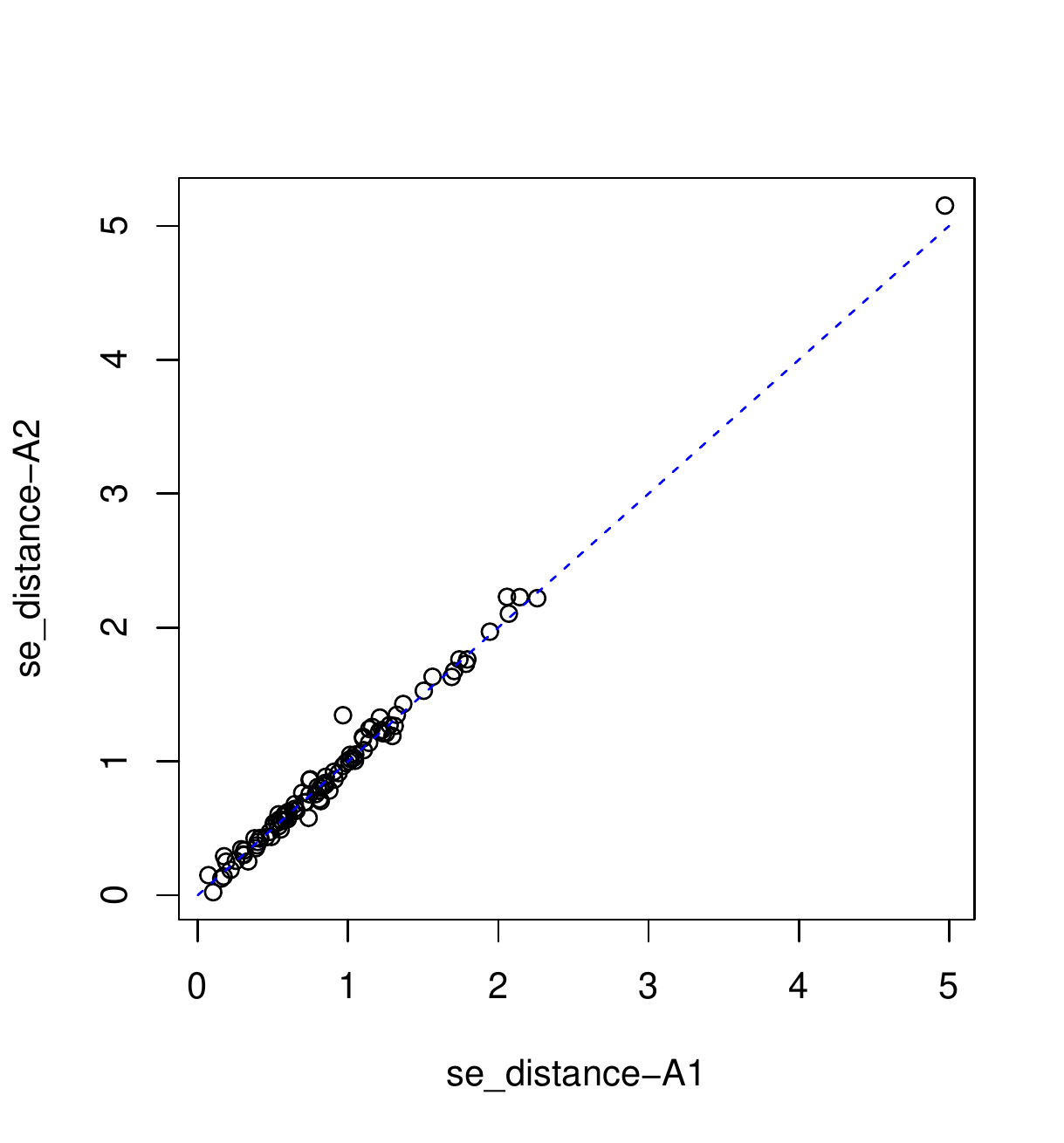}
\includegraphics[width=0.32\textwidth, height=0.28\textwidth, trim=0 10 25 50, clip=true]{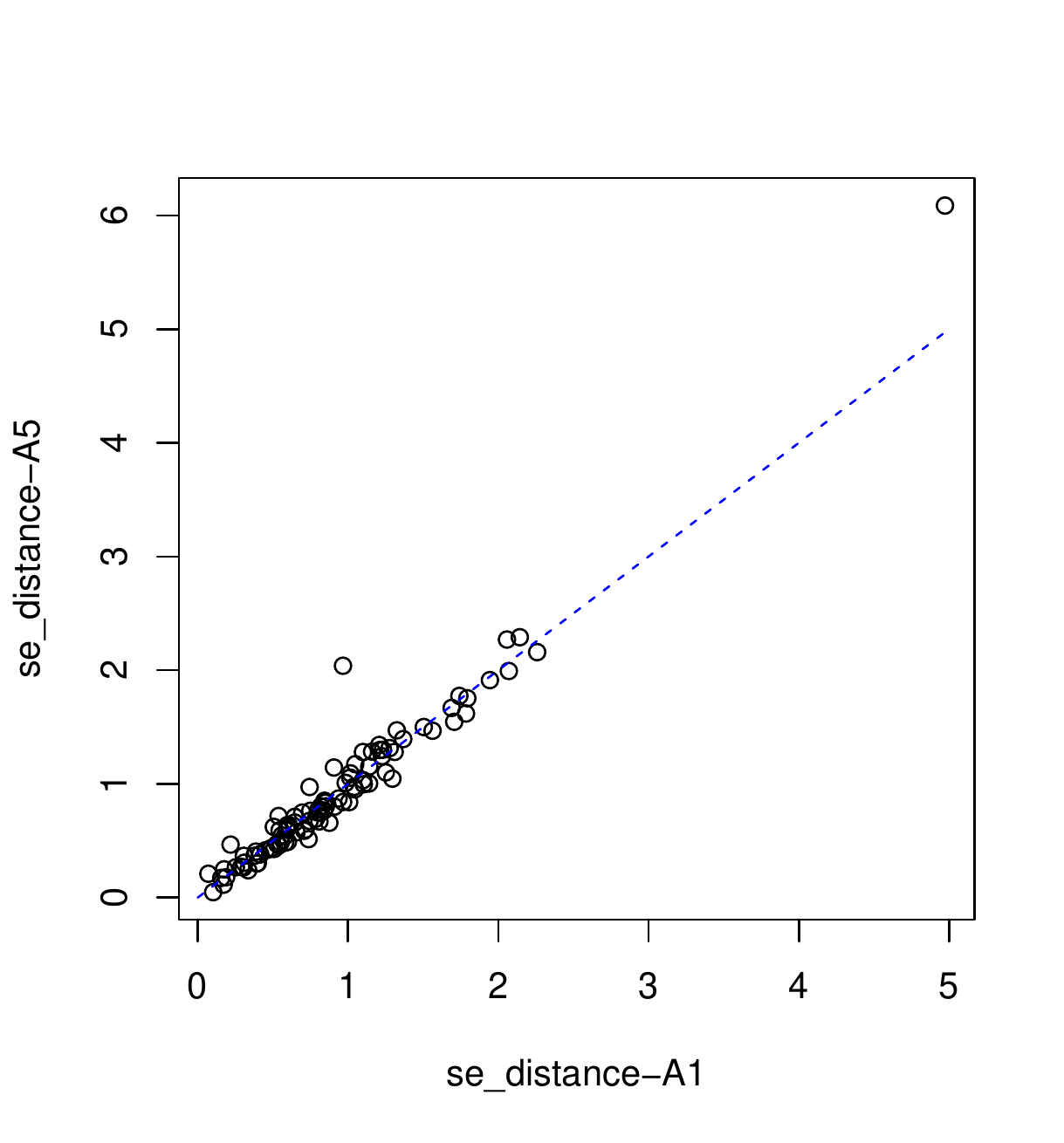}
\includegraphics[width=0.32\textwidth, height=0.28\textwidth, trim=0 10 25 50, clip=true]{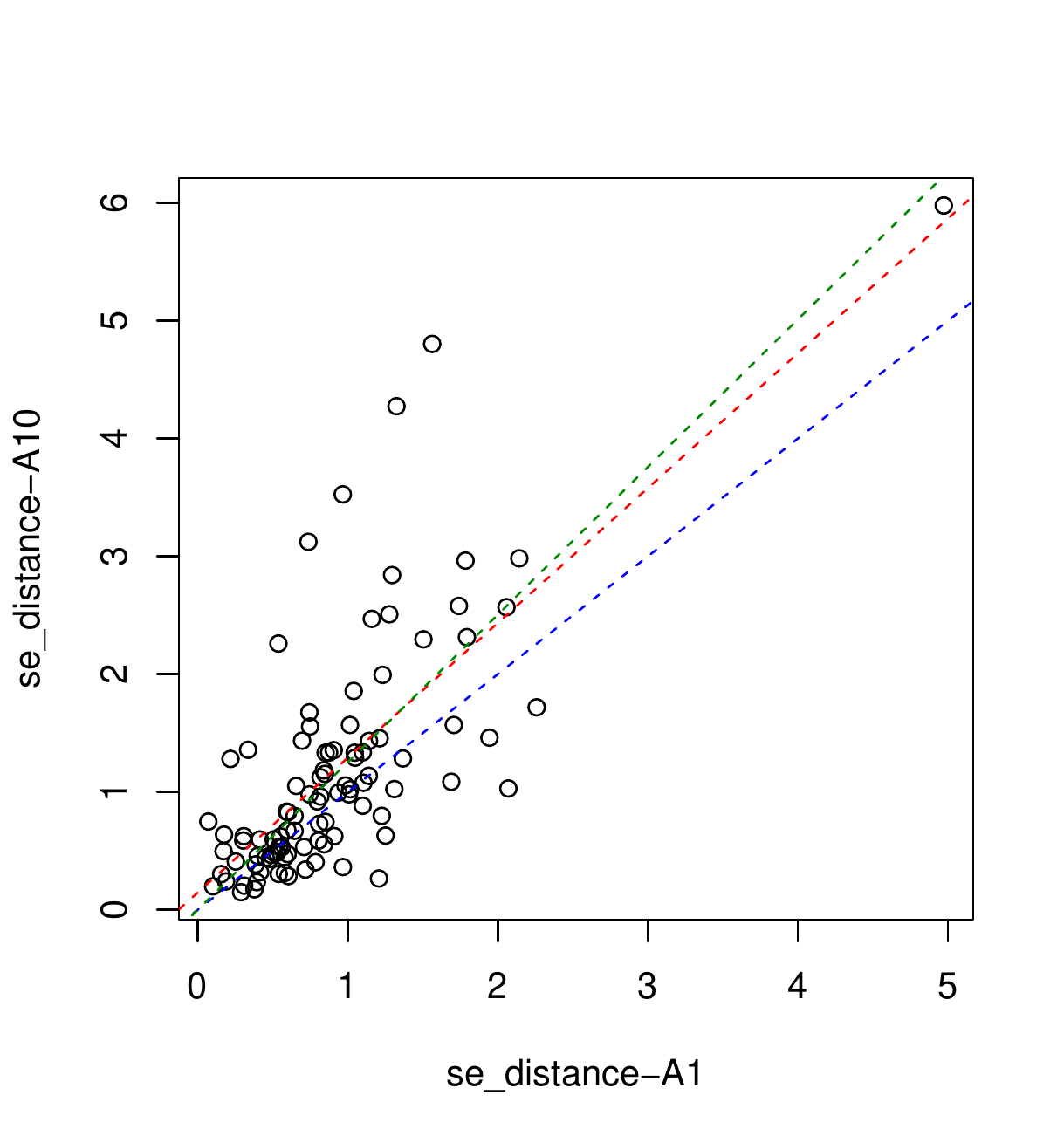}\\
\includegraphics[width=0.32\textwidth, height=0.28\textwidth, trim=0 10 25 50, clip=true]{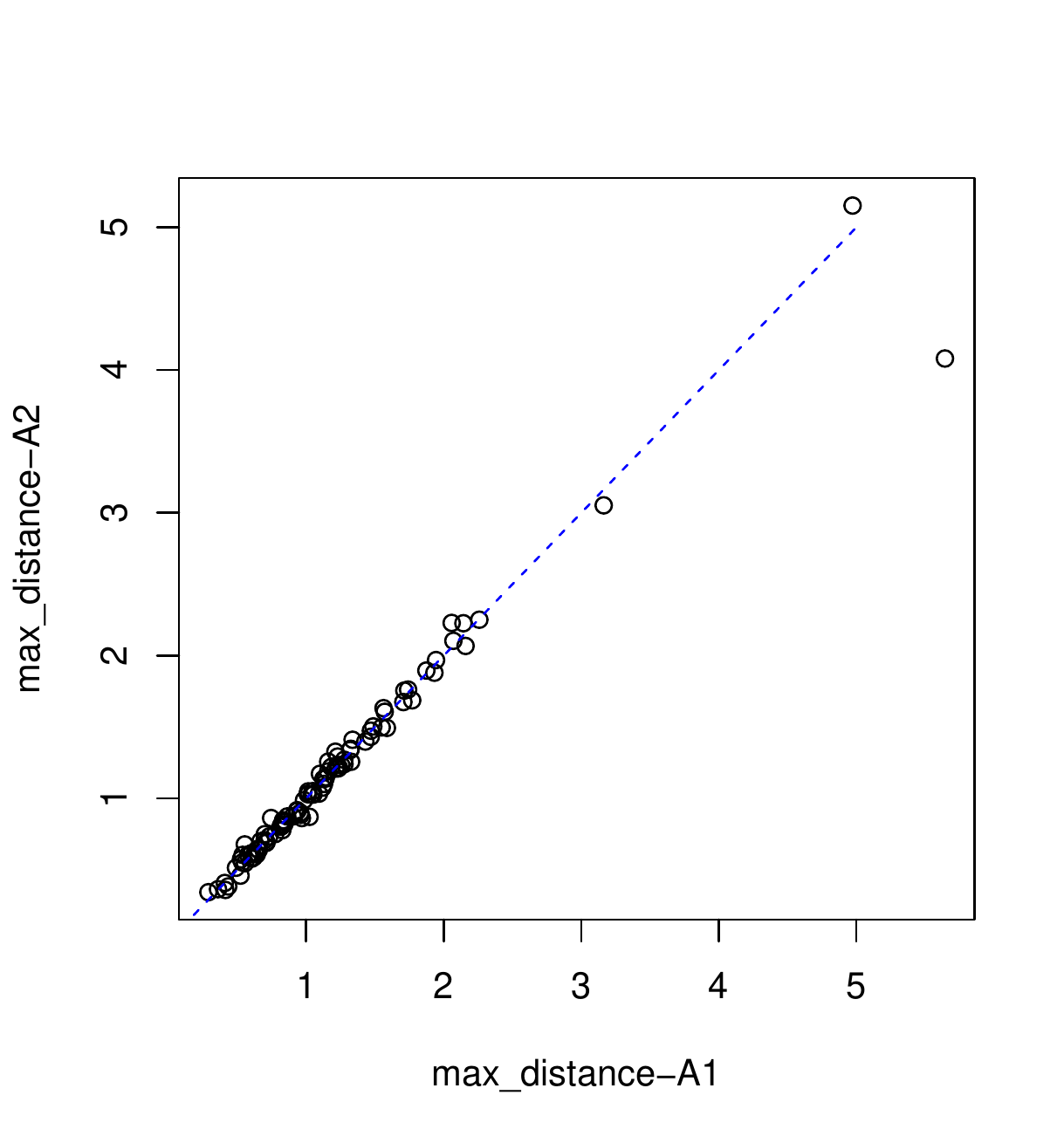}
\includegraphics[width=0.32\textwidth, height=0.28\textwidth, trim=0 10 25 50, clip=true]{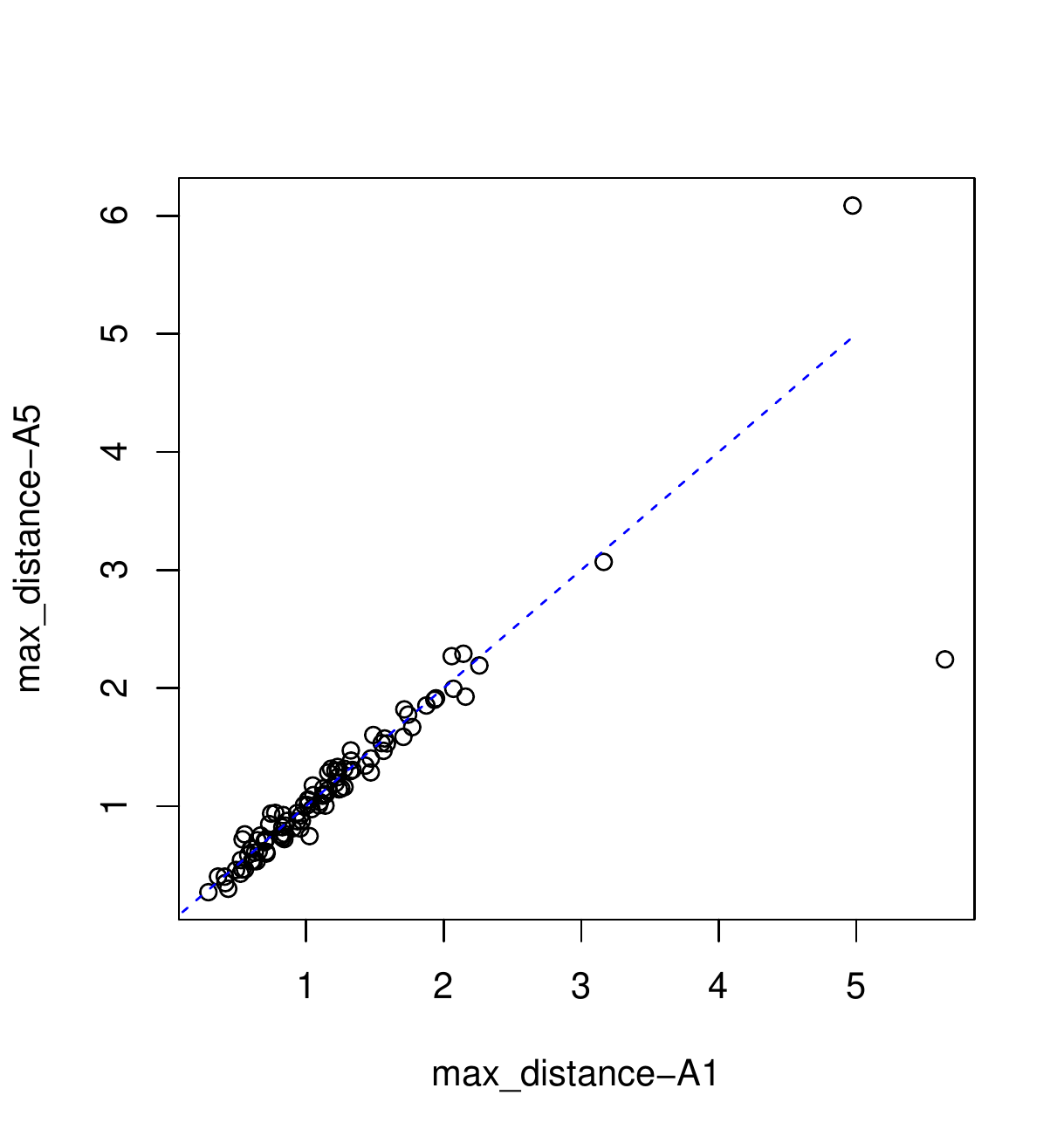}
\includegraphics[width=0.32\textwidth, height=0.28\textwidth, trim=0 10 25 50, clip=true]{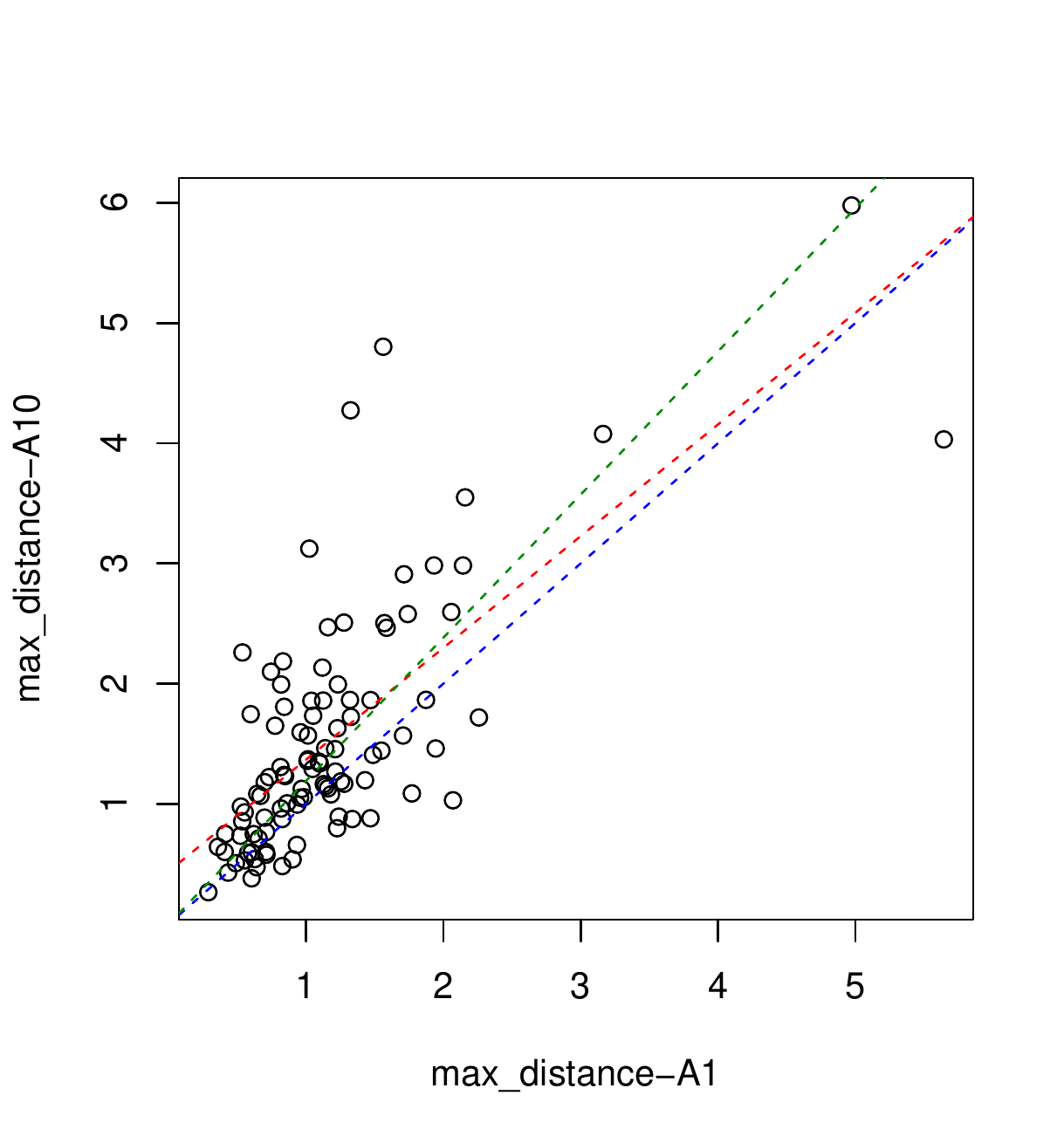}
\caption{Comparison of the diversity metrics with different $t_0$ in the dynamic network embedding. The top panels contain the scatter plots of $E_{i,t_0}$ for $t_0=1$ versus $t_0=2$ (left), $t_0=5$ (middle) and $t_0=10$ (right), where the blue dashed line is $y=x$, and the red and green dashed lines are the least-squares fittings with and without intercept. The bottom panels contain the scatter plots of $M_{i,t_0}$.} \label{fig:dynamic-t0}
\end{figure}
\spacingset{1.45}

Our approach in Section~\ref{subsec:trajectory} corresponds to $t_0=1$. We now consider $t_0\in \{2,5, 10\}$ and compare the results with those for $t_0=1$. Given $t_0$, for each $i$, we generate the trajectory using \eqref{dynamic-SCORE-variant} and calculate two diversity metrics, the {\it se\_distance} $E_i$ and the {\it max\_distance} $M_i$, in the same way as in Section~\ref{subsec:citee-diversity}. The results are shown in Figure~\ref{fig:dynamic-t0}. The top panels contain the scatter plot of $E_{i,t_0}$ versus $E_{i,t_0=1}$, for the top 100 nodes with highest degrees. When $t_0\in \{2,5\}$, the scatter plot fits the line $y=x$ very well, indicating that the diversity metrics have little changes. When $t_0=10$, for many nodes, $E_{i,t_0}$ deviates from $E_{i,t_0=1}$. This is as expected, because the eigenvectors of $A_{10}$ are quite different from the eigenvectors of $A_1$, which cause the projected subspace and the resulting research trajectories to be quite different. However, this does not really affect the assessment of authors' research diversity, because the scatter plot suggests that $E_{i,t_0=10}$ and $E_{i,t_0=1}$ are strongly positively correlated.   
We conclude that switching to a different $t_0$ won't yield any significant change of the results in Section~\ref{subsec:citee-diversity}.

\section{Supplementary results for Section~\ref{sec:coauthorship}}\label{sec:coau-supp}

In this section, we present supplementary results on the analysis of coauthorship networks. Section~\ref{subsec:interpretation-coauthor} describes how to interpret the communities using the topic modeling on paper abstracts. Section~\ref{subsec:coau_leafAuthors} presents the representative authors of each community. 
Section~\ref{subsec:NewmanGirvan} compares the current community detection results with those by the Newman-Girvan modularity. Section~\ref{subsec:coau_chooseK} explains the choice of $K$ in community detection. Section~\ref{subsec:coau_whyDCBM} justifies the assumption of no mixed memberships. Section~\ref{subsec:coau_robustness} investigates the robustness of results with respect to the way we construct the coauthorship network.

\subsection{Interpretation of the six first-layer communities}  \label{subsec:interpretation-coauthor} 
In Section~\ref{sec:coauthorship}, we constructed a coauthorship network (36 journals, 1975-2015) and performed hierarchical community detection. It gives rise to a community tree in Figure~\ref{fig:coau_tree}. 
The first layer of this tree has 6 communities. 
Similarly as in Section~\ref{subsec:interpretation-citee}, we use the topic modeling result in \cite{SCC-paper2} to interpret these communities. 

\begin{figure}[htb!]
\centering
\includegraphics[width = 0.9 \textwidth]{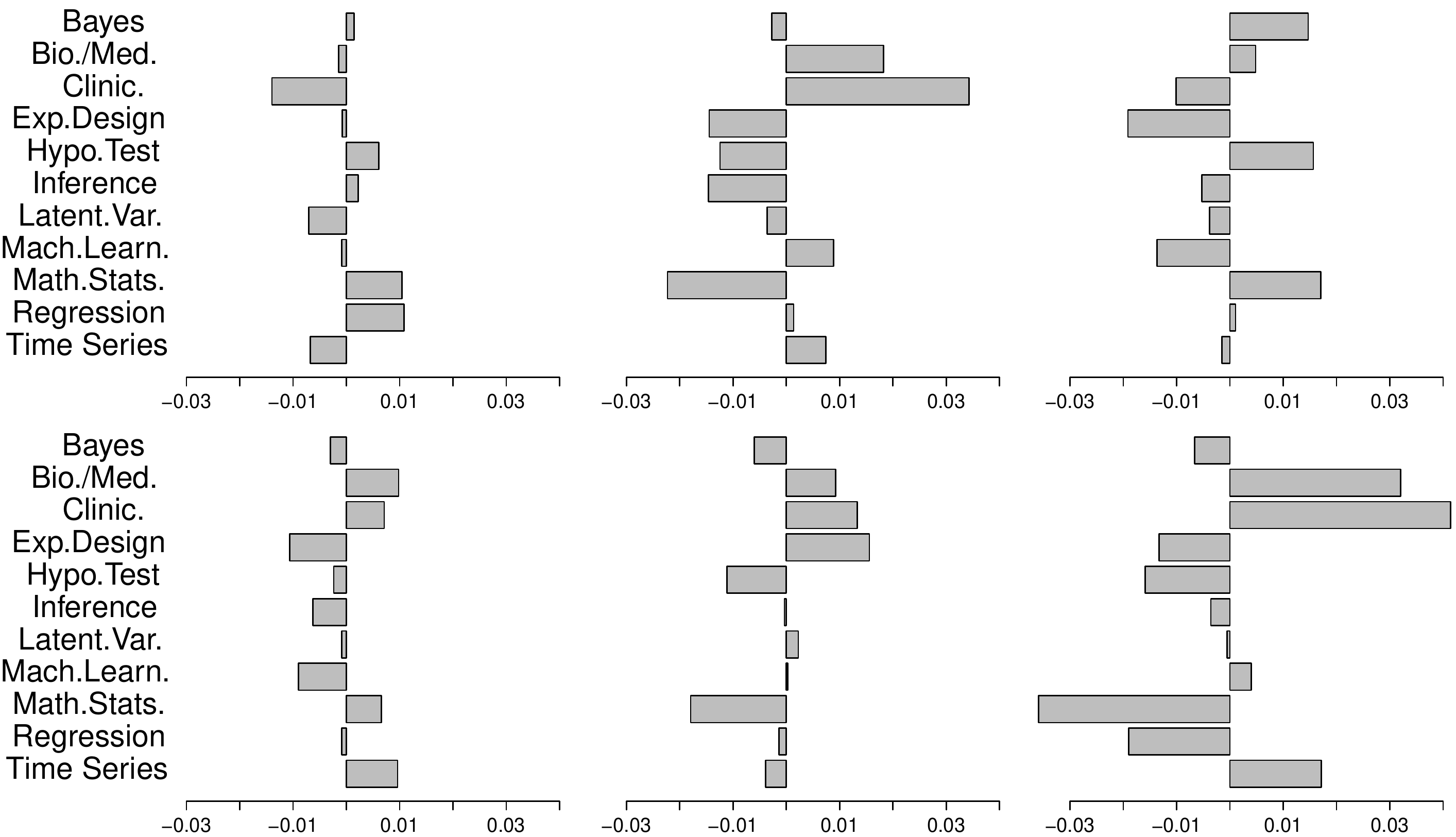}
\caption{The average topic weight vectors for the $6$ communities, C1, C2, $\ldots$, C6.}
\label{fig:coau_topic_layer1}
\end{figure}

As explained in Section~\ref{subsec:interpretation-citee}, we got a centered topic interest vector $z_a\in\mathbb{R}^{11}$ for each author $a$.  Now, for each of the $6$ communities, we take the within-community average of $z_a$.  
The $6$ resultant vectors are displayed in  Figure~\ref{fig:coau_topic_layer1}.  
Combining the figure with a careful read of the large-degree nodes in each community, we propose to label 
the $6$ communities as {\it ``C1 Non-parametric Statistics"}, {\it ``C2 Biostatistics (Europe)"}, {\it ``C3 Mathematical Statistics"}, {\it ``C4 Biostatistics (UNC)"}, {\it ``C5 Semi-parametric Statistics"}, and {\it ``C6 Biostatistics (UM)"}, where we also list some comments on each community in Table~\ref{tb:hierarchical_CommName} of the main article.

\spacingset{1.1}
\spacingset{1.45}

\subsection{Representative authors of the 26 leaf communities} \label{subsec:coau_leafAuthors}
The hierarchical tree in Figure~\ref{fig:coau_tree} has 26 leaf communities. To help for interpretation, we present the representative authors in each leaf community, ordered by their degrees within the leaf community. 
Some representative authors are shown in Table~\ref{tab:cd_names_all} of the main article. We now further expand this table.

\spacingset{1.1}
\renewcommand\theadalign{bc}
\renewcommand\theadgape{\Gape[3pt]}
\renewcommand\cellgape{\Gape[3pt]}
\setlength{\tabcolsep}{3pt}

\begin{table}[tb!]
\centering
      \scalebox{0.68}{
        \begin{tabular}{|c|c|c|l|}
        \hline
        ID & Name & Representative Authors\\
\hline
C1-1-1 & \makecell{Shen-Wong-\\-Hettmansperger\\ (144 nodes) $p=0$} & \makecell[l]{
Hannu Oja, Harvard Rue, Friedrich Gotze, Wei Pan,  \emph{Thomas P. Hettmansperger}, Jun Liu, 
\emph{Xiaotong Shen}, \\
Douglas A. Wolfe, Ishwar Basawa, Leonhard Held, Johan Lim, James J. Chen, 
Sun Young Hwang,\\ Arkady Khodursky, Ralph L. Kodell, Sara Taskinen, Xinle Wang, \emph{Wing Hung Wong}, Hongshik Ahn}\\
 \hline
C1-1-2 & \makecell{Manteiga-Fraiman\\(118 nodes)\\ $p=0.040$} & \makecell[l]{\emph{Wenceslao Gonzalez-manteiga}, Graciela Boente, Juan Antonio Cuesta, Daniel Pena, Antonio Cuevas, \\\emph{Ricardo Fraiman}, Richard Johnson, Michael Akritas, Jacobo De Una-alvarez,
Peter X. K. Song,  \\
Herve Cardot, Manuel Febrero, Carlos Matran, Christopher Morrell, Juan Romo}\\
\hline
C1-1-3 & \makecell{Mardia-Jupp\\(102 nodes)\\ $p=0$} & \makecell[l]{Christian Genest, Ian Dryden, \emph{Kanti V. Mardia}, Rainer Von Sachs, Wensheng Guo, Stephen T. Buckland,\\
John Kent, Louis-paul Rivest, Huiling Le,  Jonathan Raz, Charles C. Taylor, Ole Barnadorff-nielsen,\\ Wen Cheng, Kilani Ghoudi, David B. Hitchcock, \emph{Peter E. Jupp}}\\
 \hline
C1-1-4 & \makecell{Hall-M\"uller\\(331 nodes)\\ $p=0.34$} &\makecell[l]{\emph{Peter Hall}, James S. Marron, Jianqing Fan, Liang Peng, Byeong U. Park, \emph{Hans-Georg M\"uller,}
M. C. Jones,\\
 Laurens De Haan,  Theo Gasser, Wolfgang Hardle, Peter Muller, Bingyi Jing, 
   Giovanni Parmigiani,\\  Thomas Diciccio,  Irene Gijbels, Dominique Picard, Dimitris Politis, Alexandre Tsybakov, Jing Qin,\\ Yaacov Ritov, D. M. Titterington, Ingrid Van Keilegom, Jane-Ling Wang,
Qihua Wang, Andrew Wood,\\ Anestis Antoniadis,
Song Xi Chen, Jan Hannig,
Iain Johnstone, Gerard Kerkyacharian}\\
\hline
C1-1-5 & \makecell{Basu-Lindsay\\(68 nodes)\\ $p=0.012$} &  \makecell[l]{\emph{Bruce Lindsay}, Dankmar Bohning, Domingo Morales, Leandro Pardo,  Dongwan Shin, \emph{Ayanendranath Basu},\\
Maria Luisa Menendez, Konstantinos Zografos, Julio Angel Pardo, Noel Cressie, Tim Friede}\\
\hline
C1-1-6 & \makecell{Gao-Tong\\(189 nodes)\\ $p=0$} & \makecell[l]{Marc Hallin, Wai Keung Li, David Nualart, David Nott, \emph{Howell Tong}, Vo Anh, Ivette Gomes, \\ 
Madan Puri, Dag Tjostheim,  Jose M. Angulo, Kung Sik Chan, \emph{Jiti Gao}, Lanh Tat Tran, Shiqing Ling,\\ Davy Paindaveine, Frits Ruymgaart, Andrei Volodin, Heung Wong,
F. J. Alonso}\\
\hline
C1-2 & \makecell{Dette-Bretz\\(104 nodes)\\ $p=0.0049$} & \makecell[l]{\emph{Holger Dette}, \emph{Frank Bretz}, Axel Munk, Tony Hayter, Wei Liu, Henry Wynn, Siu Hung Cheung,\\
Werner Brannath, Feifang Hu, Wengkee Wong, Nicolai Bissantz, Mortaza Jamshidian, Franz Koenig}\\
\hline
C1-3 & \makecell{Robert-Brown\\(249 nodes)\\ $p=0$} & \makecell[l]{William Strawderman, George Casella, Kerrie Mengersen, \emph{Christian Robert}, \emph{Lawrence Brown}, Tony Cai,\\ Eric Moulines, Murad Taqqu, Anthony Pettitt, Arthur Cohen, Hongzhe Li, Jiunn Tzon Hwang,\\ France Mentre, Martin Wells, Hongyu Zhao, Dominique Fourdrinier, Elias Moreno, Judith Rousseau,\\ Gilles Celeux, Fabienne Comte, Randal Douc, Arnaud Guillin, James P. Hobert, Tatsuya Kubokawa}\\
\hline     
\end{tabular} 
}
\caption{The offspring leaf communities of C1 and the representative authors (ordered by degree within leaf community). To label each community, two or three authors are selected by node betweenness and closeness; if any of them is also a representative author, we present his/her full name in italics.}
      \label{tab:cd_names_1} 
\end{table} 
\spacingset{1.45}

Recall that each leaf community is labeled using the last names of the two nodes with largest betweenness metric \citep{freeman1977set}  and the one node with largest closeness metric \citep{bavelas1950communication} (if the latter happens to be one of the former,  we will not use the same name twice). Table~\ref{tab:cd_names_1} contains the representative authors of the offspring leaf communities of C1. Table~\ref{tab:cd_names_234} contains the representative authors of the offspring leaf communities of C2, C3 and C4. Table~\ref{tab:cd_names_6} contains the representative authors of the offspring leaf communities of C5 and C6.

\spacingset{1.1}
\begin{table}[htb!]
\centering
      \scalebox{0.68}{
        \begin{tabular}{|c|c|c|l|}
        \hline
        ID & Name & Representative Authors\\
        \hline
C2 & \makecell{Kenward-Molenberghs\\(202 nodes)\\ $p=0$} & \makecell[l]{\emph{Geert Molenberghs}, Emmanuel Lesaffre, Marc Aerts, Christophe Croux, Helena Geys, \emph{Mike Kenward}, \\     
Paddy Farrington,  Byron J. T. Morgan, Ariel Alonso, Els Goetghebeur, Geert Verbeke, Victor Yohai, \\ 
Brian Cullis, Peter Rousseeuw, Philippe Beutels, Carmen Cadarso-suarez, Christel Faes, Niel Hens} \\
\hline
C3-1 & \makecell{Balakrishnan-Gupta\\(311 nodes)\\ $p=0$} &  \makecell[l]{\emph{Narayanaswamy Balakrishnan}, \emph{Arjun Gupta}, Manlai Tang, Yasunori Fujikoshi, Ramesh Gupta,\\ 
 Victor Leiva, Gang Zheng, Alan Agresti, Jafar Ahmadi, Markos Koutras, Debasis Kundu,\\ 
 Saralees Nadarajah, Jose M. Ruiz, William R. Schucany, Ningzhong Shi,  Ming Tan, Zhi Geng,\\
Richard F. Gunst, Akimichi Takemura, Moti L. Tiku, Ram Tripathi, Mohammad Arashi}\\  
\hline
C3-2 & \makecell{Bolfarine-Cordeiro\\(58 nodes)\\ $p=0.0003$} & \makecell[l]{\emph{Gauss M. Cordeiro}, \emph{Heleno Bolfarine}, Victor H. Lachos, Reinaldo B. Arellano-valle, Silvia L. P. Ferrari,\\
Edwin Ortega, Vicente G. Cancho, Francisco Cribari-neto, 
 Manuel Galea, Jorge Alberto Achcar}\\
\hline
C3-3 & \makecell{Pepe-Leisenring-Sun\\ (86 nodes)\\ $p=0.0002$} & \makecell[l]{\emph{Jianguo Sun}, Govind S. Mudholkar, \emph{Margaret Pepe}, Liuquan Sun, \emph{Wendy Leisenring}, Yudi Pawitan,\\
  Xinyuan Song, Xingwei Tong,  Xian Zhou,  Ziding Feng, 
 Patrick Heagerty, Alan Hutson, Youngjo Lee}\\
\hline
C4-1 & \makecell{Ibrahim-Herring\\(142 nodes)\\  $p=0.003$} &\makecell[l]{\emph{Joseph Ibrahim}, David Dunson, Hongtu Zhu, Andy Lee, Ming-hui Chen, Keith E. Muller,\\ Kelvin K. W. Yau, Haitao Chu,
Wing Fung, Qi Man Shao,  Marina Vannucci, Bo Cheng Wei,\\ \emph{Amy Herring}, 
Martin Kulldorff, Zhengyan Lin, Niansheng Tang, Stephen R. Cole, Jerome Dedecker}\\
\hline
C4-2 & \makecell{Bass-Perkins\\(104 nodes)\\ $p=0$} & \makecell[l]{
Yuval Peres, \emph{Richard Bass}, Zhen Qing Chen, Frank Den Hollander, Davar Khoshnevisan,\\
Donald Dawson,  
Klaus Fleischmann, \emph{Edwin Perkins}, Jay Rosen, Itai Benjamini, J. Theodore Cox,\\ Amir Dembo, Fabio Martinelli, Carl Mueller, Cyril Roberto, Zhan Shi, Yimin Xiao, Ofer Zeitouni}\\
\hline
C4-3 & \makecell{Mason-Horvath\\(109 nodes)\\ $p=0$} & \makecell[l]{
\emph{Lajos Horvath}, Josef Steinebach, Miklos Csorgo, Luc Devroye, Piotr Kokoszka, Evarist Gine,\\
Armelle Guillou, Marie Huskova, \emph{David Mason}, Ricardas Zitikis, Sandor Csorgo, Jim Kuelbs,\\ Gabor Lugosi, Wei Biao Wu, Alexander Aue, Istvan Berkes, Endre Csaki}\\
\hline
C4-4 & \makecell{Williamson-Lipsitz\\(120 nodes)\\ $p=0.0003$} &\makecell[l]{\emph{Stuart Lipsitz}, Robert H. Lyles, Enrique Schisterman, Brian Reich,  \emph{John Williamson}, Peter Diggle,\\
 Nan Laird, Huiman X. Barnhart, Amita Manatunga, William P. Mccormick, Allan Sampson,\\
Francesca Dominici, Scott Emerson, Garrett Fitzmaurice, Henry W. Block, Somnath Datta}\\
\hline
C4-5 & \makecell{Ying-Wei\\(60 nodes)\\ $p=0.008$} & \makecell[l]{
\emph{Lee-jen Wei}, \emph{Zhiliang Ying}, Tze Leung Lai, Danyu Y. Lin, David Siegmund, Daniel Krewski, Lu Tian,\\ 
Tianxi Cai, Louis Gordon, Sin-ho Jung, W. J. Padgett, Richard Arratia, Kani Chen, Zhezhen Jin}\\
\hline
        \end{tabular}%
        }
\caption{The offspring leaf communities of C2, C3, and C4, as well as the representative authors (ordered by degree within leaf community).  Names in italics: same as in Table \ref{tab:cd_names_1}.}
      \label{tab:cd_names_234}%
\end{table}%

\spacingset{1.45}

\spacingset{1.1}
\begin{table}[htb!]
\centering
      \scalebox{0.7}{
        \begin{tabular}{|c|c|c|l|}
        \hline
        ID & Name & Representative Authors\\
\hline
C5-1 & \makecell{Tsiatis-Betensky\\(185 nodes)\\ $p=0.009$} & \makecell[l]{Paul Yip, Xiaohua Zhou, \emph{Rebecca Betensky}, John Crowley, Adrian Raftery, \emph{Anastasios Tsiatis}, Ji Zhu,\\ Richard Huggins,
George Michailidis, John Oquigley, Ajit Tamhane, Babette Brumback, Cyrus Mehta,\\ Yosef Rinott,  
 James Robins, Michael Daniels, Dianne M. Finkelstein, Xuelin Huang, K. F. Lam}\\
\hline
C5-2 & \makecell{Mukerjee-Reid\\(193 nodes)\\ $p=0$} & \makecell[l]{\emph{Rahul Mukerjee}, Zhidong Bai, Christos Koukouvinos, Kashinath Chatterjee, Sanpei Kageyama, \\
Dennis Lin, Kai Tai Fang, Ashish Das, S. Hedayat, Minqian Liu, C. Radhakrishna Rao, Chien Fu Wu,\\
 Aloke Dey, Norman Draper, Sudhir Gupta, Boxin Tang, David Cox, Angela Dean, Lih-yuan Deng}\\
\hline
C5-3-1 & \makecell{Chen-Turnbull-\\-Johnson (201 nodes)\\ $p=0.31$} & 
\makecell[l]{\emph{Wesley Johnson}, Brian Caffo, Dongchu Sun, Weichung J. Shih, \emph{Bruce Turnbull}, Richard Lockhart, \\
Richard Simon, \emph{Gemai Chen}, Mathias Drton, Galin L. Jones, Edward L. Korn, Kung Yee Liang,\\
Haiqun Lin,  Shuangge Ma,   Paul R. Rosenbaum, Edward J. Bedrick, Adam J. Branscum}\\
\hline
C5-3-2  & \makecell{Carroll-Wang\\(231 nodes)\\ $p=0$} & \makecell[l]{\emph{Raymond Carroll}, Mitchell Gail, Xihong Lin, Laurence Freedman, Hua Liang,
 Jianhua Huang,\\ David Ruppert, Suojin Wang, Kevin W. Dodd, Dean Follmann,  Victor Kipnis, Alan Welsh, \\
 Dennis W. Buckman, Michael P. Fay, Marc G. Genton, Patricia M. Guenther, Susan M. Krebs-smith}\\
\hline
C5-4 & \makecell{Buhlmann-Wellner\\(166 nodes)\\ $p=0.0013$} & \makecell[l]{Mark Van Der Laan, Aad Van Der Vaart, \emph{Peter Buhlmann}, Subhashis Ghosal, Ram Tiwari,\\ Larry Wasserman,
Bin Yu, Joseph Kadane, Thomas Kneib, \emph{Jon Wellner}, Jayanta K. Ghosh,\\ Paul Gustafson, 
 Torsten Hothorn, Harry Van Zanten, Martin Wainwright, Charmaine Dean, }\\
\hline
C5-5 & \makecell{Whilte-Higgins\\(71 nodes)\\ $p=0.016$} & \makecell[l]{Martin Schumacher, Simon Thompson, John Whitehead, Nicky Best, \emph{Ian White}, \emph{Julian P. T. Higgins},\\
Jon Wakefield, Dan Jackson, Sylvia Richardson, Patrick Royston,  Willi Sauerbrei, Douglas G. Altman}\\
\hline
C5-6 & \makecell{Ghosh-Walker\\(197 nodes)\\ $p=0$} & \makecell[l]{\emph{Stephen Walker}, \emph{Malay Ghosh}, Alan Gelfand, Pranab Kumar Sen, Robert Kohn, Gareth Roberts,\\
Thomas Mathew, Bradley Carlin, Antonio Lijoi, Adrian Smith, Mark F. J. Steel, Stefano Favaro,\\ 
Hemant Ishwaran, Ajay Jasra, K. Krishnamoorthy, Igor Prunster, Jim Zidek}\\
\hline
C5-7 & \makecell{Li-Tsai\\(159 nodes)\\ $p=0.034$} & \makecell[l]{Lixing Zhu, Robert Tibshirani, Dennis Cook, \emph{Chih-ling Tsai}, \emph{Runze Li}, Jun Shao, Trevor Hastie, \\
Shein-chung Chow, Riquan Zhang, Andreas Buja, Taizhong Hu, Bing Li, Lexin Li, Wenbin Lu, \\ 
Jerome Friedman, Roger Koenker, Gaorong Li, Lu Lin, Jonathan Taylor, Hansheng Wang}\\
\hline
C6 & \makecell{Taylor-Kalbfleisch\\(264 nodes)\\ $p=0$} & \makecell[l]{\emph{Jeremy Taylor}, Xin Tu, Daniel Commenges, Donald R. Hoover, Thomas Ten Have,
Joan Chmiel,\\ Allan Donner, Lawrence Kingsley, Gary Koch, David Madigan,
Alvaro Munoz,  Vincent Carey,\\ Vern Farewell, Donald B. Rubin,
Odd Aalen, Daniela De Angelis, Naiji Lu, Bhramar Mukherjee,\\ Marcello Pagano, Bernard Rosner,
A Saah, Daniel Sargent, You-gan Wang, Jiahua Chen}\\
\hline
        \end{tabular}%
        }
\caption{The offspring leaf communities of C5 and C6 and the representative authors (ordered by degree within leaf community). Names in italics: same as in Table \ref{tab:cd_names_1}. In C5-2, C5-3-2, and C-6, Reid, Wang and Kalbfleisch are Nancy Reid, Ching-yun Wang, and John D. Kalbfleisch, respectively.}
      \label{tab:cd_names_6}%
\end{table}%

\spacingset{1.45}

\subsection{Community detection by Nemann-Girvan modularity} \label{subsec:NewmanGirvan}
In obtaining the hierarchical community tree, we used a modification of SCORE \citep{Jin2015} for community detection. There are other existing methods for community detection. One of the popular approaches is the Newman-Girvan modularity \citep{girvan2002community}. We apply this method to the coauthorship network (36 journals, 1975-2015) and compare the results with those by our method.  

The Newman-Girvan modularity method requires an exhaustive search over all possible community assignments. However, the coauthorship network is large (4,383 nodes), making it computationally infeasible to implement the method directly. We instead use the spectral approximation in \cite{newman2006modularity}. This method also requires an input of $K$. We fix $K=6$, the same as the number of first-layer communities in our hierarchical community tree.


\spacingset{1.1}
\begin{table}[htb!]
\centering
\scalebox{.85}{
\begin{tabular}{l|l|llllll | c}
\hline 
\multicolumn{2}{c|}{\multirow{2}{*}{Communities}} & NG1  & NG2   & NG3   & NG4   & NG5  & NG6  & Majority   \\
\cline{3-8}
\multicolumn{2}{c|}{} & 86 & 837 & 538 & 131 & 39 & 2752 & ($>60\%$) \\
\hline 
C1-1-1   & 144             &   & 17\%  &    &    &   & 81\%  & NG6\\
C1-1-2   & 118             &   & 12\%  & 36\%  &   &   &  52\%  & ---\\
C1-1-3   & 102             &   & 12\%  &    &    &   & 87\%   & NG6\\
C1-1-4   & 331             &   & 83\% &    &    &   & 15\%   & NG2\\
C1-1-5   & 68              &   & 16\%  &    &    &   & 82\%   & NG6\\
C1-1-6   & 189             &   & 16\%  &   &    &   & 84\%  & NG6\\
C1-2     & 104             &  &    &   &   &  & 98\%  & NG6\\
C1-3     & 249             &  & 12\%  &   &   &  & 88\%  & NG6 \\
\hline
C2       & 202             &   &    & 69\% &   &   & 31\%  & NG3\\
\hline
C3-1     & 311             &    &    &  &     &    & 92\%  & NG6\\
C3-2     & 58              &    &     & 64\%  &     &    & 36\%   & NG3\\
C3-3     & 86              &    &     &     &     &    & 100\%   & NG6\\
\hline
C4-1     & 142             &    &     & 73\% &    &    & 23\%   & NG3\\
C4-2     & 104             &    &     &   &     &    & 99\%  & NG6\\
C4-3    & 109             &    &     & 29\%  &     &    & 71\%   & NG6\\
C4-4     & 120             &    &     & 51\%  &    &    & 44\%   & ---\\
C4-5     & 60              &    &     &     &   &    & 92\%   & NG6\\
\hline
C5-1     & 185             &    &   & 19\% &    &    & 74\%  & NG6\\
C5-2     & 193             &   &    &    &     &    & 92\%  & NG6\\
C5-3-1   & 201             &    & 31\%  &    &     &    & 66\%  & NG6\\
C5-3-2  & 231             &    & 89\% &   &     &    &    & NG2\\
C5-4     & 166             &   &    &   &     &    & 89\%  & NG6 \\
C5-5     & 71              & 38\% & 15\%  &     &     &    & 46\%  & --- \\
C5-6     & 197             &    & 28\%  &    &    &    & 68\%  & NG6 \\
C5-7    & 159             &    & 45\%  &     &     &    & 55\%   & ---\\
\hline
C6      & 264             & 13\% &    &     & 40\% & 15\% & 30\%  & ---\\
\hline
\end{tabular}}
\caption{Comparison with the community detection results by applying Newman-Girvan modularity with $K=6$. 
Rows correspond to the leaf communities in Figure~\ref{fig:coau_tree}. For each leaf community, we report its proportion of nodes in each of the six NG communities; numbers $<10\%$ are omitted.} \label{tb:NGmodule}
\end{table} 
\spacingset{1.45}

The results are shown in Table~\ref{tb:NGmodule}. Each row represents a leaf community in Figure~\ref{fig:coau_tree}, and we report its proportion of nodes in each of the six estimated communities by Newman-Girvan modularity. We aim to check whether the 26 leaf communities found by our method are indeed tight-knit clusters of nodes that is {\it non-splitting} in the Newman-Girvan method. This is obviously true for some leaf communities: e.g., C1-2 has 98\% of nodes in NG6, and C5-3-2 has 89\% of nodes in NG2. In the last column of Table~\ref{tb:NGmodule}, we indicate that a leaf community has its {\it majority} in one NG community if their intersection occupies more than $60\%$ nodes in that leaf community (we use $60\%$ as a heuristic threshold; readers may change it to a different threshold, which results can be obtained immediately from Table~\ref{tb:NGmodule}). 
We see that 21 out of 26 leaf communities have their {\it majority} in an NG community, suggesting that they are indeed tight-knit clusters. In the remaining 5 leaf communities, C4-4 and C6 have a SgnQ $p$-value $<0.001$, hence, they are supposed to be further split in our algorithm. We did not split them only because their sizes are already small (see Section~\ref{subsec:tree} for details).  
We conclude that the results by our method and the results by Newman-Girvan modularity are concordant with each other to some extent. 

However, the Newman-Girvan modularity method has no consistency under the DCBM model \citep{zhao2012consistency}, while SCORE is proved to lead to consistent community detection \citep{Jin2015}. For this reason, we stick to the approach in the main article.

\subsection{The choice of $K$}  \label{subsec:coau_chooseK}
We explain how we decided $K=6$ in the first layer of the hierarchical community detection tree. The determination of $K$ for other layers is similar. 

\begin{figure}[htb!]
    \centering
    \includegraphics[height=0.28\textwidth, width=0.4\textwidth, trim=0 10 20 0, clip=true]{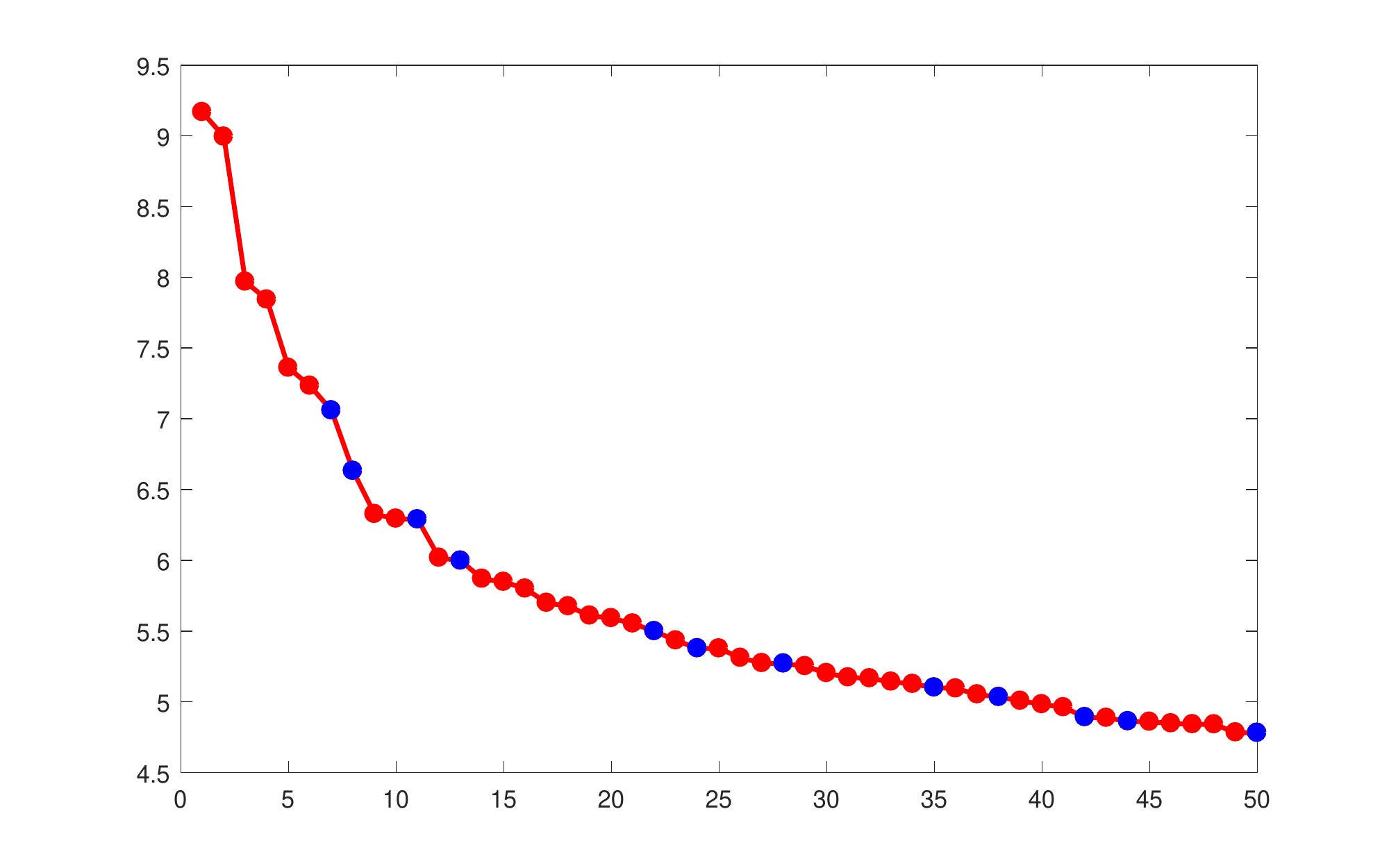}
    \includegraphics[height=0.285\textwidth, width=.4\textwidth, trim=0 15 20 0, clip=true]{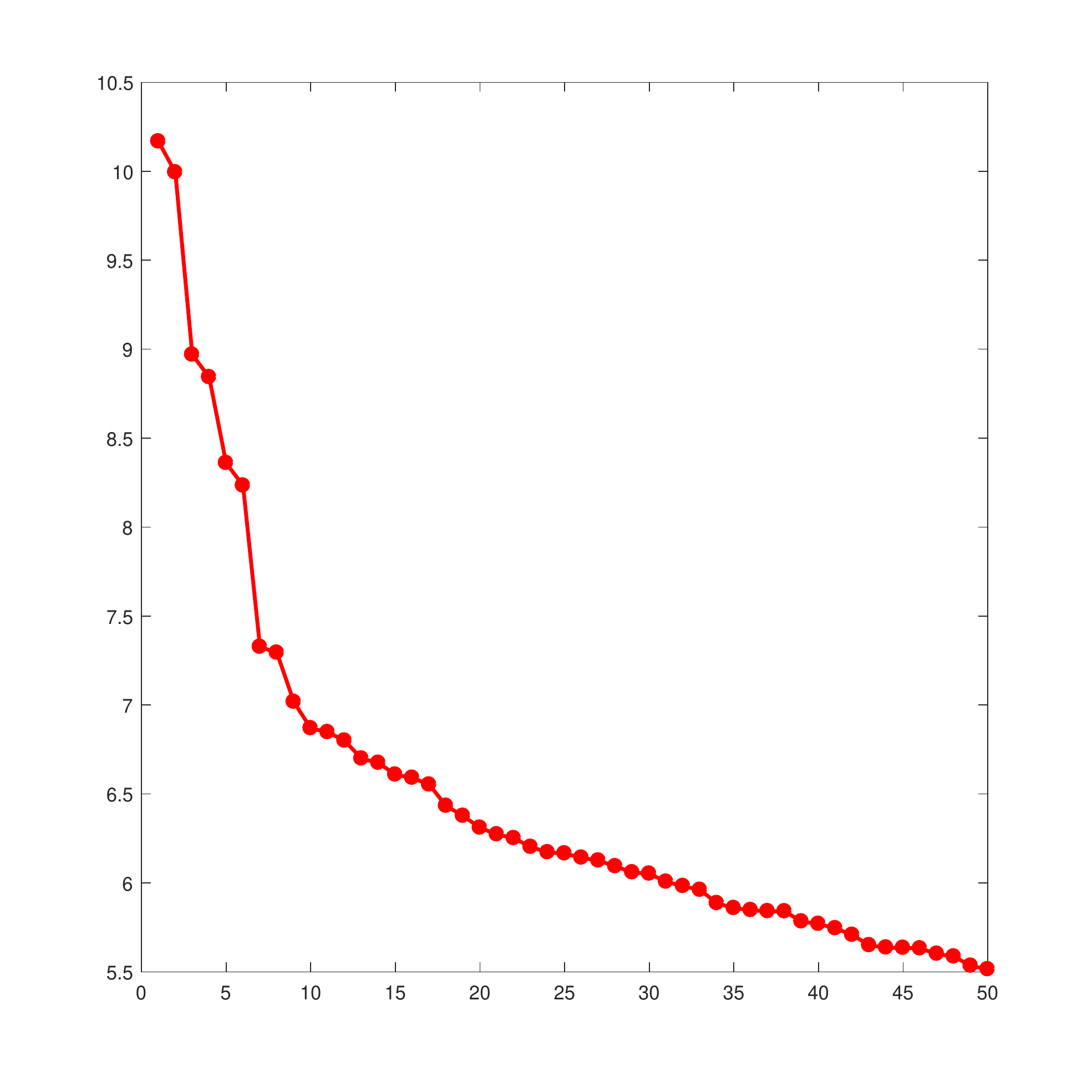} 
    \caption{Scree plots of the coauthorship network. Left: absolute eigenvalues of $A$ (red: positive eigenvalues; blue: negative eigenvalues). Right: absolute eigenvalues of $A+I_n$.}
    \label{fig:coau_scree}
\end{figure}

Given the adjacency matrix $A$ of this network, we first checked the scree plot (left panel of Figure~\ref{fig:coau_scree}). The elbow points are 4, 7, 8, and 11. We thus focused on $K\in \{4,5,\ldots, 11\}$. 
Note that the 7th, 8th, and 11th largest eigenvalue (in magnitude) are negative. Since the coauthorship network only has assortative communities, these negative eigenvalues are less likely to contain true signals. As explained in Section~\ref{subsec:tree}, we penalize negative eigenvalues by conducting eigen-decomposition on $A+I_n$, instead of $A$.  The scree plot associated with $A+I_n$ is shown on the right panel of Figure~\ref{fig:coau_scree}), suggesting that $K=6$ is appropriate. 
%

\spacingset{1.1}
\spacingset{1.45}


\spacingset{1.1}
\begin{figure}[htb!]
    \centering
    \includegraphics[width=0.325\textwidth, height =0.28\textwidth, trim=30 55 10 40, clip=true]{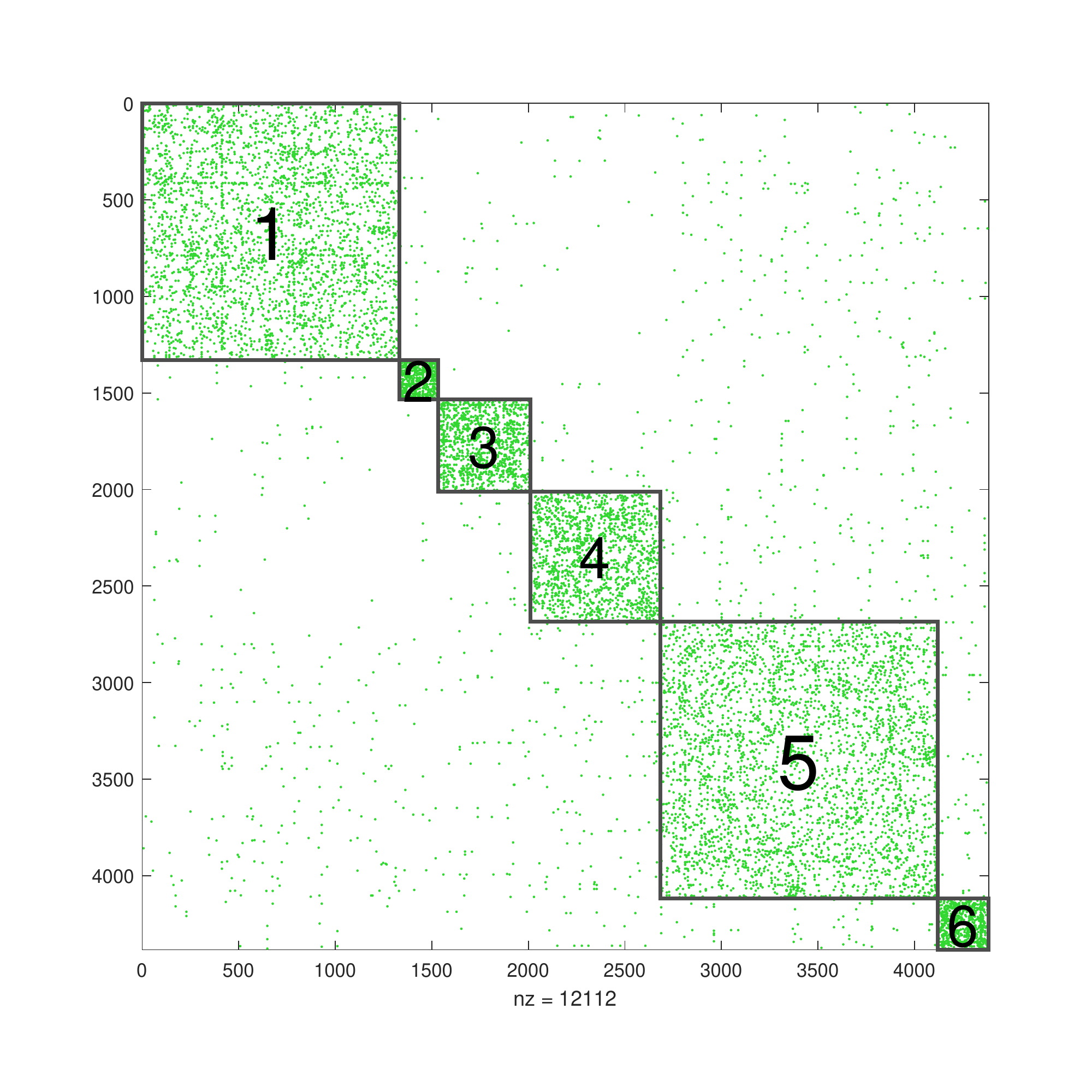} 
    \includegraphics[width=0.325\textwidth, height =0.28\textwidth, trim=30 55 10 40, clip=true]{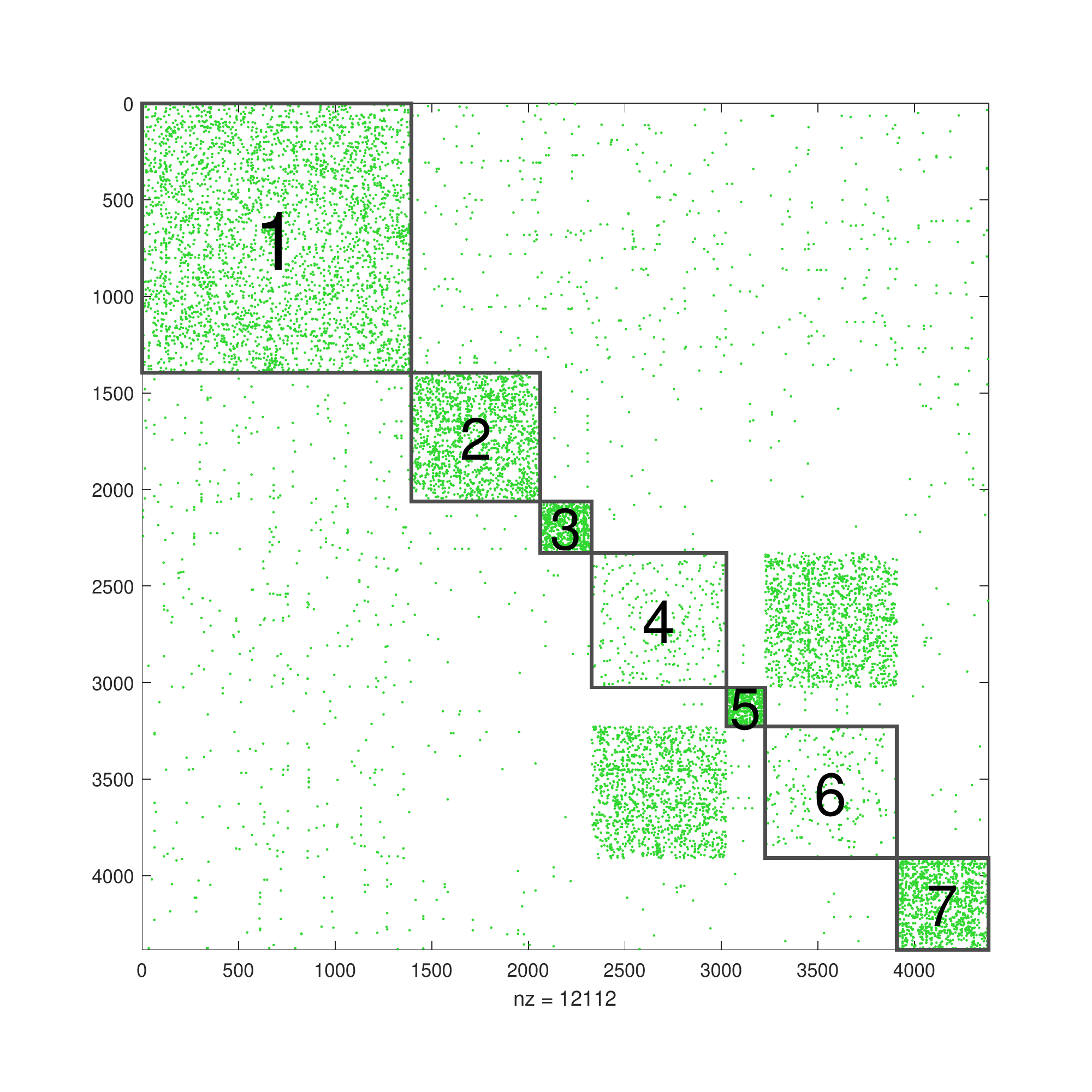}
    \includegraphics[width=0.325\textwidth, height =0.28\textwidth, trim=30 55 10 40, clip=true]{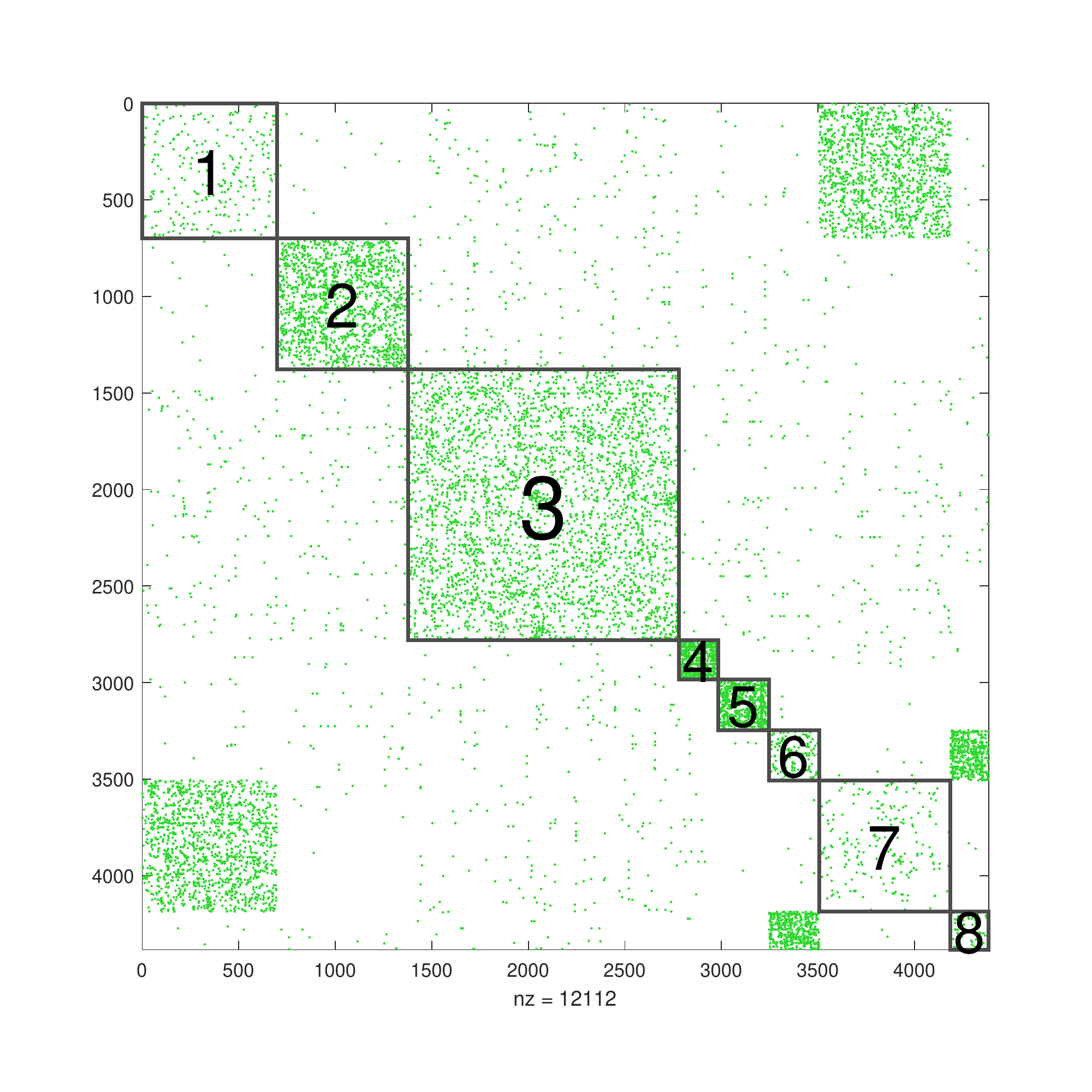}
    \caption{The community detection results of SCORE with $K=6$ (left), $K=7$ (middle) and $K=8$ (right).}
    \label{fig:coau_different_K}
\end{figure} 
\spacingset{1.45}

The above analysis alone is insufficient to determine $K=6$. We next ran the modified SCORE algorithm in Section~\ref{subsec:tree} for each $K\in\{4,5,\ldots,11\}$ and investigated the permuted adjacency matrix. 
It showed evidence of bad fitting when we increased $K$ from 6 to 7 (see Figure~\ref{fig:coau_different_K}). Note that the 7th and 8th largest eigenvalues (in magnitude) are negative; due to assortativity of coauthorship networks, they are likely to be driven by noise. This is confirmed by the fitting of SCORE. For $K=7$, the estimated communities 4 and 6 have a lot of inter-community edges; for $K=8$, the estimated communities 1, 7 and 8 have many inter-community edges. Based on these plots, we further focused on $K\leq 6$. By analyzing the author names in estimated communities (using our internal knowledge of the academic statistics society), we decided that $K=6$ was the best choice.

\subsection{Why mixed memberships are not considered}  \label{subsec:coau_whyDCBM}
\begin{figure}[htb!]
    \centering
    \includegraphics[width=0.242\textwidth, height=.22\textwidth, trim=35 55 20 25, clip=true]{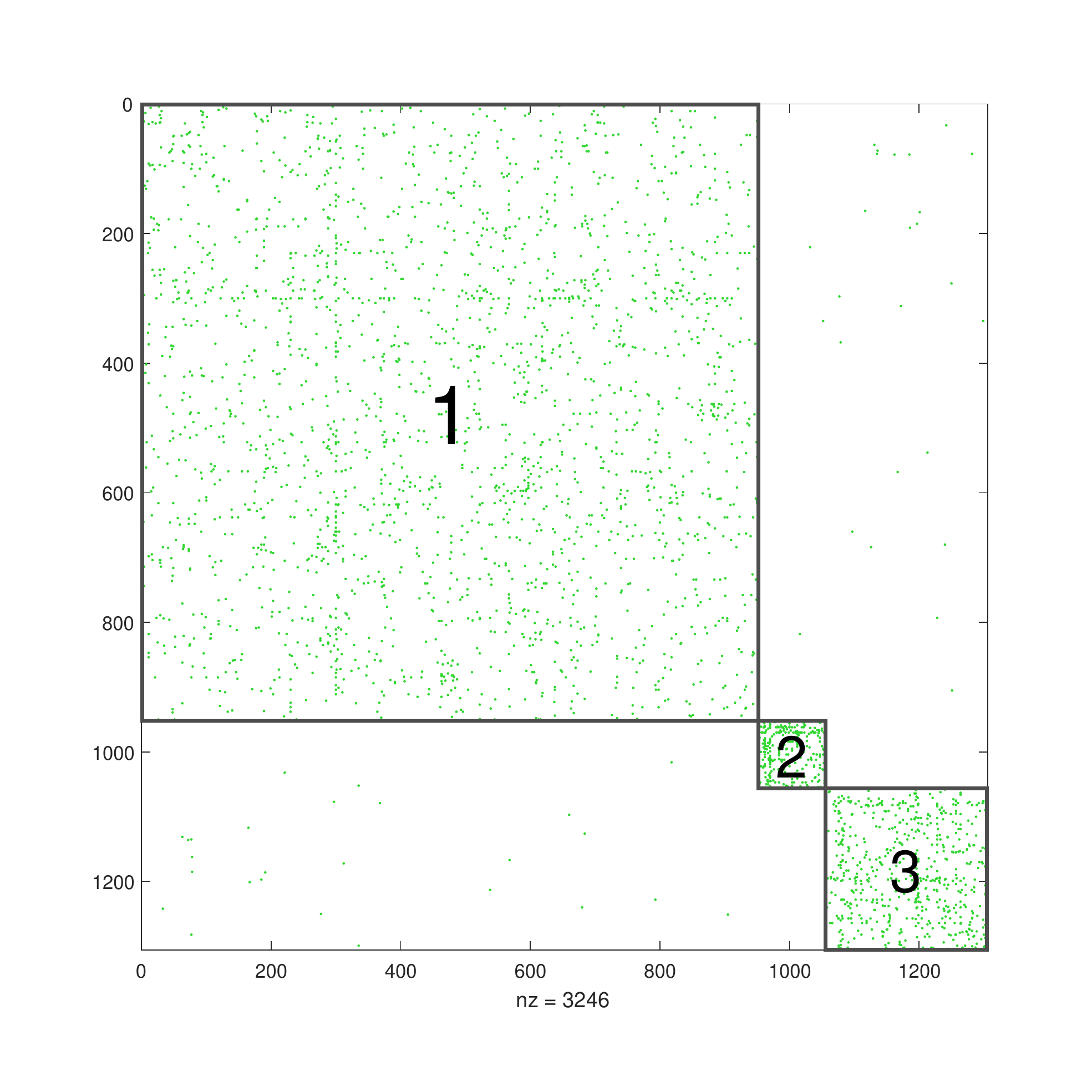} 
    \includegraphics[width=0.242\textwidth, height=.22\textwidth, trim=35 55 20 25, clip=true]{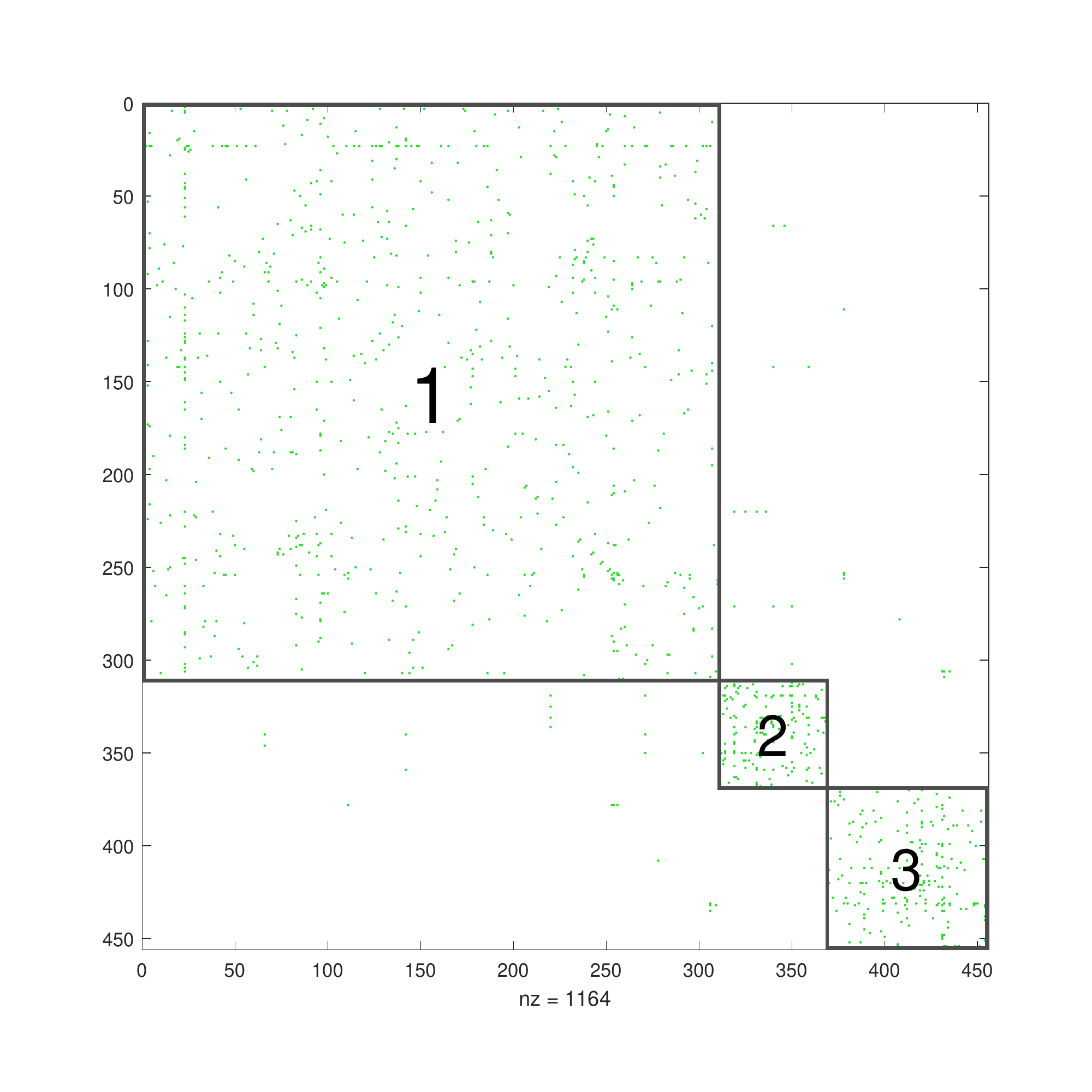}
    \includegraphics[width=0.242\textwidth, height=.22\textwidth, trim=35 55 20 25, clip=true]{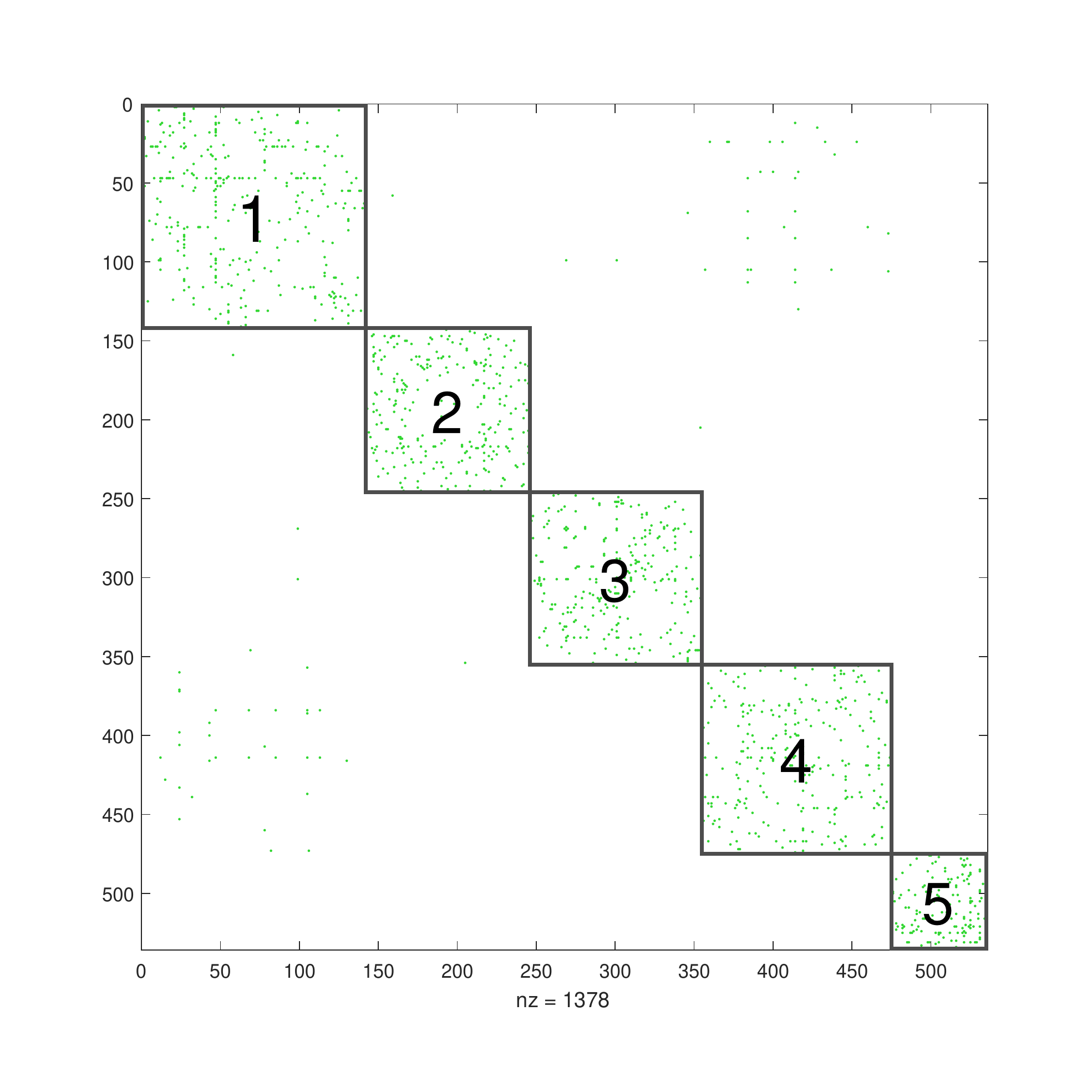}
    \includegraphics[width=0.242\textwidth, height=.22\textwidth, trim=35 55 20 25, clip=true]{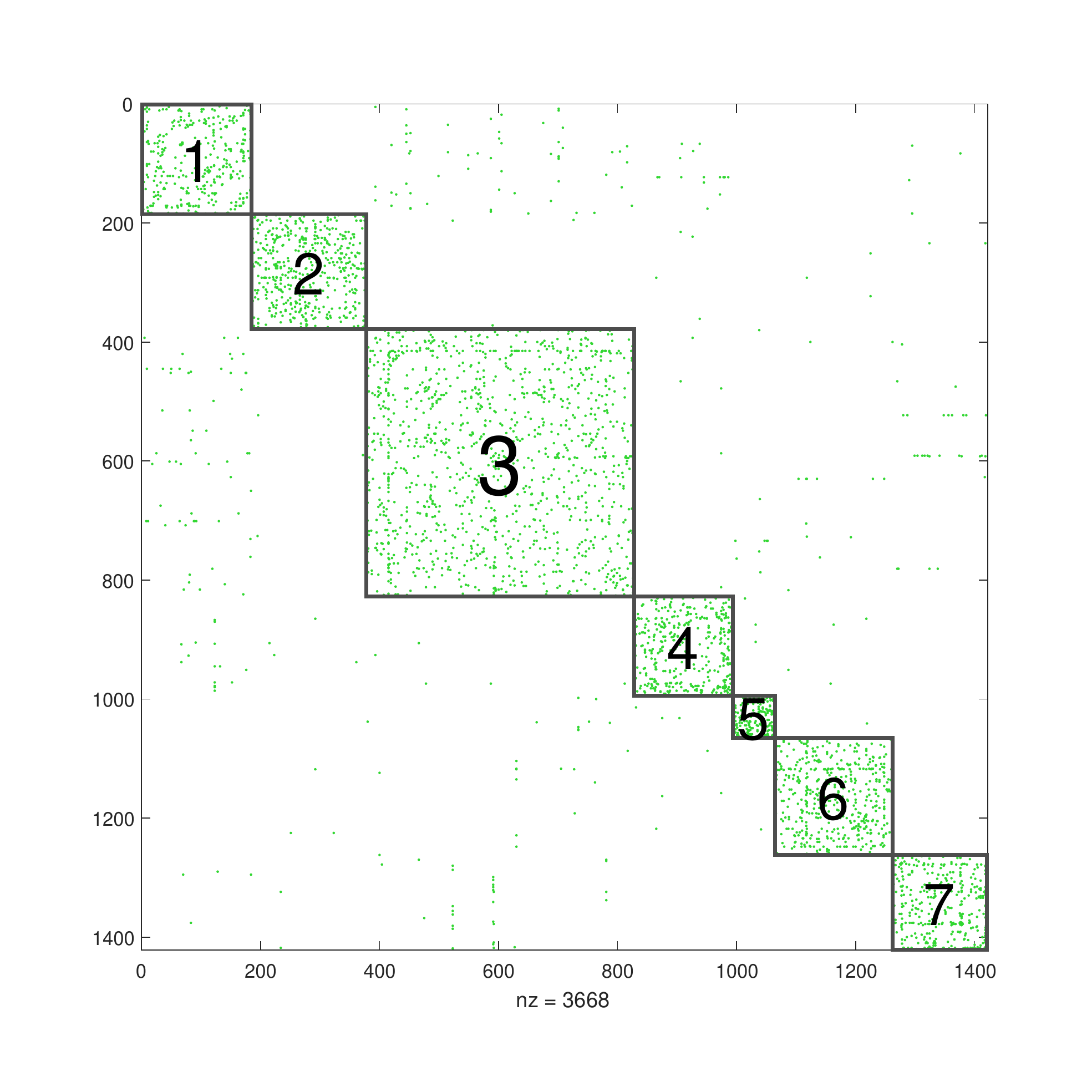}
    \caption{The community detection results of SCORE on C1, C3, C4 and C5 (these are the first-layer communities in Figure~\ref{fig:coau_tree}.}
    \label{fig:coau_no_mixed}
\end{figure}

In Section~\ref{sec:coauthorship}, we model the coauthorship networks with the DCBM model, a special case of the DCMM model by eliminating mixed memberships. We now justify why we use DCBM, instead of DCMM. 
As discussed in \cite{JKL2017}, there are two common strategies to model the communities in a large social network --- (a) a mixed membership model with a relatively small $K$, or (b) a non-mixing model with a relatively large $K$. 
When constructing the hierarchical tree in Figure~\ref{fig:coau_tree}, our goal is to divide the coauthorship network into a large number of leaf communities, each being a tight-knit cluster of authors. For this reason, we used strategy (b) and adopted the DCBM model. Furthermore, we checked the goodness-of-fit of DCBM on the whole network and each first-layer community, using the permuted adjacency matrix with estimated community labels. The plot for the whole network is on the left panel of Figure~\ref{fig:coau_different_K}. 
The permuted adjacency matrix is nearly blockwise diagonal,  suggesting that the data provides no enough evidence of mixed membeships. The plots for the first-layer communities are shown in Figure~\ref{fig:coau_no_mixed}. Among the six first-layer communities, only C1, C3, C4 and C5 are further split in the tree. For each of them, we applied SCORE and plotted the permuted adjacency matrix. 
Again, the data provides no enough evidence of mixed memberships.

\spacingset{1.1}
\spacingset{1.45}

As far as we know, there are no existing tests for testing DCMM against DCBM. In this hypothesis testing problem, both hypotheses are highly composite and have many free parameters. It is challenging to find a test statistic that simultaneously has a tractable null distribution and enjoys good power. We leave this to future work.

\subsection{Robustness to the construction of the network} \label{subsec:coau_robustness}
When constructing the coauthorship network, we define an edge between two nodes if and only if the two authors have coauthored at least $m_0$ papers in the range of our data set. 
Since we prefer to focus on long-term active researchers and solid collaborations, 
$m_0 = 1$ would be too small \citep{JaJ2016}, as the network may include too many edges between active researchers and non-actives ones (e.g., 
a Ph.D advisee who joined industry  and ceased to publish in academic journals).  
The choices of $m_0 = 2$ and $m_0 = 3$ are both acceptable. We now compare their corresponding community detection results.

In the main article, we used $m_0=3$ and obtained a coauthorship network with 4,383 nodes in the giant component. By choosing $m_0=2$, we obtained a coauthorship network with $10,741$ nodes in the giant component. We then applied the same community detection algorithm as in Section~\ref{subsec:tree} with $K=6$ and compared the resulting communities, denoted by D1-D6, with the six first-layer communities, C1-C6, in Figure~\ref{fig:coau_tree}. The results are summarized in Table~\ref{tb:m0compare}.

\spacingset{1.1}
\begin{table}[htb!]
\centering
\scalebox{.92}{
\begin{tabular}{l|l|llllll l }
\hline 
\multicolumn{2}{c|}{Communities} & D1  & D2   & D3   & D4   & D5  & D6  & All \\
\hline
\multirow{3}{*}{\#Nodes} & Total & 2178 & 611 & 1775 & 2315 & 2516 & 1346  & 10741\\
& In ${\cal N}_{m_0=3}$ & 834	& 308 & 698 & 925 & 1059 & 559 & 4383\\
& Fraction & 38\% & 50\% & 39\% & 40\% & 42\% & 42\% & 41\% \\
\hline
\multicolumn{9}{c}{Comparison with the results for $m_0=3$}\\
\hline
\multicolumn{2}{l|}{C1 (1331 nodes)}    & {\bf 49}\% &   & 12\% &  & 10\% & 36\% \\
\multicolumn{2}{l|}{C2 (202 nodes)}      &   &  {\bf 82}\%  &  &   &   &   \\
\multicolumn{2}{l|}{C3 (477 nodes)}     &    &     &  {\bf 74}\%   &  12\%   &    &  \\
\multicolumn{2}{l|}{C4 (673 nodes)}      &  12\%  &     &   12\%  & {\bf 55}\%  &  10\%  &   \\
\multicolumn{2}{l|}{C5 (1436 nodes)}      &   &    &    &  24\%   &  {\bf 46}\%  & 17\%  \\
\multicolumn{2}{l|}{C6 (264 nodes)}       &  &    &     & 28\% & {\bf 57}\% & \\
\hline
\end{tabular}}
\caption{The first-layer communities (D1-D6) for $m_0=2$ versus the first-layer communities (C1-C6) for $m_0=3$. For each of D1-D6, we report its total number of nodes and its number of nodes in ${\cal N}_{m_0=3}$ (the 4383-node network). For each of C1-C6, we report its proportion of nodes in each of D1-D6; numbers $<10\%$ are omitted.} \label{tb:m0compare}
\end{table} 
\spacingset{1.45}


When $m_0$ is decreased from 3 to 2, the size of the coauthorship network greatly expands, where only 41\% of nodes of the expanded network are in the original 4383-node network. As the network expands, there are two possible cases: (a) each of old communities absorbs new nodes and grows; (b) the new nodes form some new communities. The results in Table~\ref{tb:m0compare} suggest that it is Case (a): In this table,  for each of D1-D6, we report the fraction of nodes that are in the original 4383-node network (denoted by ${\cal N}_{m_0=3}$). These numbers range from 38\% and 50\% and are largely comparable with each other. It implies that, as $m_0$ decreases from 3 to 2, 
none of D1-D6 purely consists of newly added nodes; instead, each of them is built on one or more old communities. Furthermore, for each of C1-C6, we calculate its fraction of nodes in each of D1-D6. We then use these fractions to map this community to one of the six new communities; see Table~\ref{tb:m0compare}. C1-C4 are mapped to D1-D4 respectively, and both C5 and C6 are mapped to D5. 

We can similarly apply the recursive community detection algorithm on the 10741-node network associated with $m_0=2$. However, the resulting hierarchical tree has more layers, branches, and leaves, and it requires a lot more efforts to manually interpret each leaf. For this reason, we only report the results for $m_0 = 3$ in the main article.


\section{Proof of Theorem~\ref{thm:simplex}} \label{sec:proof}
The first bullet point follows directly from the second bullet point. It suffices to prove the second bullet point. 

We now show the second bullet point. 
Fix $t\geq 1$. Define $Y_t = \Omega_t\Xi \Lambda^{-1}$. Then, for every $1\leq i\leq n$, we have
\beq \label{proof-1}
r_i^{(t)} = \left[ \frac{Y_t(i,2)}{Y_t(i,1)},\;\; \frac{Y_t(i,3)}{Y_t(i,1)},\;\; \ldots, \;\; \frac{Y_t(i,K)}{Y_t(i,1)}\right]. 
\eeq
Under the DCMM model, $\Omega_t = \Theta^{(t)}\Pi^{(t)}P(\Pi^{(t)})'\Theta^{(t)}$. Therefore, 
\[
Y_t =  \Theta^{(t)}\Pi^{(t)}P(\Pi^{(t)})'\Theta^{(t)}\Xi\Lambda^{-1}= \Theta^{(t)}\Pi^{(t)}M_t. 
\]
It follows that
\beq \label{proof-2}
Y_t(i,k) = \theta_i^{(t)} \sum_{\ell=1}^K\pi_i^{(t)}(\ell)M_t(\ell, k), \qquad 1\leq i\leq n,1\leq k\leq K. 
\eeq
Since $\min_{1\leq k\leq K}\{M_t(1,k)\}>0$, we immediately have $Y_t(i,1)>0$ for every $1\leq i\leq n$. 
We plug \eqref{proof-2} into \eqref{proof-1} to get 
\begin{align*}
r_i^{(t)}(k) &= \frac{Y_t(i, k+1)}{Y_t(i,1)}\cr
&=   \frac{\theta_i^{(t)} \sum_{\ell=1}^K\pi_i^{(t)}(\ell) M_t(\ell, k+1)}{\theta_i^{(t)}\sum_{\ell=1}^K\pi_i^{(t)}(\ell) M_t(\ell, 1)}\cr
&= \frac{\sum_{\ell=1}^K\pi_i^{(t)}(\ell)M_t(\ell, k+1)}{\sum_{s=1}^K\pi_i^{(t)}(s) M_t(s, 1)}\cr
&= \frac{1}{\sum_{s=1}^K\pi_i^{(t)}(s) M_t(s, 1)} \sum_{\ell=1}^K\pi_i^{(t)}(\ell)M_t(\ell, 1)\frac{M_t(\ell, k+1)}{M_t(\ell,1)}. 
\end{align*}
By definition of $v_1^{(t)}, \ldots, v_K^{(t)}$, we have  $M_t(\ell,k+1)/M_t(\ell,1)=v^{(t)}_\ell(k)$. The above can be re-written as
\[
r_i^{(t)}(k) = \sum_{\ell=1}^K\; \frac{\pi_i^{(t)}(\ell)M_t(\ell,1)}{\sum_{s=1}^K\pi_i^{(t)}(s) M_t(s, 1)}\; v^{(t)}_\ell(k). 
\]
Let $h_t = (M_t(1,1), M_t(2,1), \ldots, M_t(K,1))'$ and $w_i^{(t)}=(\pi_i^{(t)}\circ h_t)/\| \pi_i^{(t)}\circ h_t\|_1$. Then, 
\[
w_i^{(t)}(\ell) = \frac{\pi_i^{(t)}(\ell)M_t(\ell,1)}{\sum_{s=1}^K \pi_i^{(t)}(s)M_t(s,1) }.
\]
Combining the above gives
\[
r_i^{(t)}(k) = \sum_{\ell=1}^K w_i^{(t)}(\ell)\ v^{(t)}_\ell(k), \qquad \mbox{for every }1\leq k\leq K-1, 
\] 
which is $r_i^{(t)}=\sum_{\ell=1}^K w_i^{(t)}(\ell)\cdot v^{(t)}_\ell$ in the vector form. \qed

\section{Access to the MADStat dataset}

Our dataset is called the {\it Multi-Attribute Dataset on Statisticians (MADStat)}. The dataset and all the code is publicly accessible at the following sites:
\begin{itemize}
\item The MADStat project: \url{http://zke.fas.harvard.edu/MADStat.html}
\item GitHub: \url{https://github.com/ZhengTracyKe/MADStat}
\end{itemize}
It contains a list of ready-to-use data matrices: 
\begin{itemize} 
\item The full data are in the file \texttt{AuthorPaperInfo.RData}. It has two variables. 
\begin{itemize}
\item \texttt{AuPapMat}: This matrix summarizes the bibtex data. It has 4 columns, where {\it idxAu} is author ID, {\it idxPap} is paper ID, {\it year} is publication year, and {\it journal} is publication journal.  
\item \texttt{PapPapMat}: This matrix summarizes the citation data. It has 5 columns, where {\it FromPap} and {\it ToPap} are the paper IDs, {\it FromYear} and {\it ToYear} are the publication years of two papers, and {\it SelfCite} is an indicator whether this is a self citation. 
\end{itemize}
Additionally, the file \texttt{author\_name.txt} contains all author names, in the same order as their IDs. 
The file  \texttt{BibtexInfo.RData} contains the bibtex information of papers. 

\item The adjacency matrices of  co-citation networks 
\begin{itemize}
\item \texttt{CiteeAdjFinal.mat}: The adjacency matrix of the citee network (1991-2000). This is the network used to produce the Statistics Triangle and Research Map in Section~\ref{subsec:triangle}. It has 2831 authors. To get the names of the 2831 authors, use the variable {\it keepNodeID} in \texttt{CiteeAdjFinal.mat}, as well as the file \texttt{author\_name.txt}. 
\item \texttt{CiteeDynamicFinal.mat}: The adjacency matrices of the 21 citee networks (1991-2015). These are the networks used to produce the Research Trajectories in Section~\ref{subsec:trajectory}.  
\end{itemize}

\item The adjacency matrices of co-authorship networks 
\begin{itemize}  
\item \texttt{CoauAdjFinal.mat}: The adjacency matrix of the co-authorship network (36 journals). This is the network used to obtain the community tree in Section~\ref{subsec:tree}. It has 4383 authors. 
The names of the 4383 authors are given by the variable {\it authorNames} in \texttt{CoauAdjFinal.mat}. 
\item \texttt{CoauSankeyFinal.mat}: The adjacency matrices of the 3 co-authorship networks (4 journals). This is the network used to produce the Sankey diagram in Section~\ref{subsec:sankey}. 
These sparse matrices are stored on the original 47311 authors. The variable {\it authorNames} contains the names of all 47311 authors. The variable {\it V} contains the indices of 1687 nodes used to draw the Sankey diagram. Restricting each adjacency matrix on $V$ gives the data matrices used in the paper.  
\end{itemize}

\end{itemize}

\end{document}